\RequirePackage{ifpdf}
\ifpdf % We are running pdfTeX in pdf mode
\documentclass[pdftex]{sigma}
\else
\documentclass{sigma}
\fi

\numberwithin{equation}{section}

\begin{document}

\allowdisplaybreaks

\renewcommand{\PaperNumber}{061}

\FirstPageHeading

\renewcommand{\thefootnote}{$\star$}

\ShortArticleName{Completely Integrable Systems Associated with
Classical Root Systems}

\ArticleName{Completely Integrable Systems\\
Associated with  Classical Root Systems\footnote{This paper is a
contribution to the Vadim Kuznetsov Memorial Issue `Integrable
Systems and Related Topics'. The full collection is available at
\href{http://www.emis.de/journals/SIGMA/kuznetsov.html}{http://www.emis.de/journals/SIGMA/kuznetsov.html}}}

\Author{Toshio OSHIMA}

\AuthorNameForHeading{T. Oshima}

\Address{Graduate School of Mathematical Sciences, University of
Tokyo, \\  3-8-1, Komaba, Meguro-ku, Tokyo 153-8914, Japan}

\Email{\href{mailto:oshima@ms.u-tokyo.ac.jp}{oshima@ms.u-tokyo.ac.jp}}

\URLaddress{\url{http://akagi.ms.u-tokyo.ac.jp/}}

\ArticleDates{Received December 14, 2006, in f\/inal form March
19, 2007; Published online April 25, 2007}

\Abstract{We study integrals of completely integrable quantum
systems associated with classical root systems. We review
integrals of the systems invariant under the corresponding Weyl
group and as their limits we construct enough integrals of the
non-invariant systems, which include systems whose complete
integrability will be f\/irst established in this paper. We also
present a conjecture claiming that the quantum systems with enough
integrals given in this note coincide with the systems that have
the integrals with constant principal symbols corresponding to the
homogeneous generators of the $B_n$-invariants. We review
conditions supporting the conjecture and give a new condition
assuring it.}

\Keywords{completely integrable systems; Calogero--Moser systems;
Toda lattices with boun\-dary conditions}

\Classification{81R12; 70H06}

\begin{flushright}
\textit{To the memory of Professor Vadim B.~Kuznetsov}
\end{flushright}

%\newpage

%\tableofcontents

%\newpage

\section{Introduction}
A Schr\"odinger operator
\begin{gather}
  P=-\frac12\sum\limits_{j=1}^n\frac{\partial^2}{\partial x_j^2} + R(x)
\label{eq:Shr}
\end{gather}
with the potential function $R(x)$ of $n$ variables
$x=(x_1,\dots,x_n)$ is called {\it completely integrable} if there
exist $n$ dif\/ferential operators $P_1,\dots,P_n$ such that
\begin{gather}
   [P_i,P_j]=0\qquad(1\le i < j\le n),
   \nonumber\\
   P\in\mathbb C[P_1,\dots,P_n],\label{eq:CI}
   \\
   P_1,\dots,P_n\quad \text{are algebraically independent}.\nonumber
\end{gather}
In this paper, we explicitly construct the integrals
$P_1,\ldots,P_n$ for completely integrable potential functions
$R(x)$ of the form
\begin{gather}
\label{eq:typeBn}
  R(x) =\sum\limits_{1\le i<j\le n}(u_{ij}^-(x_i-x_j)+u_{ij}^+(x_i+x_j)) + \sum\limits_{k=1}^n v_k(x_k)
\end{gather}
appearing in other papers. The Schr\"odinger operators with these
commuting dif\/ferential operators treated in this paper include
Calogero--Moser--Sutherland systems (cf.\ \cite{Calogero, Moser,
Olshanetsky1981, Olshanetsky1983, Sutherland}), Heckman--Opdam's
hypergeometric systems (cf.\ \cite{Sekiguchi} for type $A_{n-1}$,
\cite{Heckman} in general), their extensions (cf.\ \cite{Adler,
Diejen, Levi, Ochiai1996, Ochiai2003, Ochiai1994, Oshima2005}) and
f\/inite Toda lattices corresponding to (extended) Dynkin diagrams
for classical root systems
(cf.~\cite{Bogoyavlensky, Goodman1982, %Goodman1984,
Goodman1986, Kostant, Olshanetsky1979, Ruijsenaars, Toda}) and
those with boundary conditions (cf.~\cite{Diejen,
Inozemtsev1989-2, Kuznetsov1997, Kuznetsov1995, Kuznetsov1992,
Kuznetsov1994, Ochiai1996, Oshima2005}).

Put $\partial_j=\partial/{\partial x_j}$ for simplicity. We denote
by  $\sigma(Q)$ the principal symbol of a dif\/ferential operator
of $Q$. For example, $\sigma(P)=-(1/2)(\xi_1^2+\dots+\xi_n^2)$.

We note that \cite{Wakida} proves that the potential function is
of the form \eqref{eq:typeBn} if
\begin{gather}
\label{eq:typeBnI}
 \sigma(P_k)
 =\sum\limits_{1\le j_1<\dots<j_k\le n}\xi_{j_1}^2\cdots \xi_{j_k}^2
  \qquad\text{for}\qquad k=1,\dots,n.
\end{gather}
In this case we say that $R(x)$ is an integrable potential
function of {\it type\/} $B_n$ or of the {\it classical type}.
Moreover when $R(x)$ is symmetric with respect to the coordinate
$(x_1,\dots,x_n)$ and invariant under the coordinate
transformation $(x_1,x_2,\dots,x_n)\mapsto (-x_1,x_2,\dots,x_n)$,
then $R(x)$ is determined by~\cite{Oshima1995} for $n\ge 3$ and
by~\cite{Ochiai2003} for $n=2$ and $P_k$ are calculated by
\cite{Oshima1998}.

Classif\/ications of the integrable potential functions under
certain conditions are given in \cite{Ochiai1996, Ochiai2003,
Ochiai1994, Oshima2005, Taniguchi, Wakida} etc. In
Section~\ref{sec:class} we review them and we present Conjecture which
claims that the potential functions given in this note exhaust
those of the completely integrable systems satisfying
\eqref{eq:typeBnI}. We also give a new condition which assures
Conjecture.

If $v_k=0$ for $k=1,\dots,n$, we can expect
$\sigma(P_k)=\sum\limits_{1\le j_1<\dots<j_k\le
n}\xi_{j_1}^2\cdots \xi_{j_k}^2$ for $k=1,\dots,n-1$ and
$\sigma(P_n)=\xi_1\xi_2\cdots\xi_n$ and we say the integrable
potential function is of {\it type\/} $D_n$. If $v_k=0$ and
$u_{ij}^+=0$ for $k=1,\dots,n$ and $1\le i<j\le n$, we can expect
$P_1=\partial_1+\cdots+\partial_n$, $\sigma(P_k)=\sum\limits_{1\le
j_1<\dots<j_k\le n}\xi_{j_1}\cdots \xi_{j_k}$ for $k=2,\dots,n$
and we say that the integrable potential function is of {\it
type\/} $A_{n-1}$. Note that the integrable potential function of
type $A_{n-1}$ or $D_n$ is of type $B_n$.

The elliptic potential function of type $A_{n-1}$ with
\begin{gather*}
  u_{ij}^-(t) = C\wp(t;2\omega_1,2\omega_2)+C',\qquad u_{ij}^+(t)=v_k(t) = 0\qquad
  (C,\,C'\in\mathbb C)
\end{gather*}
(cf.~\cite{Olshanetsky1983}) and that of type $B_n$ with
\begin{gather*}
  u_{ij}^-(t) = v_{ij}^+(t) = A\wp(t;2\omega_1,2\omega_2),
 \nonumber \\
  v_k(t) = \displaystyle\sum\limits_{j=0}^3 C_j\wp(t+\omega_j;2\omega_1,2\omega_2)
  -\frac C2,
\qquad
  (A,\,C_i,\,C\in\mathbb C)
\end{gather*}
introduced by \cite{Inozemtsev1989-1} are most fundamental and
their integrability and the integrals of higher order are
established by \cite{Ochiai1994, Oshima1998, Oshima1995}. Here
$\wp(t;2\omega_1,2\omega_2)$ is the Weierstrass elliptic function
whose fundamental periods are $2\omega_1$ and $2\omega_2$ and
\begin{gather*}
  \omega_0=0,\qquad \omega_1+\omega_2+\omega_3=0.
\end{gather*}

Other potential functions are suitable limits of these elliptic
potential functions, which is shown in \cite{Diejen,
Inozemtsev1989-2, Ruijsenaars} etc. We will study integrable
systems by taking analytic continuations of the integrals given in
\cite{Ochiai1994} with respect to a suitable parameter, which is
done for the invariant systems (of type $A_{n-1}$) by
\cite{Oshima1995} and (of types $B_n$ and $D_n$) by
\cite{Oshima1998} and for the systems of type~$A_{n-1}$
by~\cite{Ruijsenaars}. The main purpose of this note is to give
the explicit expression of the operators $P_1,\dots,P_n$
in~\eqref{eq:CI} in this unif\/ied way. Namely we construct enough
commuting integrals of the non-invariant systems from those of the
invariant systems given by \cite{Ochiai2003, Oshima1998,
Oshima1995}. Such study of the systems of types $A_{n-1}$, $B_2$,
$B_n$ ($n\ge3$) and $D_n$ are explained in Sections~\ref{sec:An},
\ref{sec:B2}, \ref{sec:Bn}, \ref{sec:Dn}, respectively.

Since the integrals of the system of type $A_{n-1}$ are much
simpler than those of type $B_n$, we review the above analytic
continuation for the systems of type $A_{n-1}$ in Section~\ref{sec:An}
preceding to the study for the systems of type $B_n$. There are
many series of completely integrable systems of type $B_2$, which
we review and classify in Section~\ref{sec:B2} with taking account of
the above unif\/ied way.

We present 8 series of potential functions of type $B_n$ in
Section~\ref{sec:Bn}. There are 3 (elliptic, trigonometric or
hyperbolic and rational) series of the invariant potentials of
type $B_n$ whose enough integrals are constructed by
\cite{Ochiai1994} and \cite{Oshima1998}. The complete
integrability of the remaining 5 series of the potential functions
is shown in Section~\ref{sec:Bn}, which is conjectured by~\cite{Diejen}
(4 series), partially proved by \cite{Kuznetsov1997,
Kuznetsov1995, Kuznetsov1994} (3~series) and announced by
\cite{Oshima2005} (5~series). The complete integrability of two
series among them seems to be f\/irst established in this note.
Note that when $n\ge3$, our systems which do not belong to these 8
series of type $B_n$ are the Calogero--Moser systems with elliptic
potentials and the f\/inite Toda lattices of type $A_{n-1}^{(1)}$,
whose complete integrability is known.

The main purpose of our previous study in \cite{Ochiai1996,
Ochiai2003, Ochiai1994, Oshima2005} is
 classif\/ication of the completely integrable systems associated with
classical root systems. In this note we explicitly give integrals
of all the systems classif\/ied in our previous study with
reviewing known integrals.

Since our expression of $P_k$ is natural, we can easily def\/ine
their classical limits without any ambiguity and get completely
integrable Hamiltonians of dynamical systems together with their
enough integrals. This is clarif\/ied in Section~\ref{sec:dynam}.

In Section~\ref{sec:ODE} we examine  ordinary dif\/ferential operators
which are analogues of the Schr\"odinger operators studied in this
note.

\section{Notation and preliminary results}
Let $\{e_1,\dots,e_n\}$ be the natural orthonormal base of the
Euclidean space $\mathbb R^n$ with the inner product
\begin{gather*}
 \langle x,y \rangle = \sum\limits_{j=1}^n x_jy_j
 \qquad\text{for}\qquad x=(x_1,\dots,x_n),\  y=(y_1,\dots,y_n)\in\mathbb R^n.
\end{gather*}
Here $e_j=(\delta_{1j},\dots,\delta_{nj})\in\mathbb R^n$ with
Kronecker's delta $\delta_{ij}$.

Let $\alpha\in\mathbb R^n\setminus\{0\}$. The ref\/lection
$w_\alpha$ with respect to $\alpha$ is a linear transformation of
$\mathbb R^n$ def\/ined by $w_\alpha(x)=x
-\frac{2\langle\alpha,x\rangle}{\langle
\alpha,\alpha\rangle}\alpha$ for $x\in\mathbb R^n$. Furthermore we
def\/ine a dif\/ferential operator $\partial_\alpha$ by
\begin{gather*}
 (\partial_\alpha \varphi)(x) = \frac d{dt}\varphi(x+t\alpha)\biggl|_{t=0}
\end{gather*}
and then $\partial_j=\partial_{e_j}$.

The root system $\Sigma(B_n)$ of type $B_n$ is realized in
$\mathbb R^n$ by
\begin{gather*}
 \Sigma(A_{n-1})^+ = \{e_i-e_j;\, 1 \le i<j\le n\},
 \\
 \Sigma(D_n)^+ = \{e_i \pm e_j;\, 1\le i<j\le n\},
 \\
 \Sigma(B_n)_S^+ = \{e_k;\, 1\le k\le n\},
 \\
 \Sigma(B_n)^+ = \Sigma(D_n)^+\cup\Sigma(B_n)^+_S,
 \\
 \Sigma(F) = \{\alpha,-\alpha;\,\alpha\in\Sigma(F)^+\}\qquad\text{for}\quad
 F=A_{n-1},\quad D_n\quad\text{or}\quad B_n.
\end{gather*}
The Weyl groups $W(B_n)$ of type $B_n$, $W(D_n)$ of type $D_n$ and
$W(A_{n-1})$ of type $A_{n-1}$ are the groups generated by
$w_\alpha$ for $\alpha\in\Sigma(B_n)$, $\Sigma(D_n)$ and
$\Sigma(A_{n-1})$, respectively. The Weyl group $W(A_{n-1})$ is
naturally identif\/ied with the permutation group $\mathfrak S_n$
of the set $\{1,\dots,n\}$ with $n$ elements. Let $\epsilon$ be
the group homomorphism of $W(B_n)$ def\/ined by
\begin{gather*}
  \epsilon(w) =
  \left\{ \begin{array}{rll}
      1&\text{if}& w\in W(D_n),
      \\[.5ex]
      -1&\text{if}& w\in W(B_n)\setminus W(D_n).
   \end{array}\right.
\end{gather*}

The potential function \eqref{eq:typeBn} is of the form
\begin{gather*}
  R(x) = \sum\limits_{\alpha\in\Sigma(D_n)^+}u_\alpha(\langle\alpha,x\rangle)
        + \sum\limits_{\alpha\in\Sigma(B_n)_S^+}v_\beta(\langle\beta,x\rangle)
\end{gather*}
with functions $u_\alpha$ and $v_\beta$ of one variable. For
simplicity we will denote
\begin{gather*}
  u_\alpha(x) = u_{-\alpha}(x) = u_\alpha(\langle\alpha,x\rangle)
\qquad\text{for}\quad\alpha\in\Sigma(D_n)^+,
   \\
  v_\beta(x)  = v_{-\beta}(x) = v_\beta(\langle\beta,x\rangle)
\qquad\text{for}\quad
  \beta\in\Sigma(B_n)_S^+,
  \\
  u^{\pm}_{ij}(x) = u_{e_i\pm e_j}(x),\qquad
  v_k(x) = v_{e_k}(x).
\end{gather*}
\begin{lemma}
\label{lem:shcont} For a bounded open subset $U$ of $\mathbb C$,
there exists an open neighborhood $V$ of $0$ in $\mathbb C$ such
that the following statements hold.

{\rm i)} The function $\lambda \sinh^{-1}\lambda z$ is
holomorphically extended to $(z,\lambda)\in
(U\setminus\{0\})\times V$ and the function is $1/z$ when
$\lambda=0$.

{\rm ii)} Suppose $\mathrm{Re}\, \lambda>0$. Then the functions
\begin{gather*}
e^{2\lambda t}\sinh^{-2}\lambda(z\pm t)\qquad\text{and}\qquad
e^{4\lambda t}\bigl(\sinh^{-2}\lambda(z\pm
t)-\cosh^{-2}\lambda(z\pm t)\bigr)
\end{gather*}
are holomorphically  extended to $(z,q)\in U\times V$ with
$q=e^{-2\lambda t}$ and the functions are $4e^{\mp 2\lambda z}$
and $16e^{\mp 4\lambda z}$, respectively, when $q=0$.
\end{lemma}
\begin{proof}
The claims are clear from
\begin{gather*}
 \lambda^{-1}\sinh\lambda z = z +
 \sum\limits_{j=1}^\infty\frac{\lambda^{2j}z^{2j+1}}{(2j+1)!},
 \\
 4e^{-2\lambda t}\sinh^2\lambda(z\pm t) =
  e^{\pm 2\lambda z}(1 - e^{-2\lambda t}e^{\mp2\lambda z})^2,
  \\
 \sinh^{-2}\lambda z - \cosh^{-2}\lambda z = 4\sinh^{-2}2\lambda z.\tag*{\qed}
\end{gather*}\renewcommand{\qed}{}
\end{proof}

The elliptic functions $\wp$ and $\zeta$ of Weierstrass type are
def\/ined by
\begin{gather*}
\wp(z) =
\wp(z;2\omega_1,2\omega_2)=\frac1{z^2}+\sum\limits_{\omega\neq0}
 \biggl(\frac1{(z-\omega)^2}-\frac1{\omega^2}\biggr),
 \\
\zeta(z) = \zeta(z;2\omega_1,2\omega_2)=\frac1z +
\sum\limits_{\omega\neq0}\biggl(\frac1{z-\omega} + \frac1\omega +
\frac z{\omega^2}\biggr),
\end{gather*}
where the sum ranges over all non-zero periods $2m_1\omega_1 +
2m_2\omega_2$
 ($m_1$, $m_2\in\mathbb Z$) of $\wp$.
The following are some elementary properties of these functions
(cf.\ \cite{Whittaker}).
\begin{gather}
\wp(z)  = \wp(z+2\omega_1) = \wp(z+2\omega_2), \label{eq:wpperiod}
\\
\zeta'(z)  = -\wp(z),
\\
(\wp')^2
=4\wp^3-g_2\wp-g_3=4(\wp-e_1)(\wp-e_2)(\wp-e_3),\nonumber
\\
e_\nu =\wp(\omega_\nu)\qquad\text{for}\qquad\nu=1,2,3,\qquad
\omega_3=-\omega_1-\omega_2\qquad\text{and}\qquad  \omega_0=0,
\label{eq:wpdifeq}
\\
 \wp(2z)  = \frac14\sum\limits_{\nu=0}^4\wp(z+\omega_\nu)
= \frac{(12\wp(z)^2-g_2)^2}{16\wp'(z)^2}-2\wp(z),\nonumber
%\label{eq:wp2}
\\
\wp(z; 2\omega_2, 2\omega_1)  = \wp(z; 2\omega_1, 2\omega_2),
\\
\wp(z+\omega_1; 2\omega_1, 2\omega_2)
 =e_1+\frac{(e_1-e_2)(e_1-e_3)}{\wp(z; 2\omega_1, 2\omega_2)-e_1},
\label{eq:wprat}
\\
 \wp(z;\sqrt{-1}\lambda^{-1}\pi,\infty)  = \lambda^2\sinh^{-2}\lambda z + \frac13{\lambda^2},
 \label{eq:wpt}
\\
 \wp(z;\infty,\infty)  = z^{-2},
 \label{eq:wpr}
\\
\wp(z; \omega_1, 2\omega_2) = \wp(z; 2\omega_1, 2\omega_2)+
\wp(z+\omega_1; 2\omega_1, 2\omega_2)-e_1, \label{eq:Landen}
\\
 \left|\begin{matrix}
  \wp(z_1)   &\wp'(z_1)   &1\\
  \wp(z_2)   &\wp'(z_2)   &1\\
  \wp(z_3)   &\wp'(z_3)   &1
 \end{matrix}\right| =0\qquad\text{if}\qquad z_1+z_2+z_3=0,
 \label{eq:wpDadd}
 \\
\wp(z;2\omega_1,2\omega_2) =-\frac{\eta_1}{\omega_1} +
\lambda^2\sinh^{-2}\lambda z + \sum\limits_{n=1}^\infty
\frac{8n\lambda^2e^{-4n\lambda\omega_2}}{1-e^{-4n\lambda\omega_2}}
\cosh2n\lambda z,\nonumber
\\
\eta_1  = \zeta(\omega_1;2\omega_1,2\omega_2)=
\frac{\pi^2}{\omega_1}\left(
\frac1{12}-2\sum\limits_{n=1}^\infty\frac{ne^{-4n\lambda\omega_2}}
{1-e^{-4n\lambda\omega_2}}\right),\nonumber
\\
 \tau  = \frac{\omega_2}{\omega_1},\qquad q=e^{\pi i\tau}=e^{-2\lambda\omega_2}
\qquad\text{and}\qquad\lambda=\frac{\pi}{2\sqrt{-1}\omega_1}.
\label{eq:wpFourier}
\end{gather}
Here the sums in \eqref{eq:wpFourier} converge if
\begin{gather*}
  2\mathrm{Im}\,\frac{\omega_2}{\omega_1} > \frac{|z|}{|\omega_1|}.
\end{gather*}
%%%
Let $0\le k<2m$.  Then \eqref{eq:wpFourier} means
\begin{gather*}
\wp\left(z+\frac
km\omega_2;2\omega_1,2\omega_2\right)=-\frac{\eta_1}{\omega_1}
+4\lambda^2\left(\frac{q^{k/m}e^{-2\lambda
z}}{\bigl(1-e^{-2\lambda z}q^{k/m}\bigr)^2}
 + \sum\limits_{n=1}^\infty \frac{q^{n(2-k/m)} e^{2n\lambda z}}{1- q^{2n}}\right),
 \\
-\frac{\eta_1}{\omega_1}= 4\lambda^2\left(\frac1{12} -
2\sum\limits_{n=1}^\infty \frac{nq^{2n}}{1-q^{2n}}\right).
\end{gather*}
\begin{lemma}
\label{lem:wpcont} Let $k$ and $m$ be integers satisfying
$0<k<2m$. Put
\begin{gather*}
\wp_0(z;2\omega_1,2\omega_2) = \wp(z;2\omega_1,2\omega_2) +
\frac{\eta_1}{\omega_1},\nonumber
\\
\lambda=\frac{\pi}{2\sqrt{-1}\omega_1} \qquad\text{and}\qquad
t=q^{1/m} = e^{{\pi i\omega_2}/(m\omega_1)}.
\end{gather*}
Then for any bounded open set $U$ in $\mathbb C\times \mathbb C$,
there exists a neighborhood $V$ of the origin of $\mathbb C$ such
that the following statements hold.

{\rm i)} $\wp_0(z;2\omega_1,2\omega_2) -
\lambda^2\sinh^{-2}\lambda z$ and
$\wp_0(z+\omega_1;2\omega_1,2\omega_2) +
\lambda^2\cosh^{-2}\lambda z$ are holomorphic functions of
$(z,\lambda, q)\in U\times V$ and vanish when $q=0$.

{\rm ii)} $\wp_0(z+(k/m)\omega_2;2\omega_1,2\omega_2)$ is
holomorphic for $(z,\lambda, t)\in U\times V$ and has zeros of
order $\min\{k, 2m-k\}$ along the hyperplane defined by $t=0$ and
satisfies
\begin{alignat*}{3}
& t^{-k}\wp_0\biggl(z+\dfrac
km\omega_2;2\omega_1,2\omega_2\biggr)\bigg|_{t=0}
 =4\lambda^2 e^{-2\lambda z} \qquad && (0<k<m),&
 \\
& t^{-k}\wp_0\biggl(z+\dfrac
km\omega_2;2\omega_1,2\omega_2\biggr)\bigg|_{t=0}
 =8\lambda^2 \cosh2\lambda z \qquad && (k=m),&
 \\
 & t^{k-2m}\wp_0\biggl(z+\dfrac km\omega_2;2\omega_1,2\omega_2\biggr)\bigg|_{t=0}
 =4\lambda^2 e^{2\lambda z} \qquad && (m<k<2m).&
\end{alignat*}
\end{lemma}

For our later convenience we list up some limiting formula
discussed above. Fix $\omega_1$ with $\sqrt{-1}\omega_1>0$ and let
$\omega_2\in\mathbb R$ with $\omega_2>0$. Then
$\lambda={\pi}/({2\sqrt{-1}\omega_1})>0$ and
\begin{gather}
 \sinh^2\lambda(z+\omega_1) = -\cosh^2\lambda z,\qquad
 \cosh2\lambda(z+\omega_1) = -\cosh2\lambda z,
\\
 \lim_{\lambda\to0}\lambda^2\sinh^{-2}\lambda z = \frac1{z^2},
\\
 \lim_{N\to\pm\infty}e^{2\lambda|N|}\sinh^{-2}{\lambda(z+N)}
 = 4e^{\mp2\lambda z},
 \label{eq:sh-2rat}
\\
 \lim_{\omega_2\to+\infty}\wp_0(z;2\omega_1,2\omega_2)
 = \lambda^2\sinh^{-2}\lambda z
 \label{eq:wp2sh-2},
\\
 \lim_{\omega_2\to+\infty}\wp_0(z+\omega_1;2\omega_1,2\omega_2)
 = -\lambda^2\cosh^{-2}\lambda z,
 \label{eq:wp2ch-2}
\\
\lim_{\omega_2\to\infty}e^{2r\lambda\omega_2}
\wp_0(z+r\omega_2;2\omega_1,2\omega_2) =4\lambda^2e^{-2\lambda z}
\qquad\text{if}\qquad 0<r<1,
\\
 \lim_{\omega_2\to\infty}e^{2\lambda\omega_2}
  \wp_0(z+\omega_2;2\omega_1,2\omega_2)
  =8\lambda^2\cosh2\lambda z
  \label{eq:wp2ch},
\\
 \lim_{\omega_2\to\infty}e^{2(2-r)\lambda\omega_2}
  \wp_0(z+r\omega_2;2\omega_1,2\omega_2) =4\lambda^2e^{2\lambda z}
 \qquad\text{if}\qquad 1<r<2.
\end{gather}
%%%%%%%%%%%%%%%%%%%%%%%%%%%%%%%%%%%%%%%%%%%%%%%%%%%%%%%%%%%%%%%%%
\section[Type $A_{n-1}$ $(n\ge 3)$]{Type $\boldsymbol{A_{n-1}}$ $\boldsymbol{(n\ge 3)}$}\label{sec:An}

The completely integrable Schr\"odinger operator of type $A_{n-1}$
is of the form
\begin{gather*}
 P = -\frac12\sum\limits_{j=1}^n\frac{\partial^2}{\partial x_j^2} + \sum\limits_{1\le i<j\le n}
   u_{ij}^-(x_i-x_j).
\end{gather*}
Denoting
\begin{gather*}
  u_{e_i-e_j}(x) = u_{e_j-e_i}(x) = u_{ij}^-(x_i-x_j),
\end{gather*}
we put
\begin{gather}
 P_k = \!\sum\limits_{0\le \nu\le k/2}\!\!\frac{1}{2^\nu \nu!(k\!-\!2\nu)!(n\!-\!k)!}
\!\!\sum\limits_{w\in\mathfrak S_n}\!\!u_{w(e_1\!-
e_2)}u_{w(e_3\!- e_4)}\cdots u_{w(e_{2\nu\!-\!1}\!-
e_{2\nu})}\partial_{w(e_{2\nu\!+\!1})}\cdots\partial_{w(e_k)}\nonumber
\\ \phantom{P_k}{}
 = \sum\limits_{0\le \nu\le k/2}
\sum u_{i_1i_2}^- \cdots u_{i_{2\nu-1}i_{2\nu}}^-
\partial_{i_{2\nu+1}}\cdots\partial_{i_k}
\label{eq:AnHInt}
\end{gather}
according to the integrals given in \cite{Ochiai1994, Oshima1995}.
We will examine the functions $u^-_{ij}(t)$ which satisfy
\begin{gather}
\label{eq:AnInt}
  [P_i,P_j] = 0\qquad\text{for}\qquad 1\le i<j\le n.
\end{gather}
Here we note that
\begin{gather*}
  P   = P_2 - \dfrac12P_1^2,
\qquad
  P_1 = \partial_1+\cdots+\partial_n,
\\
  P_2 = \sum\limits_{1\le i<j\le n}\partial_i\partial_j+\sum\limits_{1\le i<j\le n}u_{ij}^-(x_i-x_j),
\\
  P_3 = \sum\limits_{1\le i<j<k\le n}\partial_i\partial_j\partial_k
    + \sum\limits_{k=1}^n\sum\limits_{\substack{1\le i<j\le n\\i\ne k,\ j\ne k}}
       u_{ij}^-(x_i-x_j)\partial_k.
\\
  P_4 = \!\!\sum\limits_{1\le i<j<k<\ell\le n}\!\!\!\!\!\!\partial_i\partial_j\partial_k\partial_\ell+
       \sum\limits_{1\le k<\ell\le n}\!
       \sum\limits_{\substack{1\le i<j\le n\\i\ne k,\ell,\ j\ne k,\ell}}\!\!\!\!\!
       u_{ij}^-\partial_k\partial_\ell+\!\!\sum\limits_{1\le i<j<k<\ell\le n}\!\!\!\!
     (u_{ij}^-u_{k\ell}^-+u_{ik}^-u_{j\ell}^-+u_{i\ell}^-u_{jk}^-)
\\
\phantom{P_4}{}
      =\sum\partial_i\partial_j\partial_k\partial_\ell+\sum u^-_{ij}\partial_k\partial_\ell
        + \sum u^-_{ij}u^-_{k\ell},
\\
  P_5 = \sum \partial_i\partial_j\partial_k\partial_\ell\partial_m
        +\sum u^-_{ij}\partial_k\partial_\ell\partial_m+\sum u^-_{ij}u^-_{k\ell}\partial_m,
\\
 P_6 = \sum \partial_i\partial_j\partial_k\partial_\ell\partial_m\partial_\nu
        +\sum u^-_{ij}\partial_k\partial_\ell\partial_m\partial_\nu
        +\sum u^-_{ij}u^-_{k\ell}\partial_m\partial_\nu
        +\sum u^-_{ij}u^-_{k\ell}u^-_{m\nu},\\
  \cdots \cdots \cdots \cdots \cdots\cdots
\end{gather*}
Since $W(A_{n-1})$ is naturally isomorphic to the
permutation group $\mathfrak S_n$ of the set $\{1,\dots,n\}$, we
will identify them. In \cite{Ochiai1994, Oshima1995}, the
integrable potentials of type $A_{n-1}$ which are invariant under
the action of $\mathfrak S_n$ are determined and moreover
\eqref{eq:AnInt} with \eqref{eq:AnHInt} is proved. They are
\begin{gather}
\label{eq:uij}
  u_{e_i-e_j}(x) = u(\langle e_i-e_j,x\rangle)
\end{gather}
with an even function $u$ and \vspace{1ex}

\noindent (Ellip-$A_{n-1}$) {\it Elliptic potential of type
$A_{n-1}$}:
\begin{gather*}
 u(t) = C\wp_0(t;2\omega_1,2\omega_2),
 \\
 R_E(A_{n-1};x_1,\ldots,x_n;C,2\omega_1,2\omega_2)
  = C\sum\limits_{1\le i<j\le n}\wp_0(x_i-x_j;2\omega_1,2\omega_2),
\end{gather*}
(Trig-$A_{n-1}$) {\it Trigonometric potential of type $A_{n-1}$}:
\begin{gather*}
 u(t) = C\sinh^{-2}\lambda t,
 \qquad
 R_T(A_{n-1};x_1,\ldots,x_n;C,\lambda) = C\sum\limits_{1\le i<j\le n}\sinh^{-2}\lambda(x_i-x_j),
\end{gather*}
(Rat-$A_{n-1}$) {\it Rational potential of type $A_{n-1}$}:
\begin{gather*}
 u(t) = \frac C{t^2},
 \qquad
 R_R(A_{n-1};x_1,\ldots,x_n;C)
  = \sum\limits_{1\le i<j\le n}\frac{C}{(x_i-x_j)^2}.
\end{gather*}

We review how the integrability of (Ellip-$A_{n-1}$) implies the
integrability of other systems.

Since it follows from \eqref{eq:wp2sh-2} that
\begin{gather*}
\lim_{\omega_2\to\infty}
R_E\biggl(A_{n-1};x;\frac{C}{\lambda^2},2\omega_1,2\omega_2\biggr)=R_T(A_{n-1};x;C,\lambda),\nonumber
\\
u(t) =
\lim_{\omega_2\to\infty}\frac{C}{\lambda^2}\wp_0(t;2\omega_1,2\omega_2)=
C\sinh^{-2}\lambda t,
\end{gather*}
the integrability \eqref{eq:AnInt} for (Trig-$A_{n-1}$) follows
from that for (Ellip-$A_{n-1}$) by the analytic conti\-nuation of
$u_{e_i-e_j}(x)$ and $P_k$ with respect to $q$
(cf.~\eqref{eq:wpFourier}, \eqref{eq:AnHInt}, \eqref{eq:uij} and
Lemma~\ref{lem:wpcont} i)).

The integrability for (Rat-$A_{n-1}$) is similarly follows in view
of Lemma~\ref{lem:shcont} with
\begin{gather*}
\lim_{\lambda\to0}R_T(A_{n-1};x;\lambda^2 C,\lambda)=
R_R(A_{n-1};x;C),\nonumber
\\
u(t) = \lim_{\lambda\to0}\lambda^2 C\sinh^{-2}\lambda t.
\end{gather*}
This argument using the analytic continuation for the proof of
\eqref{eq:AnInt} is given in \cite{Oshima1995}.

As is shown \cite{Ruijsenaars}, there are two other integrable
potentials of type $A_{n-1}$ to which this argument can be
applied.  \vspace{1ex}

\noindent (Toda-$A_{n-1}^{(1)}$) {\it Toda potential of type
$A_{n-1}^{(1)}$}:
\begin{gather*}
R_L(A_{n-1}^{(1)};x;C,\lambda)=
\sum\limits_{i=1}^{n-1}Ce^{\lambda(x_i-x_{i+1})} + Ce^{\lambda
(x_n-x_1)},
\end{gather*}
(Toda-$A_{n-1}$) {\it Toda potential of type $A_{n-1}$}:
\begin{gather*}
 R_L(A_{n-1};x;C,\lambda) = \sum\limits_{i=1}^{n-1}Ce^{\lambda(x_i-x_{i+1})}.
\end{gather*}
The integrability \eqref{eq:AnInt} for these potentials similarly
follows in view of Lemmas~\ref{lem:shcont}, \ref{lem:wpcont} and
\begin{gather*}
 \lim_{\omega_2\to\infty}
 R_E\biggl(A_{n-1}; x_1 - \frac{2\omega_2}n,\ldots,x_k - \frac{2k\omega_2}n,
  \ldots,x_n -2\omega_2;\frac{e^{(4/n)\lambda\omega_2}}{4\lambda^2}C,2\omega_1,2\omega_2\biggr)
  \\ \qquad
  =R_L(A_{n-1}^{(1)};x;C,-2\lambda),
 \\
 u_{e_i-e_j}(x) =\lim_{\omega_2\to\infty}
 \frac{e^{(4/n)\lambda\omega_2}}{4\lambda^2}C\wp_0\biggl(x_i-x_j+\frac{2(j-i)\omega_2}n;2\omega_1,2\omega_2\biggr)
 \\ \phantom{ u_{e_i-e_j}(x)}{}
  =\begin{cases}
    Ce^{-2\lambda(x_i-x_{i+1})} &\text{if}\quad 1< j=i+1\le n,
    \\
    Ce^{-2\lambda(x_n-x_1)} &\text{if}\quad i=1\quad\text{and}\quad j=n,
    \\
    0 &\text{if}\quad 1\le i<j\le n\quad\text{and}\quad j-i\ne1,\,n-1
   \end{cases}
\end{gather*}
and
\begin{gather*}
 \lim_{N\to\infty}
 R_T\biggl(A_{n-1}; x_1-N,\ldots,x_n-nN;\frac{e^{2\lambda N}}4C,\lambda\biggr)
 = R_L(A_{n-1};x;C,-2\lambda),
 \\
 u_{e_i-e_j}(x)  =  \lim_{N\to\infty}\frac{e^{2\lambda N}}4C
 \sinh^{-2}\lambda(x_i-x_j+(j-i)N)
 \\ \phantom{u_{e_i-e_j}(x)}{}
  =\begin{cases}
    Ce^{-2\lambda(x_i-x_{i+1})} &\text{if}\quad 1< j=i+1\le n,
\\
    0 &\text{if}\quad 1\le i<j\le n\quad \text{and}\quad j\ne i+1,
   \end{cases}
\end{gather*}
respectively, if $\mathrm{Re}\,\lambda >0$. The restriction
$\mathrm{Re}\,\lambda >0$ is removed also by the analytic
continuation.

Thus the following theorem is obtained by the analytic
continuation of the integrals \eqref{eq:AnHInt} of
(Ellip-$A_{n-1}$) whose commutativity \eqref{eq:AnInt} is assured
by \cite{Oshima1995}.
\begin{theorem}
[$\boldsymbol{A_{n-1}}$, \cite{Goodman1986, Heckman,
Olshanetsky1983, Oshima1995, Ruijsenaars, Sekiguchi}, etc.] The
Schr\"odinger operators with the potential functions {\rm
(Ellip-$A_{n-1}$), (Trig-$A_{n-1}$), (Rat-$A_{n-1}$),
(Toda-$A_{n-1}^{(1)}$)} and {\rm(Toda-$A_{n-1}$)} are completely
integrable and their integrals are given by \eqref{eq:AnHInt} with
$u_{e_i-e_j}(x)$ in the above.
\end{theorem}
%%%
\begin{remark}\label{rem:An}
{\rm i)} There are quite many papers studying these Schr\"odinger
operators of type $A_{n-1}$. The proof of this theorem using
analytic continuation is explained in \cite{Ruijsenaars}.

{\rm ii)} The complete integrability \eqref{eq:AnInt} for
(Ellip-$A_{n-1}$) is f\/irst established by
\cite[Theorem~5.2]{Oshima1995}, whose proof is as follows. The
equations $[P_1,P_k]=[P_2,P_k]=0$ for $k=1,\dots,n$ are easily
obtained by direct calculations with the formula \eqref{eq:wpDadd}
(cf.~\cite[Lemma~5.1]{Oshima1995}). Then the relation
$[P_2,[P_i,P_j]]=0$ and periodicity and symmetry of $P_k$ imply
$[P_i,P_j]=0$ (cf.\ \cite[Lemma~3.5]{Oshima1995}). Note that the
proof of the integrability given in \cite[\S~5 Proposition 2 and
Appendix E]{Olshanetsky1983} is not correct as is clarif\/ied in
\cite[Remark 3.7]{Oshima1995} (cf.~\cite[\S~4.2]{Ruijsenaars}).

Note that the complete integrability for (Trig-$A_{n-1}$) is shown
in \cite{Sekiguchi}.

{\rm iii)} If $\mathrm{Re}\,\lambda>0$, we also have
\begin{gather*}
 \lim_{N\to\infty} R_L\big(A_{n-1}^{(1)};x_k+kN;Ce^{-2\lambda N};-2\lambda\big)
 =R_L(A_{n-1};x;C,-2\lambda).
\end{gather*}

{\rm iv)} Note that
\begin{gather*}
 R_L(A_{n-1};x_k + N_k;C,\lambda)
 =\sum\limits_{i=1}^{n-1}Ce^{\lambda(N_i-N_{i+1})}\cdot e^{\lambda(x_i-x_{i+1})}.
\end{gather*}
Hence $e^{\lambda(x_1-x_2)}-e^{\lambda(x_2-x_3)}$ gives the
potential function of a completely integrable system of
type~$A_{n-1}$ with $n=3$ but the potential function{\samepage
\begin{gather*}
\lim_{\lambda\to0}\lambda^{-1}\big(e^{\lambda(x_1-x_2)}-e^{\lambda(x_2-x_3)}\big)
= x_1-2x_2+x_3
\end{gather*}
does not give such a system because it does not satisfy
\eqref{eq:A2} in Remark~\ref{rem:A2}.}

Considering the limit of the parameters of the integrable
potential function, we should take care of the limit of integrals.

{\rm v)} Let $P_n(t)$ denote the dif\/ferential operator $P_n$ in
\eqref{eq:AnHInt} def\/ined by replacing $u_{ij}^-$ by $\tilde
u_{ij}^-=u_{ij}^-+t$ with a constant $t\in\mathbb C$. Then
\begin{gather}
 P_n(t) = \sum\limits_{0\le k\le n/2}\frac{(2k)!}{2^kk!} P_{n-2k}t^k
 \qquad\text{with}\qquad P_0=1,\nonumber
 \\
 [P_n(s), P_n(t)] = 0\qquad\text{for}\qquad s,\,t\in\mathbb C.
\label{eq:Bnexp}
\end{gather}
In fact, the term $u_{12}^-u_{34}^-\cdots u_{2j-1,
2j}^-\partial_{2j+1}\partial_{2j+2}\cdots\partial_{n-2k}$ appears
only in the coef\/f\/icient of $t^k$ in the right hand side of
\eqref{eq:Bnexp} and it is contained in the terms
\begin{gather*}
 \tilde u_{i_{n-2k+1}i_{n-2k+2}}^-\cdots \tilde u_{i_{n-1}i_{n}}^-
 \tilde u_{12}^-\cdots \tilde u_{2j-1,2j}^-\partial_{2j+1}\cdots\partial_{n-2k}
\end{gather*}
of $P_n(t)$, where the number of the possibilities of these
$\tilde u_{i_{n-2k+1}i_{n-2k+2}}\cdots \tilde u_{i_{n-1}i_{n}}^-$
is ${(2k)!}/(2^k k!)$ because
$\{i_{n-2k+1},i_{n-2k+2},\dots,i_n\}=\{n-2k+1,\dots,n\}$.

{\rm vi)} Since
\begin{gather*}
 P_{k-1}=(n-k+1)[P_k,x_1+\dots+x_n]\qquad\text{for}\qquad k=2,\dots.n,
\end{gather*}
$[P_k,P_2]=0$ implies $[P_{k-1},P_2]=0$ by the Jacobi identity.
Here we note that
\begin{gather*}
  [u_{12}^-u_{34}^-\cdots u_{2j-1,2j}^-\partial_{2j+1}\cdots\partial_{k-1}\partial_\nu, x_\nu]
  = u_{12}^-u_{34}^-\cdots u_{2j-1,2j}^-\partial_{2j+1}\cdots\partial_{k-1}
\end{gather*}
for $\nu=k,k+1,\dots,n$.

{\rm vi)} The potential functions of (Trig-$A_{n-1}$),
(Rat-$A_{n-1}$) and (Toda-$A_{n-1}$) are specializations of more
general integrable potential functions of type $B_n$
(cf.~Def\/inition~\ref{def:restrict}).
\end{remark}
In the following diagram we show the relations among integrable
potentials of type $A_{n-1}$ by taking limits.
\begin{gather*}
\text{\bf Hierarchy of Integrable Potentials of Type
$\boldsymbol{A_{n-1}}$
$\boldsymbol{(n\ge3)}$}\notag\\
\begin{matrix}
&&\text{\rm Toda-$A_{n-1}^{(1)}$}&\to
  &\text{\rm Toda-$A_{n-1}$} \\
 &\nearrow& &\text{$\nearrow$}&\\
\text{\rm Ellip-$A_{n-1}$} &\to &\text{\rm Trig-$A_{n-1}$}&\to
  &\text{\rm Rat-$A_{n-1}$}
\end{matrix}
\end{gather*}

\section[Type $B_2$]{Type $\boldsymbol{B_2}$}\label{sec:B2}

In this section we study the following commuting dif\/ferential
operators $P$ and $P_2$.
\begin{gather}
 P=-\dfrac12\left(\dfrac{\partial ^2}{\partial x^2}+\dfrac{\partial ^2}{\partial y^2}\right) + R(x,y),
 \nonumber\\
 P_2=\dfrac{\partial^4}{\partial x^2\partial y^2} + S \qquad\text{with}\qquad\mathrm{ord}\, S<4,
 \nonumber\\
 [P,P_2]=0.\label{eq:b2-0}
\end{gather}
The Schr\"odinger operators $P$ of type $B_2$ in this section are
known to be completely integrable. They are listed in
\cite{Ochiai1996, Ochiai2003, Oshima2005}. We review them and give
the explicit expression of $P_2$.

First we review the arguments given in \cite{Ochiai1996,
Ochiai2003}. Since $P$ is self-adjoint, we may assume $P_2$ is
also self-adjoint by replacing $P_2$ by its self-adjoint part if
necessary. Here for
\begin{gather*}
A=\sum a_{ij}(x,y)\frac{\partial^{i+j}}{\partial x^i\partial y^j}
\end{gather*}
we def\/ine
\begin{gather*}
{}^t\!A=\sum (-1)^{i+j}\frac{\partial^{i+j}}{\partial x^i\partial
y^j}a_{ij}(x,y)
\end{gather*}
and $A$ is called self-adjoint if ${}^t\!A=A$.
\begin{lemma}[\cite{Ochiai1996}]\label{lem:b2-inv}
Suppose $P$ and $P_2$ are self-adjoint operators satisfying
\eqref{eq:b2-0}. Then
\begin{gather}
R(x,y) = u^+(x+y)+u^-(x-y)+v(x)+w(y),\nonumber
\\
P_2=\left(\frac{\partial ^2}{\partial x\partial y}\! +\!
u^-(x\!-\!y) \!-\! u^+(x\!+\!y)\right)^2 \!\!\!-
2w(y)\frac{\partial^2}{\partial x^2} \!-\! 2v(x)\frac{\partial
^2}{\partial y^2}\!+\! 4v(x)w(y)\!+ \!T(x,y) \label{eq:b2-1}
\end{gather}
and the function $T(x,y)$ satisfies
\begin{gather}
 \frac{\partial T(x,y)}{\partial x} = 2(u^+(x+y)-u^-(x-y))\frac{\partial w(y)}{\partial y}
    + 4w(y)\frac{\partial}{\partial y}(u^+(x+y)-u^-(x-y)),\nonumber
    \\
  \frac{\partial T(x,y)}{\partial y} = 2(u^+(x+y)-u^-(x-y))\frac{\partial v(x)}{\partial x}
    + 4v(x)\frac\partial{\partial x}(u^+(x+y)-u^-(x-y)).
\label{eq:b2-2}
\end{gather}
Conversely, if a function $T(x,y)$ satisfies \eqref{eq:b2-2} for
suitable functions $u^\pm(t)$, $v(t)$ and $w(t)$, then
\eqref{eq:b2-0} is valid for $R(x,y)$ and $P_2$ defined by
\eqref{eq:b2-1}.
\end{lemma}
\begin{remark}
\label{rem:B2w0} {\rm i)} If $w(y)=0$, then $T(x,y)$ does not
depend on $x$.

{\rm ii)} The self-adjointness of $P_2$ and the vanishing of the
third order term of $[P,P_2]$ imply that $P_2$ should be of the
form \eqref{eq:b2-1} with a suitable function $T(x,y)$. Then the
vanishing of the f\/irst order term implies \eqref{eq:b2-2}. The
last claim in Lemma~\ref{lem:b2-inv} is obtained by direct
calculation.
\end{remark}

Since $T(x,y)$ satisfying \eqref{eq:b2-2} is determined by
$(u^-,u^+;v,w)$ up to the dif\/ference of constants, we will write
$T(u^-,u^+;v,w)$ for the corresponding $T(x,y)$ which is an
element of the space of meromorphic functions of $(x,y)$ modulo
constant functions and def\/ine $Q(u^-,u^+;v,w)$ by
\begin{gather}
\label{eq:defQ}
 T(u^-, u^+;v, w)= 2(u^-(x-y)+u^+(x+y))(v(x)+w(y)) -4Q(u^-, u^+;v, w).
\end{gather}
Note that $T(u^-,u^+;v,w)$ and $Q(u^-,u^+;v,w)$ are def\/ined only
if the function $T(x,y)$ satis\-fying~\eqref{eq:b2-2} exists. The
following lemma is a direct consequence of~\eqref{eq:b2-2} and
this def\/inition of~$Q$.
\begin{lemma}[\cite{Ochiai1996}]\label{lem:B2Tsym}
Suppose $T(u^-,u^+;v,w)$ and $T(u^-_i,u^+_i;v_j,w_j)$ are defined.
Then for any $C$, $C_i$, $C'_j\in\mathbb C$ the left hand sides of
the following identities are also defined and
\begin{gather*}
 T(u^-(t)+C, u^+(t)+C; v(t), w(t))) =  T(u^-(t), u^+(t); v(t), w(t)),
  \\[1ex]
 Q(u^-(t), u^+(t); v(t)+C, w(t)+C)) =  Q(u^-(t), u^+(t); v(t), w(t)),
  \\[1ex]
 Q(u^-(Ct), u^+(Ct); v(Ct), w(Ct)) =  Q(u^-(t), u^+(t); v(t), w(t))|_{x\to Cx,\,y\to Cy},
  \\
 Q\left(\sum\limits_{i=1}^2A_iu^-_i,\sum\limits_{i=1}^2A_iu^+_i;
  \sum\limits_{j=1}^2C_jv_j,\sum\limits_{j=1}^2C_jw_j\right)
  =\sum\limits_{i=1}^2\sum\limits_{j=1}^2 A_iC_jQ(u^-_i,u^+_i;v_j,w_j).
\end{gather*}
Hence the left hand sides give pairs $P$ and $P_2$ with
$[P,P_2]=0$.
\end{lemma}
For simplicity we will use the notation
\begin{gather}
\label{eq:convention}
 Q(u^-,u^+;v ) = Q(u^-,u^+;v,v),\ \
 Q(u;v,w) = Q(u,u;v,w),\ \
 Q(u;v) = Q(u,u;v,v).
\end{gather}

The same convention will be also used for $T(u^-,u^+;v,w)$. The
integrable potentials of type~$B_2$ in this note are classif\/ied
into three kinds. The potentials of the f\/irst kind are the
unif\/ied integrable potentials which are in the same form as
those of type $B_n$ with $n\ge 3$, which we call {\em normal
integrable potentials\/} of type $B_2$.

The integrable potentials of type $B_2$ admit a special
transformation called {\em dual\/} which does not exist in $B_n$
with $n\ge 3$. Hence there are normal potentials and their dual in
the invariant integrable potentials of type $B_2$. Because of this
duality, there exist another kind of invariant integrable
potentials of type $B_2$, which we call {\em special integrable
potentials\/} of type $B_2$.

In this section we present $\bigl(R(x,y), T(x,y)\bigr)$ as
suitable limits of elliptic functions as in the previous section
since it helps to study the potentials of type $B_n$ in
Section~\ref{sec:Bn}.  We reduce the complete integrability of the
limits to that of a systems with elliptic potentials. But we can
also check~\eqref{eq:b2-2} by direct calculations (cf.\
Remark~\ref{rem:B2direct}).
%%%%%%%%%%%%%%%%%%%%%%%%%%%%%%%%%%%%%%%%%%%%%%%%%%%%%%%%%%%%%%%%%%
\subsection{Normal case}\label{sec:B2N}
In this subsection we study the integrable systems \eqref{eq:b2-0}
with \eqref{eq:b2-1} which have natural extension to type $B_n$
for $n\ge 3$ and have the form
\begin{gather*}
  u^-(t) = Au^-_0(t),\qquad u^+(t) = Au^+_0(t),
  \qquad
  v(t) = \sum\limits_{j=0}^3C_jv_j(t),\qquad w(t) = \sum\limits_{j=0}^3C_jw_j(t)
\end{gather*}
with any $A,\,C_0,\,C_1,\,C_2,\,C_3 \in\mathbb C$. These systems
are expressed by the symbol
\begin{gather*}
(\langle u^-_0\rangle,\langle u^+_0\rangle; \langle
v_0,v_1,v_2,v_3\rangle,\langle w_0,w_1,w_2,w_3\rangle).
\end{gather*}
%%%
The most general system is the following (Ellip-$B_2$) def\/ined
by elliptic functions, which is called Inozemtsev model
\cite{Inozemtsev1989-1}.

\begin{theorem}[$\boldsymbol{B_2}$: Normal case, \cite{Inozemtsev1989-1, Ochiai2003, Ochiai1996, Oshima2005} etc.]
\label{thm:B2Normal} The operators $P$ and $P_2$ defined by the
following pairs of $R(x,y)$ and $T(x,y)$ satisfy \eqref{eq:b2-0}
and \eqref{eq:b2-1}.\vspace{1ex}
%%%%%%%%%%%%%%%%%%%%%%%%%%%%%%%%%%%%%%%%%%%%%%%%%%%%%%%%%%%%%%%%

\noindent {\rm (Ellip-$B_2$):}\qquad
$(\langle\wp(t)\rangle;\langle\wp(t),\wp(t+\omega_1),\wp(t+\omega_2),\wp(t+\omega_3)\rangle)$
\begin{gather*}
  v(x)  = \sum\limits_{j=0}^3C_j\wp(x+\omega_j),\qquad
  w(y) = \sum\limits_{j=0}^3C_j\wp(y+\omega_j),
  \\
 u^-(x-y)  = A\wp(x-y),\qquad
 u^+(x+y) = A\wp(x+y),
\\
 R(x,y) = A(\wp(x-y)+\wp(x+y))+\sum\limits_{j=0}^3C_j(\wp(x+\omega_j)+\wp(y+\omega_j)),
\\
 T(x,y) = 2A(\wp(x-y)+\wp(x+y))\Biggl(\sum\limits_{j=0}^3C_j\bigl(\wp(x+\omega_j)+\wp(y+\omega_j)\Biggr)
 \\ \phantom{T(x,y) =}{}
-4A\sum\limits_{j=0}^3C_j\wp(x+\omega_j)\wp(y+\omega_j).
\end{gather*}
{\rm (Trig-$B_2$):}\qquad $(\langle\sinh^{-2}\lambda t\rangle;
\langle\sinh^{-2}\lambda t,\sinh^{-2}2\lambda t,
 \sinh^2\lambda t,\sinh^22\lambda t\rangle)$
\begin{gather*}
  v(x)  =C_0\sinh^{-2}\lambda x + C_1\cosh^{-2}\lambda x +C_2\sinh^2\lambda x + \frac14C_3\sinh^22\lambda x,
  \\
  w(y)  = C_0\sinh^{-2}\lambda y + C_1\cosh^{-2}\lambda y +C_2\sinh^2\lambda y + \frac14C_3\sinh^22\lambda y,
  \\
 u^-(x-y)  = A\sinh^{-2}\lambda(x-y),\qquad
 u^+(x+y) = A\sinh^{-2}\lambda(x+y),
\\
 R(x,y)  = A\bigl(\sinh^{-2}\lambda(x-y)+\sinh^{-2}\lambda(x+y)\bigr)
 +C_0\bigl(\sinh^{-2}\lambda x+\sinh^{-2}\lambda y\bigr)
\\ \phantom{R(x,y)  =}{}
+C_1\bigl(\cosh^{-2}\lambda x+\cosh^{-2}\lambda y\bigr)
+C_2\bigl(\sinh^2\lambda x+\sinh^2\lambda y\bigr)\\
\phantom{R(x,y)  =}{} +\frac14C_3\bigl(\sinh^22\lambda
x+\sinh^22\lambda y\bigr),
\\
T(x,y)  = 2A(\sinh^{-2}\lambda(x-y)+\sinh^{-2}\lambda(x+y))
\\ \phantom{T(x,y)  =}{}
\times\biggl(C_0\bigl(\sinh^{-2}\lambda x+\sinh^{-2}\lambda
y\bigr) +C_1(\cosh^{-2}\lambda x+\cosh^{-2}\lambda y)
\\ \phantom{T(x,y)  =}
{}+C_2(\sinh^2\lambda x+\sinh^2\lambda y)
+\frac14C_3(\sinh^22\lambda x+\sinh^22\lambda y)\biggr)
\\ \phantom{T(x,y)  =}{}-4A\Bigl(C_0\sinh^{-2}\lambda x\cdot\sinh^{-2}\lambda y -
C_1\cosh^{-2}\lambda x\cdot\cosh^{-2}\lambda y
\\ \phantom{T(x,y)  =}{}+ C_3(\sinh^2\lambda x + \sinh^2\lambda y + 2\sinh^2\lambda x\cdot\sinh^2\lambda y)\Bigr).
\end{gather*}
{\rm (Rat-$B_2$):}\qquad $(\langle t^{-2}\rangle;\langle
t^{-2},t^2,t^4,t^6\rangle)$
\begin{gather*}
  v(x)  = C_0x^{-2} + C_1x^2 +C_2x^4+ C_3x^6,\qquad
  w(y)  = C_0y^{-2} + C_1y^2 +C_2y^4+ C_3y^6,
  \\
 u^-(x-y)  = \frac A{(x-y)^2},\qquad
 u^+(x+y) = \frac A{(x+y)^2},
\\
 R(x,y)  = \frac A{(x\!-\!y)^2}+\frac A{(x\!+\!y)^2}
\! +\!C_0(x^{-2}+y^{-2}) \!+\!C_1(x^2\!+\!y^2)
\!+\!C_2(x^4\!+\!y^4)+C_3(x^6\!+\!y^6),
\\
 T(x,y)  = 2(u^-(x-y)+u^+(x+y))(v(x)+w(y))
-4A\biggl(\dfrac{C_0}{x^2y^2} + C_2(x^2+y^2)
\\ \phantom{T(x,y)  =}
{}+C_3(x^4+y^4+3x^2y^2)\biggr) =
8A\frac{2C_0+2C_1x^2y^2+C_2x^2y^2(x^2+y^2)+2C_3x^4y^4}{(x^2-y^2)^2}.
\end{gather*}
{\rm(Toda-$D_2^{(1)}$-bry):}\qquad $(\langle \cosh2\lambda
t\rangle;\langle \sinh^{-2}\lambda t,\sinh^{-2}2\lambda t\rangle,
\langle  \sinh^{-2}\lambda t,\sinh^{-2}2\lambda t\rangle)$
\begin{gather*}
  v(x)  = C_0\sinh^{-2}\lambda x+C_1\sinh^{-2}2\lambda x,\qquad
  w(y) = C_2\sinh^{-2}\lambda y +C_3\sinh^{-2}2\lambda y,
  \\
 u^-(x-y)  = A\cosh2\lambda(x-y),\qquad
 u^+(x+y) = A\cosh2\lambda(x+y),
 \\
 R(x,y)  = A\cosh2\lambda(x-y)+A\cosh2\lambda(x+y)
 \\ \phantom{R(x,y)  =}{}
 +C_0\sinh^{-2}\lambda x +C_1\sinh^{-2}2\lambda x
 +C_2\sinh^{-2}\lambda y +C_3\sinh^{-2}2\lambda y,
 \\
 T(x,y)  = 8A(C_0\cosh2\lambda y+C_2\cosh2\lambda x).
\end{gather*}
{\rm (Toda-$B_2^{(1)}$-bry):}\qquad $(\langle  e^{-2\lambda
t}\rangle;\langle e^{2\lambda t},e^{4\lambda t}\rangle, \langle
\sinh^{-2}\lambda t,\sinh^{-2}2\lambda t\rangle)$
\begin{gather*}
  v(x)  = C_0e^{2\lambda x}+C_1e^{4\lambda x},\qquad
  w(y) = C_2\sinh^{-2}\lambda y+C_3\sinh^{-2}2\lambda y,
  \\
 u^-(x-y)  = Ae^{-2\lambda(x-y)},\qquad
 u^+(x+y) = Ae^{-2\lambda(x+y)},
 \\
 R(x,y)  = Ae^{-2\lambda(x-y)}+Ae^{-2\lambda(x+y)}
 +C_0e^{2\lambda x}+C_1e^{4\lambda x}
 +C_2\sinh^{-2}\lambda y+C_3\sinh^{-2}2\lambda y,
 \\
 T(x,y)  =4A\bigl(C_0\cosh2\lambda y+2C_2e^{-2\lambda x}\bigr).
 \end{gather*}
{\rm(Trig-$A_1^{(1)}$-bry):}\qquad
 $(\langle  \sinh^{-2}\lambda t\rangle,0;
 \langle e^{-2\lambda t},e^{-4\lambda t},e^{2\lambda t}, e^{4\lambda t} \rangle)$
 \begin{gather*}
v(x)  = C_0e^{-2\lambda x}+ C_1e^{-4\lambda x} + C_2e^{2\lambda x}
+ C_3e^{4\lambda x},
\\
w(y)  = C_0e^{-2\lambda y}+ C_1e^{-4\lambda y} + C_2e^{2\lambda y}
+ C_3e^{4\lambda y},
\\
u^-(x-y)  = A\sinh^{-2}\lambda(x-y),\qquad  u^+(x+y) = 0,
\\
R(x,y)  = A\sinh^{-2}\lambda(x-y) +C_0(e^{-2\lambda
x}+e^{-2\lambda y}) +C_1(e^{-4\lambda x}+e^{-4\lambda y})
 \\ \phantom{R(x,y)  =}
{} +C_2(e^{2\lambda x}+e^{2\lambda y}) +C_3(e^{4\lambda
x}+e^{4\lambda y}),
\\
 T(x,y)  = 2A \sinh^{-2}\lambda(x-y)(C_0\bigl(e^{-2\lambda x}+e^{-2\lambda y}\bigr)
  \\ \phantom{T(x,y)  =}{}
{} + 2C_1e^{-2\lambda(x+y)}+ C_2\bigl(e^{2\lambda x}+e^{2\lambda
y}\bigr)+ 2C_3e^{2\lambda(x+y)}).
 \end{gather*}
{\rm(Toda-$C_2^{(1)}$):}\qquad $(\langle  e^{-2\lambda
t}\rangle,0; \langle e^{2\lambda t}, e^{4\lambda t}\rangle,
\langle e^{-2\lambda t}, e^{-4\lambda t}\rangle)$
\begin{gather*}
  v(x)  = C_0e^{2\lambda x} + C_1^{4\lambda x},\qquad
  w(y) = C_2e^{-2\lambda y}+C_3e^{-4\lambda y},
  \\
 u^-(x-y)  = Ae^{-2\lambda(x-y)},\qquad
 u^+(x+y) = 0,
 \\
 R(x,y)  = Ae^{-2\lambda(x-y)} +C_0e^{2\lambda x} + C_1^{4\lambda x}
 +C_2e^{-2\lambda y}+C_3e^{-4\lambda y},
 \\
 T(x,y) =2A\bigl(C_0e^{2\lambda y} +C_2e^{-2\lambda x}\bigr).
 \end{gather*}
{\rm (Rat-$A_1$-bry):}\qquad $(\langle  t^{-2}\rangle,0;\langle t,
t^2,t^3,t^4\rangle)$
\begin{gather*}
  v(x)  = C_0x+C_1x^2+C_2x^3+C_3x^4,\qquad
  w(y) = C_0y+C_1y^2+C_2y^3+C_3y^4,
  \\
 u^-(x-y)  = \frac A{(x-y)^2},\qquad
 u^+(x+y) = 0,
 \\
 R(x,y)  = \frac{A}{(x-y)^2}+C_0(x+y)+C_1(x^2+y^2)+C_2(x^3+y^3)+C_3(x^4+y^4),
 \\
 T(x,y)  = \frac{2A}{(x-y)^2}(C_0(x+y)+C_1(x^2+y^2)+C_2xy(x+y)+2C_3x^2y^2).
\end{gather*}
\end{theorem}
%%%%%%%%%%%
\begin{remark}
For example, $(\langle  t^{-2}\rangle,0;\langle t,
t^2,t^3,t^4\rangle)$ in the above (Rat-$A_1$-bry) means
\begin{gather*}
  u^-(t)=A t^{-2},\qquad u^+(t)=0,\qquad
  v(t)=w(t)=C_0 t + C_1 t^2+C_2 t^3 + C_3 t^4
\end{gather*}
with a convention similar to that in \eqref{eq:convention}.
\end{remark}
%%%%%%%%%%%
We will review the proof of the above theorem after certain
remarks.
%%%%%%%%%%%
\begin{remark}\label{rem:B2uni}
All the invariant integrable potentials of type $B_2$ together
with $P_2$ are determined by \cite{Ochiai2003, Ochiai1994}. They
are classif\/ied into three cases. In the normal case they are
(Ellip-$B_2$), \mbox{(Trig-$B_2$)} and (Rat-$B_2$) which have the
following unif\/ied expression of the invariant potentials given
by \cite[Lemma~7.3]{Oshima1995}, where the periods $2\omega_1$ and
$2\omega_2$ may be inf\/inite (cf.~\eqref{eq:wpt} and
\eqref{eq:wpr})
\begin{gather*}
 R(x,y) = A\wp(x-y)+A\wp(x+y)
  + \frac{C_4\wp(x)^4+C_3\wp(x)^3+C_2\wp(x)^2+C_1\wp(x)+C_0}{\wp'(x)^2}
\\ \phantom{R(x,y) =}
 {} + \frac{C_4\wp(y)^4+C_3\wp(y)^3+C_2\wp(y)^2+C_1\wp(y)+C_0}{\wp'(y)^2},
 \\
 T(x,y) = 4A(\wp(x)-\wp(y))^{-2}\biggl(C_4\wp(x)^2\wp(y)^2+\dfrac{C_3}2\wp(x)^2\wp(y)
\\ \phantom{T(x,y) =}
{}+\dfrac{C_3}2\wp(x)\wp(y)^2 +
       C_2\wp(x)\wp(y)+ \dfrac{C_1}2\wp(x)+\dfrac{C_1}2\wp(y) + C_0\biggr).
\end{gather*}
This is the original form we found in the classif\/ication of the
invariant integrable systems of type $B_n$
(cf.~\cite{Ochiai1994}). Later we knew Inozemtzev model and in
fact, when the periods are f\/inite, \eqref{eq:wpdifeq} and
\eqref{eq:wprat} show that the above potential function
corresponds to (Ellip-$B_2$).

When $\omega_1=\omega_2=\infty$, $\wp(t)=t^{-2}$ and
$(\wp(x)-\wp(y))^{-2}=x^4y^4(x^2-y^2)^{-2}$ and
\begin{gather*}
  R(x,y) = \dfrac{A}{(x-y)^2}+\dfrac A{(x+y)^2}
    + \dfrac14(C_4x^{-2}+C_3+C_2x^2+C_1x^4+C_0x^6)\\
\phantom{R(x,y) =}{}
  + \dfrac14(C_4y^{-2}+C_3+C_2y^2+C_1y^4+C_0y^6),
  \\
  T(x,y) = 2A(x^2-y^2)^{-2}(2C_4+C_3(x^2+y^2)+2C_2x^2y^2+C_1x^2y^2(x^2+y^2)+2C_0x^4y^4).
\end{gather*}
\end{remark}

We review these invariant cases discussed in
\cite{Ochiai2003,Oshima1995}. Owing to the identity
\begin{gather*}
 2(u^{-}-u^{+})v'+4v((u^-)'-(u^+)')+\partial_y(2(u^-+u^+)(v+w)-4vw)
 \\ \qquad
 =2\left|\begin{matrix}
  v   &  v'  & 1\\
  w   & -w'  & 1\\
  u^- & -(u^-)'& 1
 \end{matrix}\right|
 +
  2\left|\begin{matrix}
  v   & -v'  & 1\\
  w   & -w'  & 1\\
  u^+ & (u^+)'& 1
 \end{matrix}\right|
\end{gather*}
and \eqref{eq:wpDadd}, the right hand side of the above is zero
and we have \eqref{eq:b2-2} when
\begin{gather*}
 u^-=C\wp(x\!-\!y)+C',\quad u^+=C\wp(x\!+\!y)+C',\quad v=C\wp(x)+C'\quad\text{and}\quad w=C\wp(y)+C'
\end{gather*}
with $T(x,y) = 2(u^-+u^+)(v+w)-4vw$. Hence with
$Q\bigl(\wp(t);\wp(t)\bigr)=\wp(x)\wp(y)$, the function $T(x,y)$
given by \eqref{eq:defQ} satisf\/ies \eqref{eq:b2-2} with the
above $u^{\pm}$, $v$ and $w$. Using the transformations
$(x,y)\mapsto(x+\omega_j,y+\omega_j)$, we have
\begin{gather}
\label{eq:B2Q} \wp(x+\omega_j) \wp(y+\omega_j) =
Q(\wp(t+\omega_j);\wp(t+\omega_j)) = Q(\wp(t);\wp(t+\omega_j))
\end{gather}
for $j=0,1,2,3$ because the function $\wp(x\pm y)$ does not change
under these transformations (cf.~\eqref{eq:wpperiod}). Thus we
have Theorem~\ref{thm:B2Normal} for (Ellip-$B_2$) in virtue of
Lemma~\ref{lem:b2-inv} and Lemma~\ref{lem:B2Tsym}.

Here we note that $\wp$ may be replaced by $\wp_0$.

By the limit under $\omega_2\to\infty$, we have the following
(Trig-$B_2$) from (Ellip-$B_2$). See the proof of
\cite[Proposition~6.1]{Oshima1998} for the precise
argument.\vspace{2ex}

\noindent (Trig-$B_2$):
\begin{gather}
Q(\sinh^{-2}\lambda t;\sinh^{-2}\lambda t)=\sinh^{-2}\lambda
x\cdot \sinh^{-2}\lambda y, \label{eq:B23}
\\
Q(\sinh^{-2}\lambda t;\cosh^{-2}\lambda t) = -\cosh^{-2}\lambda
x\cdot \cosh^{-2}\lambda y, \label{eq:B23c}
\\
Q(\sinh^{-2}\lambda t;\sinh^2\lambda t)=0, \label{eq:B23b}
\\
Q\biggl(\sinh^{-2}\lambda t;\frac14\sinh^22\lambda t\biggr)
  = \sinh^2\lambda x+\sinh^2\lambda y+2\sinh^2\lambda x\cdot\sinh^2\lambda y.
\label{eq:B23d}
\end{gather}
The equations \eqref{eq:B23}, \eqref{eq:B23c} and \eqref{eq:B23b}
correspond to \eqref{eq:wp2sh-2}, \eqref{eq:wp2ch-2} and
\eqref{eq:wp2ch}, respectively. Moreover \eqref{eq:Landen} should
be noted and \eqref{eq:B23d} corresponds to \eqref{eq:wp2ch} with
replacing $(\omega_1,\lambda)$ by $(\omega_1/2,2\lambda)$.

By the limit under $\lambda\to 0$, we have the following
(Rat-$B_2$) from (Trig-$B_2$) as was shown in the proof of
\cite[Proposition~6.3]{Oshima1998}. Here we note
\eqref{eq:sh-2rat} and
\begin{gather*}
\cosh^{-2}\lambda t\cdot\sinh^2\lambda t=1-\cosh^{-2}\lambda t,
\\
\cosh^{-2}\lambda t\cdot\sinh^4\lambda t=-1+\cosh^{-2}\lambda
t+\sinh^2\lambda t,
\\
\cosh^{-2}\lambda t\cdot\sinh^6\lambda t=1-\cosh^{-2}\lambda
t-2\sinh^2\lambda t+\frac14\sinh^22\lambda t,
\\
\lim_{\lambda\to 0}\lambda^{-2j} \cosh^{-2}\lambda
t\cdot\sinh^{2j}\lambda t=t^{2j}\qquad\text{for}\qquad j=1,2,3,
\\
\frac1{(x-y)^2}+\frac1{(x+y)^2} = \frac{2(x^2+y^2)}{(x^2-y^2)^2}.
\end{gather*}
The result is as follows.\vspace{2ex}

\noindent (Rat-$B_2$):
\begin{gather*}
Q(t^{-2};{t^{-2}}) =x^{-2}y^{-2},
\\
T(t^{-2};{t^{-2}})
=4(x^{-2}+y^{-2})((x-y)^{-2}+(x+y)^{-2})-4x^{-2}y^{-2}
\\ \phantom{T(t^{-2};{t^{-2}})}
{}=\frac{4(x^2+y^2)^2-4(x^2-y^2)^2}{x^2y^2(x^2-y^2)^2}=\frac{16}{(x^2-y^2)^2},
\\
Q(t^{-2};{t^2}) = 0,
\\
T(t^{-2};{t^2}) = \frac{4(x^2+y^2)^2}{(x^2-y^2)^2} =
\frac{16x^2y^2}{(x^2-y^2)^2}+4,
\\
Q(t^{-2};{t^4})  =x^2+y^2,
\\
T(t^{-2},t^4) =
\frac{4(x^2+y^2)(x^4+y^4)}{(x^2-y^2)^2}-4(x^2+y^2)=\frac{8x^2y^2(x^2+y^2)}{(x^2-y^2)^2},
\\
Q(t^{-2};{t^6})= x^4+y^4+3x^2y^2,
\\
T(t^{-2},t^6) =
\frac{4(x^2+y^2)(x^6+y^6)}{(x^2-y^2)^2}-4(x^4+3x^2y^2+y^4)=\frac{16x^4y^4}{(x^2-y^2)^2}.
\end{gather*}
This expression of $T(x,y)$ for (Rat-$B_2$) is also given in
Remark~\ref{rem:B2uni}. Note that we ignore the dif\/ference of
constants for $Q$ and $T$.
%%%%%%%%%%%%%%%%%%
\begin{proof}[Proof of Theorem~\ref{thm:B2Normal}]
The three cases (Ellip-$B_2$), (Trig-$B_2$) and (Rat-$B_2$) have
been explai\-ned.
Note that if $u^\pm$, $v$, $w$ and $T(u^-,u^+;v,w)$ (or
$Q(u^-,u^+;v,w)$) are def\/ined and they have an analytic
parameter, Lemma~\ref{lem:b2-inv} assures that their analytic
continuations also def\/ine $P$ and $P_2$ satisfying $[P,P_2]=0$.

(Toda-$D_2^{(1)}$-bry) $\leftarrow$ (Ellip-$B_2$): Replacing
$(x,y)$ by $(x+\omega_2,y)$, we have
\begin{gather*}
 Q(\cosh 2\lambda t;0,\sinh^{-2}\lambda t)
=\lim_{\omega_2\to\infty}Q\biggl(\frac{e^{2\lambda\omega_2}}{8\lambda^2}
\wp_0(t+\omega_2);
  \frac1{\lambda^2}\wp_0(t+\omega_2), \frac1{\lambda^2}\wp_0(t)\biggr)
\\ \phantom{Q(\cosh 2\lambda t;0,\sinh^{-2}\lambda t)}
{}=\lim_{\omega_2\to\infty}\frac{e^{2\lambda\omega_2}}{8\lambda^4}\wp_0(x+\omega_2)\wp_0(y)
=\cosh2\lambda x\cdot \sinh^{-2}\lambda y,
\\
 Q(\cosh 2\lambda t;0,\cosh^{-2}\lambda t)=\!\!\lim_{\omega_2\to\infty}Q\biggl(
  \frac{e^{2\lambda\omega_2}}{8\lambda^2}\wp_0(t\!+\!\omega_2);
 -\frac1{\lambda^2}\wp_0(t\!+\!\omega_1+\omega_2);
 -\frac1{\lambda_2}\wp_0(t\!+\!\omega_1)\!\biggr)
\\ \phantom{Q(\cosh 2\lambda t;0,\cosh^{-2}\lambda t)}
{}=-\lim_{\omega_2\to\infty}\frac{e^{2\lambda\omega_2}}{8\lambda^4}
  \wp_0(x+\omega_1+\omega_2)\cdot\wp_0(y+\omega_1)\\
\phantom{Q(\cosh 2\lambda t;0,\cosh^{-2}\lambda t)}
=-\cosh2\lambda x\cdot\cosh^{-2}\lambda y,
\\
 Q(\cosh 2\lambda t;\sinh^{-2}\lambda t,0)
 = \sinh^{-2}\lambda x\cdot \cosh^2\lambda y,
\\
Q(\cosh 2\lambda t;\cosh^{-2}\lambda t,0) = -\cosh^{-2}\lambda
x\cdot \sinh^2\lambda y.
\end{gather*}
Hence
\begin{gather*}
 T(\cosh 2\lambda t;0,\sinh^{-2}\lambda t) =
2\bigl(\cosh2\lambda(x+y)+\cosh2\lambda(x-y)\bigr)\cdot
\sinh^{-2}\lambda y
\\ \phantom{T(\cosh 2\lambda t;0,\sinh^{-2}\lambda t) =}{}
{}-4\cosh2\lambda x\cdot \sinh^{-2}\lambda y  = 8\cosh2\lambda x,
\\
T(\cosh 2\lambda t;0,\cosh^{-2}\lambda t) =
2\bigl(\cosh2\lambda(x+y)+\cosh2\lambda(x-y)\bigr)\cdot
\cosh^{-2}\lambda y
\\ \phantom{T(\cosh 2\lambda t;0,\cosh^{-2}\lambda t) =}
{}+4\cosh2\lambda x\cdot \cosh^{-2}\lambda y  = 8\cosh2\lambda x,
\\
 T(\cosh 2\lambda t;0,\sinh^{-2}2\lambda t) = T(\cosh 2\lambda t;\sinh^{-2}2\lambda t,0)=0,
\\
 T(\cosh 2\lambda t;\sinh^{-2}\lambda t,0) = 8\cosh2\lambda y.
\end{gather*}

\noindent
(Toda-$B_2^{(1)}$-bry)${}\leftarrow{}$(Toda-$D_2^{(1)}$-bry):
Replacing $(x,y)$ by $(x-N,y)$, we have
\begin{gather*}
 T(e^{-2\lambda t};0,\sinh^{-2}\lambda t)
=\lim_{N\to\infty} T(2e^{-2\lambda N}\cosh2\lambda
(t-N);0,\sinh^{-2}\lambda t)
\\ \phantom{T(e^{-2\lambda t};0,\sinh^{-2}\lambda t)}
{}=\lim_{N\to\infty}16e^{-2\lambda N}\cosh2\lambda(x-N)=
8e^{-2\lambda x},
\\
T(e^{-2\lambda t};0,\sinh^{-2}2\lambda t)= \lim_{N\to\infty}
 T(2e^{-2\lambda N}\cosh2\lambda (t-N);0,\sinh^{-2}2\lambda t)=0,
\\
T(e^{-2\lambda t};e^{2\lambda t},0)
=\!\!\lim_{N\to\infty}T\biggl(2e^{-2\lambda N}\cosh2\lambda
(t\!-\!N);
 \frac14e^{2\lambda N}\sinh^{-2}\lambda(t\!-\!N),0\biggr)\!
=4\cosh2\lambda y,
\\
 T\bigl(e^{-2\lambda t};e^{4\lambda t},0\bigr)
=\lim_{N\to\infty}T\biggl(2e^{-2\lambda N}\cosh2\lambda (t-N);
 \frac14e^{4\lambda N}\sinh^{-2}2\lambda(t-N),0\biggr)=0.
\end{gather*}

\noindent (Trig-$C_2^{(1)}$) $\leftarrow$ (Trig-$B_2^{(1)}$-bry):
Replacing $(x,y)$ by $(x+N,y+N)$,
\begin{gather*}
 T(e^{-2\lambda t},0;e^{2\lambda t},0)
=\lim_{N\to\infty}T(e^{-2\lambda
t},e^{-2\lambda(t+2N)};e^{-2\lambda N}e^{2\lambda(t+N)},0)
\\ \phantom{T(e^{-2\lambda t},0;e^{2\lambda t},0)}
{}=\lim_{N\to\infty}e^{-2\lambda N}\cosh2\lambda(y+N)=2e^{2\lambda
y},
\\
 T(e^{-2\lambda t},0;e^{4\lambda t},0)
=\lim_{N\to\infty}T(e^{-2\lambda
t},e^{-2\lambda(t+2N)};e^{-4\lambda N}e^{4\lambda(t+N)},0)=0.
\end{gather*}
By the transformation $(x,y,\lambda)\mapsto(y,x,-\lambda)$, we
have
\begin{gather*}
T(e^{-2\lambda t},0;0,e^{-2\lambda t})=2e^{-2\lambda x},\qquad
T(e^{-2\lambda t},0;0,e^{-4\lambda t})=0.
\end{gather*}

\noindent (Trig-$A_1^{(1)}$-bry) $\leftarrow$ (Trig-$B_2$):
Replacing $(x,y)$ by $(x+N,y+N)$,
\begin{gather*}
 Q\bigl(\sinh^{-2}\lambda t, 0;e^{2\lambda t} \bigr)=
\lim_{N\to\infty} Q\biggl(\sinh^{-2}\lambda
t,\sinh^{-2}\lambda(t+2N);
 \frac14e^{2\lambda N}\sinh^{-2}\lambda(t+N)\biggr)
\\ \phantom{Q\bigl(\sinh^{-2}\lambda t, 0;e^{2\lambda t} \bigr)}
{}= \lim_{N\to\infty}\frac14e^{2\lambda
N}\sinh^{-2}\lambda(x+N)\sinh^{-2}\lambda(y+N)=0,
\\
 T\bigl(\sinh^{-2}\lambda t, 0;e^{2\lambda t} \bigr)
 =2(e^{2\lambda x}+e^{2\lambda y})\sinh^{-2}\lambda(x-y),
\\
 Q(\sinh^{-2}\lambda t, 0;e^{4\lambda t})\lim_{N\to\infty}
 Q(\sinh^{-2}\lambda t,\sinh^{-2}\lambda(t+2N);
 4e^{-4\lambda N}\sinh^22\lambda(t+N))
 \\ \phantom{Q\bigl(\sinh^{-2}\lambda t, 0;e^{2\lambda t} \bigr)}
{}=16\lim_{N\to\infty}e^{-4\lambda
N}(\sinh^2\lambda(x+N)+\sinh^2\lambda(y+N)
\\
\phantom{Q\bigl(\sinh^{-2}\lambda t, 0;e^{2\lambda t} \bigr)=}{} +
2\sinh^2\lambda(x+N)\sinh^2\lambda(y+N))=2e^{2\lambda(x+y)},
\\
 T(\sinh^{-2}\lambda t, 0;e^{4\lambda t})=
2\sinh^{-2}\lambda(x-y)(e^{4\lambda x}+e^{4\lambda
y})-8e^{2\lambda(x+y)}
\\ \phantom{T(\sinh^{-2}\lambda t, 0;e^{4\lambda t})}{}
{} =2{\sinh^{-2}\lambda(x-y)}(e^{4\lambda x}+e^{4\lambda y}
  -e^{2\lambda(x+y)}(e^{\lambda(x-y)}-e^{-\lambda(x-y)})^2)
\\ \phantom{T(\sinh^{-2}\lambda t, 0;e^{4\lambda t})}{}
{}=4e^{2\lambda(x+y)}{\sinh^{-2}\lambda(x-y)}.
\end{gather*}

\noindent (Rat-$A_1$-bry) $\leftarrow$ (Trig-$A_1$-bry): Taking
the limit $\lambda\to0$,
\begin{gather*}
 Q(t^{-2},0;t)=\lim_{\lambda\to0}Q\left(\lambda^2\sinh^{-2}\lambda t,0;
 \frac{1}{2\lambda}(e^{2\lambda t}-1)\right)=0,
 \\
 T(t^{-2},0;t)=\frac{2(x+y)}{(x-y)^2},
 \\
 Q(t^{-2},0;t^2)=\lim_{\lambda\to0}Q\biggl(\lambda^2\sinh^{-2}\lambda t,0;
  \frac1{4\lambda^2}(e^{2\lambda t}+e^{-2\lambda t}-2)\biggr)=0,
 \\
 T(t^{-2},0;t^2)=2\frac{x^2+y^2}{(x-y)^2},
 \\
 Q(t^{-2},0;t^3)=\lim_{\lambda\to0}Q\biggl(\lambda^2\sinh^{-2}\lambda t,0;
\frac1{8\lambda^3}(e^{4\lambda t} - 3e^{2\lambda t}-e^{-2\lambda
t}+3)\biggl)
\\ \phantom{Q(t^{-2},0;t^3)}{}
=\lim_{\lambda\to0}\frac2{8\lambda}(e^{2\lambda(x+y)}-1)
=\frac12(x+y),
\\
T(t^{-2},0;t^3) = 2\frac{x^3+y^3}{(x-y)^2}-2(x+y)
=2\frac{xy(x+y)}{(x-y)^2},
\\
Q(t^{-2},0;t^4)=\lim_{\lambda\to0}Q\biggl(\lambda^2\sinh^{-2}\lambda
t,0; \frac1{16\lambda^4}(e^{4\lambda t}+e^{-4\lambda
t}-4e^{2\lambda t}-4e^{-2\lambda t}+6)\biggr)
\\ \phantom{Q(t^{-2},0;t^4)}{}
=\lim_{\lambda\to0}\frac2{16\lambda^2}(e^{2\lambda(x+y)}+e^{-2\lambda(x+y)}-2)
=\frac12(x+y)^2,
\\
T(t^{-2},0;t^4) =
2\frac{x^4+y^4}{(x-y)^2}-2(x+y)^2=\frac{4x^2y^2}{(x-y)^2}.
\end{gather*}
Thus we have completed the proof of Theorem~\ref{thm:B2Normal} by
using Lemma~\ref{lem:b2-inv} and Lemma~\ref{lem:B2Tsym}.
\end{proof}
%%%
\begin{remark}\label{rem:B2direct}
Theorem~\ref{thm:B2Normal} can be checked by direct calculations.
For example, Remark~\ref{rem:B2w0} and the equations
\begin{gather*}
 2(\varepsilon e^{-2\lambda(x+y)}-e^{-2\lambda(x-y)})(e^{2\lambda x})'
+4e^{2\lambda x}\frac{\partial}{\partial x} (\varepsilon
e^{-2\lambda(x+y)}-e^{-2\lambda(x-y)})
\\ \qquad{}
 =4\lambda(\varepsilon e^{-2\lambda y}-e^{2\lambda y})
 -8\lambda(\varepsilon e^{-2\lambda y}-e^{2\lambda y})=
\frac{\partial}{\partial y}(2(\varepsilon e^{-2\lambda
y}+e^{2\lambda y})),
\\
2(\varepsilon e^{-2\lambda(x+y)}-e^{-2\lambda(x-y)})(e^{4\lambda
x})' +4e^{4\lambda x}\frac{\partial}{\partial x} (\varepsilon
e^{-2\lambda(x+y)}-e^{-2\lambda(x-y)})=0
\end{gather*}
with $\varepsilon=1$ give $T(e^{-2\lambda t};e^{2\lambda t},0)$
and $T(e^{-2\lambda t};e^{4\lambda t},0)$ for (Trig-$B_2$-bry).
Moreover the functions $T(e^{-2\lambda t},0;e^{2\lambda t},0)$ and
$T(e^{-2\lambda t},0;e^{4\lambda t},0)$ for (Toda-$C_2^{(1)}$)
also follow from these equations with $\varepsilon=0$.
\end{remark}

\subsection{Special case}\label{sec:B2S}
In this subsection we study the integrable systems \eqref{eq:b2-0}
with \eqref{eq:b2-1} which are of the form
\begin{gather*}
R(x,y) =u^-(x-y) + u^+(x+y) + v(x) + w(y),
\\
u^-(t) = \sum\limits_{j=0}^1A_ju^-_j(t),\qquad u^+(t) =
\sum\limits_{j=0}^1A_ju^+_j(t),
\\
  v(t) = \sum\limits_{j=0}^1C_jv_j(t),\qquad w(t) = \sum\limits_{j=0}^1C_jw_j(t)
\end{gather*}
with $A_0,\,A_1,\,C_0,\,C_1\in\mathbb C$. The most general system
(Ellip-$B_2$-S) in the following theorem is presented by
\cite{Ochiai2003} as an elliptic generalization of (Trig-$B_2$-S)
found by \cite{Oshima1995}.
\begin{theorem}[$\boldsymbol{B_2}$ : Special case, \cite{Ochiai1996, Ochiai2003,
Oshima2005} etc.]\label{thm:B2S} The operators $P$ and $P_2$
defined by the following pairs of $R(x,y)$ and $T(x,y)$ satisfy
\eqref{eq:b2-0} and \eqref{eq:b2-1}.\vspace{1ex}

\noindent
{\rm(Ellip-$B_2$-S):}\qquad$(\langle\wp(t;2\omega_1,2\omega_2),\wp(t;\omega_1,2\omega_2)\rangle;
 \langle\wp(t;\omega_1,2\omega_2),\wp(t;\omega_1,\omega_2)\rangle)$
\begin{gather*}
v(x)  = C_0\wp(x;\omega_1,2\omega_2)+C_1\wp(x;\omega_1,\omega_2),
\qquad w(y)  =
C_0\wp(y;\omega_1,2\omega_2)+C_1\wp(y;\omega_1,\omega_2),
\\
u^-(x-y)  =
A_0\wp(x-y;2\omega_1,2\omega_2)+A_1\wp(x-y;\omega_1,2\omega_2),
\\
u^+(x+y)  =
A_0\wp(x+y;2\omega_1,2\omega_2)+A_1\wp(x+y;\omega_1,2\omega_2),
\\
R(x,y)  =
A_0\wp(x-y;2\omega_1,2\omega_2)+A_0\wp(x+y;2\omega_1,2\omega_2)
\\ \phantom{R(x,y)  =}
{} +A_1\wp(x-y;\omega_1,2\omega_2)+A_1\wp(x+y;\omega_1,2\omega_2)
\\ \phantom{R(x,y)  =}
 {}+C_0\wp(x;\omega_1,2\omega_2)+C_0\wp(y;\omega_1,2\omega_2)
 +C_1\wp(x;\omega_1,\omega_2)+C_1\wp(y;\omega_1,\omega_2),
\\
T(x,y)  =
2(A_0\wp(x-y;2\omega_1,2\omega_2)+A_0\wp(x+y;2\omega_1,2\omega_2)
\\ \phantom{T(x,y)  =}
 {}+A_1\wp(x-y;\omega_1,2\omega_2)+A_1\wp(x+y;\omega_1,2\omega_2))
 \\ \phantom{T(x,y)  =}
{}\times
(C_0\wp(x;\omega_1,2\omega_2)+C_0\wp(y;\omega_1,2\omega_2)
 +C_1\wp(x;\omega_1,\omega_2)+C_1\wp(y;\omega_1,\omega_2))
\\ \phantom{T(x,y)  =}
{} -4A_0C_0\sum\limits_{j=0}^1
  \wp(x+\omega_j;2\omega_1,2\omega_2)\cdot\wp(y+\omega_j;2\omega_1,2\omega_2)
\\ \phantom{T(x,y)  =}
{} - 4A_0C_1\sum\limits_{j=0}^3\wp(x+\omega_j;2\omega_1,2\omega_2)
\wp(y+\omega_j;2\omega_1,2\omega_2)
\\ \phantom{T(x,y)  =}
{} -4A_1C_0 \wp(x;\omega_1,2\omega_2) \wp(y;\omega_1,2\omega_2) \\
\phantom{T(x,y)  =}{} -4A_1C_1\sum\limits_{j=0}^1
  \wp(x+\omega_{2j};\omega_1,2\omega_2) \wp(y+\omega_{2j};\omega_1,2\omega_2).
\end{gather*}
{\rm(Trig-$B_2$-S):}\qquad$( \langle\sinh^{-2}\lambda
t,\sinh^{-2}2\lambda t\rangle;
 \langle\sinh^{-2}2\lambda t,\sinh^22\lambda t\rangle)$
 \begin{gather*}
  v(x)  = C_0\sinh^{-2}2\lambda x+C_1\sinh^22\lambda x,\qquad
  w(y)  = C_0\sinh^{-2}2\lambda y+C_1\sinh^22\lambda y,
  \\
 u^-(x-y)  = A_0\sinh^{-2}\lambda(x-y)+A_1\sinh^{-2}2\lambda(x-y),
 \\
 u^+(x+y)  = A_0\sinh^{-2}\lambda(x+y)+A_1\sinh^{-2}2\lambda(x+y),
\\
 R(x,y)  = A_0\sinh^{-2}\lambda(x+y)+A_0\sinh^{-2}\lambda(x-y)
 + A_1\sinh^{-2}2\lambda(x+y)
 \\ \phantom{R(x,y)  =}
{}+A_1\sinh^{-2}2\lambda(x-y) + C_0\sinh^{-2}2\lambda x+C_0\sinh^{-2}2\lambda y\\
 \phantom{R(x,y)  =}{}
 + C_1\sinh^22\lambda x+C_1\sinh^22\lambda y,
\\
T(x,y)=  2(A_0\sinh^{-2}\lambda(x+y)+A_0\sinh^{-2}\lambda(x-y) +
A_1\sinh^{-2}2\lambda(x+y)
 \\ \phantom{T(x,y)=}
{}+A_1\sinh^{-2}2\lambda(x-y))(C_0\sinh^{-2}2\lambda
x+C_0\sinh^{-2}2\lambda y
 + C_1\sinh^22\lambda x
 \\ \phantom{T(x,y)=}
 {}+C_1\sinh^22\lambda y)- A_0C_0\bigl(\sinh^{-2}\lambda x\cdot \sinh^{-2}\lambda y
 + \cosh^{-2}\lambda x\cdot \cosh^{-2}\lambda y\bigr)
\\ \phantom{T(x,y)=}
{} - 4A_0C_1(\sinh^2\lambda x+\sinh^2\lambda y+ 2\sinh^2\lambda x\cdot\sinh^2\lambda y)\\
\phantom{T(x,y)=}{}
 - 4A_1C_0\sinh^{-2}2\lambda x\cdot\sinh^{-2}2\lambda y.
\end{gather*}
{\rm(Rat-$B_2$-S):}\qquad$(\langle t^{-2},t^2\rangle;
 \langle t^{-2},t^2\rangle)$
 \begin{gather*}
  v(x)  = C_0x^{-2}+C_1x^2,\qquad
  w(y) = C_0y^{-2}+C_1y^2,
  \\
 u^-(x-y)  = \frac{A_0}{(x-y)^2}+A_1(x-y)^2,\qquad
 u^+(x+y) = \frac{A_0}{(x+y)^2}+A_1(x+y)^2,
\\
 R(x,y)  =\frac{A_0}{(x-y)^2}+\frac{A_0}{(x+y)^2}+A_1(x-y)^2+A_1(x+y)^2
 +\frac{C_0}{x^2}+\frac{C_0}{y^2}+C_1x^2+C_1y^2,
\\
T(x,y)  =\frac{16A_0C_0+16A_0C_1x^2y^2}{(x^2-y^2)^2}
+16A_1C_1x^2y^2.
\end{gather*}
{\rm(Toda-$D_2^{(1)}$-S-bry):}\qquad$(\langle \cosh2\lambda,
\cosh4\lambda t\rangle;
 \langle \sinh^{-2}2\lambda t\rangle,\langle\sinh^{-2}2\lambda t\rangle)$
 \begin{gather*}
  v(x)  = C_0\sinh^{-2}2\lambda x,\qquad
  w(y) = C_1\sinh^{-2}2\lambda y,
  \\
 u^-(x-y)  = A_0\cosh2\lambda(x-y)+ A_1\cosh4\lambda(x-y),
 \\
 u^+(x+y)  = A_0\cosh2\lambda(x+y)+ A_1\cosh4\lambda(x+y),
 \\
 R(x,y)  =A_0\cosh2\lambda(x-y)+A_0\cosh2\lambda(x+y)+ A_1\cosh4\lambda(x-y)
 \\ \phantom{R(x,y)  =}
{}+A_1\cosh4\lambda(x+y) +C_0\sinh^{-2}2\lambda x + C_1\sinh^{-2}2\lambda y,\\
T(x,y)  =8A_1(C_0\cosh4\lambda y+C_1\cosh4\lambda x).
\end{gather*}
{\rm(Toda-$B_2^{(1)}$-S-bry):}\qquad$(\langle e^{-2\lambda t},
e^{-4\lambda t}\rangle;
 \langle e^{4\lambda t}\rangle, \langle\sinh^{-2}2\lambda t \rangle)$
 \begin{gather*}
v(x)  = C_0e^{4\lambda x},\qquad w(y) = C_1\sinh^{-2}2\lambda y,
\\
u^-(x-y)  = A_0e^{-2\lambda(x-y)}+A_1e^{-4\lambda(x-y)},\qquad
u^+(x+y) = A_0e^{-2\lambda(x+y)}+A_1e^{-4\lambda(x+y)},
\\
R(x,y)  =A_0e^{-2\lambda(x-y)}\!+\!A_0e^{-2\lambda(x+y)}
\!+\!A_1e^{-4\lambda(x-y)}\!+\!A_1e^{-4\lambda(x+y)}
    \!+\! C_0e^{4\lambda x}\!+\!C_1\sinh^{-2}2\lambda y,
    \\
T(x,y)  =4A_1\bigl(C_0\cosh4\lambda y+2C_1e^{-4\lambda x}\bigr).
\end{gather*}
{\rm(Toda-$C_2^{(1)}$-S):}\qquad$( \langle e^{-2\lambda t},
e^{-4\lambda t}\rangle,0;
 \langle e^{4\lambda t}\rangle, \langle e^{-4\lambda t}\rangle)$
\begin{gather*}
v(x)  = C_0e^{4\lambda x},\qquad w(y) = C_1e^{-4\lambda y},
\\
u^-(x-y)  = A_0e^{-2\lambda(x-y)} + A_1e^{-4\lambda(x-y)},\qquad
u^+(x+y) =0,
\\
R(x,y)  =A_0e^{-2\lambda(x-y)} + A_1e^{-4\lambda(x-y)}
+C_0e^{4\lambda x}+C_1e^{-4\lambda y},
\\
T(x,y)  =2A_1\bigl(C_0e^{4\lambda y}+C_1e^{-4\lambda x} \bigr).
\end{gather*}
{\rm(Trig-$A_1$-S-bry):}\qquad$(\langle\sinh^{-2}\lambda t,
\sinh^{-2}2\lambda t\rangle,0;
 \langle e^{-4\lambda t},e^{4\lambda t}\rangle)$
\begin{gather*}
 v(x)  = C_0e^{-4\lambda x}+C_1e^{4\lambda x},\qquad
w(y) = C_0e^{-4\lambda y}+C_1e^{4\lambda y},
\\
u^-(x-y)  = A_0\sinh^{-2}\lambda(x-y) +
A_1\sinh^{-2}2\lambda(x-y),\qquad u^+(x+y) =0,
\\
R(x,y)  =A_0\sinh^{-2}\lambda(x-y)\!+\!A_1\sinh^{-2}2\lambda(x-y)
  \!+ \!C_0e^{-4\lambda x}\!+\!C_0e^{-4\lambda y}\! +\! C_1e^{4\lambda x}\!+\!C_1e^{4\lambda y},
\\
T(x,y)  =   2A_1\sinh^{-2}2\lambda(x-y)
 (C_0e^{-4\lambda x}+C_0e^{-4\lambda y} +C_1e^{4\lambda x}+C_1e^{4\lambda y} )
\\ \phantom{T(x,y)  =}{}
+4A_0\sinh^{-2}\lambda(x-y)(C_0e^{-2\lambda(x+y)}+C_1e^{2\lambda(x+y)}).
\end{gather*}
{\rm(Rat${}^d$-$D_2$-S-bry):}\qquad$(\langle t^2,
t^4\rangle;\langle t^{-2}\rangle, \langle t^{-2}\rangle)$
\begin{gather*}
v(x)  = C_0x^{-2},\qquad w(y) = C_1y^{-2},
\\
 u^-(x-y)  = A_0(x-y)^2+A_1(x-y)^4,\qquad
 u^+(x+y) =A_0(x+y)^2+A_1(x+y)^4,
 \\
 R(x,y)  =2A_0(x^2+y^2)+2A_1(x^4+6x^2y^2+y^4)+ \frac{C_0}{x^2}+\frac{C_1}{y^2},
 \\
T(x,y)  =32A_1(C_0y^2+C_1x^2).
\end{gather*}
\end{theorem}
%%%
\begin{proof}
(Ellip-$B_2$-S): We have the following from \eqref{eq:B2Q},
Lemma~\ref{lem:B2Tsym} and \eqref{eq:Landen}.
\begin{gather*}
  Q(\wp(t;\omega_1,2\omega_2);\wp(t;\omega_1,2\omega_2))=
 \wp(x;\omega_1,2\omega_2)\wp(y;\omega_1,2\omega_2),
\\
  Q(\wp(t;2\omega_1,2\omega_2);\wp(t;\omega_1,2\omega_2))
= Q(\wp(t;2\omega_1,2\omega_2);\wp(t;2\omega_1,2\omega_2)
\\ \qquad
{}+\wp(t
2A_1\sinh^{-2}2\lambda(x-y)(C_0+\omega_1;2\omega_1,2\omega_2))
\\ \qquad
{}=\wp(x;2\omega_1,2\omega_2)\wp(y;2\omega_1,2\omega_2)+
\wp(x+\omega_1;2\omega_1,2\omega_2)\wp(y+\omega_1;2\omega_1,2\omega_2),
\\
Q(\wp(t;\omega_1,2\omega_2);\wp(t;\omega_1,\omega_2))
=\wp(x;\omega_1,2\omega_2)\wp(y;\omega_1,2\omega_2)
\\ \qquad
{} +
\wp(x+\omega_2;\omega_1,2\omega_2)\wp(y+\omega_2;\omega_1,2\omega_2),
\\
Q(\wp(t;2\omega_1,2\omega_2);\wp(t;\omega_1,\omega_2))
 =\sum\limits_{j=0}^3\wp(x+\omega_j;2\omega_1,2\omega_2)\wp(y+\omega_j;2\omega_1,2\omega_2).
\end{gather*}

(Rat-$B_2$-S) is given in \cite[(7.13)]{Oshima1995} but it is easy
to check \eqref{eq:b2-2} or prove the result as a~limit of
(Trig-$B_2$-S). (Rat${}^d$-$D_2$-S-bry) follows from
$T(t^2;t^{-2},0)=0$ and $T(t^4;t^{-2},0)=32y^2$
(cf.~Remark~\ref{rem:B2w0} i)). Moreover (Trig-$B_2$-S),
(Toda-$D_2^{(1)}$-S-bry), (Toda-$C_2^{(1)}$-S) and
(Trig-$A_1$-S-bry) are obtained from the corresponding normal
cases together with Lemma~\ref{lem:B2Tsym}. For example, $Q$ for
(Trig-$B_2$-S) is given by \eqref{eq:B23}, \eqref{eq:B23b} and
\begin{gather*}
  Q(\sinh^{-2}\lambda t;\sinh^{-2}2\lambda t)= Q\biggl(\sinh^{-2}\lambda t;
\frac14\sinh^{-2}\lambda t-\frac14\cosh^{-2}\lambda t\biggr)
\\  \phantom{Q(\sinh^{-2}\lambda t;\sinh^{-2}2\lambda t)}
{} =\frac14(\sinh^{-2}\lambda t\sinh^{-2}\lambda
y+\cosh^{-2}\lambda x\cosh^{-2}\lambda y),
\\
 Q(\sinh^{-2}2\lambda t;\sinh^{-2}2\lambda t) = \sinh^{-2}2\lambda x\cdot \sinh^{-2}2\lambda y,
\\
 Q(\sinh^{-2}2\lambda t;\sinh^22\lambda t) = 0.
\end{gather*}
Thus we get Theorem~\ref{thm:B2S} from Theorem~\ref{thm:B2Normal}.
\end{proof}
%%%
%%%%%%%%%%%%%%%
\subsection{Duality}
\begin{definition}[Duality in $\boldsymbol{B_2}$, \cite{Ochiai1996}]\label{defn:dual}
Under the coordinate transformation
\begin{gather*}
 (x,y)\mapsto(X,Y)=\biggl(\frac{x+y}{\sqrt 2},\frac{x-y}{\sqrt 2}\biggr)
\end{gather*}
the pair $(P,P^2 - P_2)$ also satisf\/ies \eqref{eq:b2-0}, which
we call the {\it duality\/} of the commuting dif\/ferential
operators of type $B_2$.
\end{definition}

Denoting $\partial_x=\partial/{\partial x}$,
$\partial_y=\partial/{\partial y}$ and put
\begin{gather*}
 L=P^2 - P_2 - \left(\frac12\partial_x^2-\frac12\partial_y^2 +w-v\right)^2 +u^-
 (\partial_x+\partial_y)^2+u^+(\partial_x-\partial_y)^2.
\end{gather*}
Then the order of $L$ is at most 2 and the second order term of
$L$ is
\begin{gather*}
-(u^++u^-+v+w)(\partial_x^2+\partial_y^2) -
2(u^--u^+)\partial_x\partial_y + 2w\partial_x^2 + 2v\partial_y^2
-(w-v)(\partial_x^2-\partial_y^2)
\\ \qquad
{}  +u^-(\partial_x+\partial_y)^2+u^+(\partial_x-\partial_y)^2 =
0.
\end{gather*}
Since $L$ is self-adjoint, $L$ is of order at most 0 and the 0-th
order term of $L$ is
\begin{gather*}
 -\frac12(\partial_x^2+\partial_y^2)(u^++u^-+v+w) + (u^++u^-+v+w)^2 - 4vw-T
  - \partial_x\partial_y(u^--u^+)
\\ \qquad
{} -\frac12(\partial_x^2-\partial_y^2)(w-v) = (u^++u^-+v+w)^2 -
4vw-T
\end{gather*}
and therefore we have the following proposition.
%%%%%%%%%%
\begin{proposition}[\cite{Ochiai1996, Ochiai2003}]
 {\rm i)}
By the duality in Definition~{\rm \ref{defn:dual}} the pair
$\bigl(R(x,y),T(x,y)\bigr)$ changes into $\bigl(\tilde
R(x,y),\tilde T(x,y)\bigr)$ with
\begin{gather*}
 \tilde R(x,y) = v\biggl(\dfrac{x+y}{\sqrt2}\biggr) + w\biggl(\dfrac{x-y}{\sqrt2}\biggr)
 + u^+(\sqrt{2}x) + u^-(\sqrt{2}y),
\\
 \tilde T(x,y) = \tilde R(x,y)^2
 - 4v\biggl(\dfrac{x+y}{\sqrt2}\biggr)w\biggl(\dfrac{x-y}{\sqrt2}\biggr)
 - T\biggl(\dfrac{x+y}{\sqrt2},\dfrac{x-y}{\sqrt2}\biggr).
\end{gather*}

{\rm ii)} Combining the duality with the scaling map
$R(x,y)\mapsto c^{-2}R(cx,cy)$, the following pair
$\bigl(R^d(x,y),T^d(x,y)\bigr)$ defines commuting differential
operators if so does $\bigl(R(x,y),T(x,y)\bigr)$. This $R^d(x,y)$
is also called the {\it dual} of $R(x,y)$,
\begin{gather*}
R^d(x,y) =  v(x+y) + w(x-y) + u^+(2x) + u^-(2y),
\\
T^d(x,y) = R^d(x,y)^2 - 4v(x+y)w(x-y) - T(x+y,x-y).
\end{gather*}
\end{proposition}

\begin{remark} {\rm i)}
We list up the systems of type $B_2$ given in Sections~\ref{sec:B2N} and
\ref{sec:B2S}:
\begin{gather*}
(u^-(t),u^+(t);v(t),w(t)) \tag*{\it Symbol}
\\
(\langle \wp(t)\rangle;\langle
\wp(t),\wp(t+\omega_1),\wp(t+\omega_2),\wp(t+\omega_3)\rangle),
\tag{\rm Ellip-$B_2$}
\\
(\langle
\wp(t;2\omega_1,2\omega_2),\wp(t;\omega_1,2\omega_2)\rangle;
\langle
\wp(t;\omega_1,2\omega_2),\wp(t;\omega_1,\omega_2)\rangle),
\tag{\rm Ellip-$B_2$-S}
\\
(\langle \sinh^{-2}\lambda t\rangle;\langle \sinh^{-2}\lambda t,
\sinh^{-2}2\lambda t,\sinh^2\lambda t,\sinh^22\lambda t\rangle),
\tag{\rm Trig-$B_2$}
\\
(\langle \sinh^{-2}\lambda t,\sinh^{-2}2\lambda t\rangle;
\langle\sinh^{-2}2\lambda t,\sinh^22\lambda t\rangle) \tag{\rm
Trig-$B_2$-S},
\\
(\langle t^{-2}\rangle;\langle t^{-2},t^2,\ t^4,\ t^6\rangle),
\tag{\rm Rat-$B_2$}
\\
(\langle t^{-2},t^2\rangle;\langle t^{-2},t^2\rangle), \tag{\rm
Rat-$B_2$-S}
\\
(\langle \cosh2\lambda t\rangle;\langle \sinh^{-2}\lambda
t,\sinh^{-2}{2\lambda t}\rangle, \langle \sinh^{-2}\lambda
t,\sinh^{-2}{2\lambda t}\rangle), \tag{\rm Toda-$D_2^{(1)}$-bry}
\\
(\langle\cosh\lambda t,\cosh2\lambda t\rangle;\langle
\sinh^{-2}\lambda t\rangle, \langle \sinh^{-2}\lambda t\rangle),
\tag{\rm Toda-$D_2^{(1)}$-S-bry}
\\
(\langle e^{-2\lambda t}\rangle;\langle e^{2\lambda t},\
e^{4\lambda t}\rangle,
 \langle \sinh^{-2}\lambda t,\ \sinh^{-2}{2\lambda t}\rangle)
\tag{\rm Toda-$B_2^{(1)}$-bry},
\\
(\langle e^{-\lambda t},e^{-2\lambda t}\rangle; \langle
e^{2\lambda t}\rangle,\langle \sinh^{-2}\lambda t\rangle),
\tag{\rm Toda-$B_2^{(1)}$-S-bry}
\\
(\langle e^{-2\lambda t}\rangle,0;\langle e^{2\lambda
t},e^{4\lambda t}\rangle, \langle e^{-2\lambda t},\ e^{-4\lambda
t}\rangle), \tag{\rm Toda-$C_2^{(1)}$}
\\
(\langle e^{-2\lambda t},e^{-4\lambda t}\rangle,0;\langle
e^{4\lambda t}\rangle, \langle e^{-4\lambda t}\rangle), \tag{\rm
Toda-$C_2^{(1)}$-S}
\\
(\langle\sinh^{-2}\lambda t\rangle,0;\langle e^{-2\lambda t},
e^{-4\lambda t}, e^{2\lambda t}, e^{4\lambda t}\rangle), \tag{\rm
Trig-$A_1$-bry}
\\
(\langle \sinh^{-2}\lambda t,\sinh^{-2}{2\lambda t}\rangle,0;
\langle e^{-4\lambda t},\ e^{4\lambda t}\rangle), \tag{\rm
Trig-$A_1$-S-bry}
\\
(\langle t^{-2}\rangle,0;\langle t,t^2,t^3,t^4\rangle), \tag{\rm
Rat-$A_1$-bry}
\\
(\langle t^{-2}\rangle,\langle t^{-2}\rangle;\langle
t^2,t^4\rangle) \tag{\rm Rat-$D_2$-S-bry}.
\end{gather*}
The f\/irst 6 cases above are classif\/ied by \cite{Ochiai2003} as
invariant systems. The systems such that at least two of $u^\pm$,
$v$ and $w$ are not entire are classif\/ied by \cite{Ochiai1996}.
(Trig-$*$) and (Toda-$*$) in the above are classif\/ied by
\cite{Oshima2005} as certain systems with periodic potentials.

We do not put $(\langle t^{-2}, t^2\rangle,0;\langle
t,t^2\rangle)$ in the list which corresponds to
(Rat-$A_1^{}$-S-bry) because its dual def\/ines a direct sum of
trivial operators (cf.~Section~\ref{sec:class}).

{\rm ii)} Since $1-\sinh^{-2}t+4\sinh^{-2}2t=\coth^2 t=t^2 -(2/3)
t^4 + o(t^4)$ and $\sinh^2 t = t^2 +(2/3) t^4 + o(t^4)$, we have
$\sinh^22\lambda x+\sinh^22\lambda y - 2\coth^2\lambda(x-y) -
2\coth^2\lambda(x+y)=8\lambda^4(x^2+y^2)^2+o(\lambda^4)$. Hence
the potential function
\begin{gather*}
\frac{A_0}{(x-y)^2}+\frac{A_0}{(x+y)^2}
+\frac{C_0}{x^2}+\frac{C_0}{y^2}+C_1(x^2+y^2)+ A_1(x^2+y^2)^2
\end{gather*}
is an analytic continuation of that of (Trig-$B_2$-S) but this is
not a completely integrable potential function of type $B_2$

{\rm iii)} The dual is indicated by superf\/ix ${}^d$. For
example, the dual of (Ellip-$B_2$) is denoted by
(Ellip${}^d$-$B_2$) whose potential function is
\begin{gather*}
  R(x,y) = A\wp(2x) + A\wp(2y)+\sum\limits_{j=0}^3C_j(\wp(x-y+\omega_j)+\wp(x+y+\omega_j))
\end{gather*}
and the dual of (Toda-$C_2^{(1)}$) is
\begin{gather*}
(\langle e^{-2\lambda t},e^{-4\lambda t}\rangle, \langle
e^{2\lambda t},e^{4\lambda t}\rangle;0,\langle e^{-4\lambda
t}\rangle) \tag{\rm Toda${}^d$-$C_2^{(1)}$}
\end{gather*}
since the dual of $\bigl(u^-(t),u^+(t);v(t),w(t)\bigr)$ is
$\bigl(w(t),v(t);u^+(2t),u^-(2t)\bigr)$.  Similarly
\begin{gather*}
(\langle\wp(t;\omega_1,2\omega_2),\wp(t;\omega_1,\omega_2)\rangle;
  \langle\wp(2t;2\omega_1,2\omega_2),\wp(2t;\omega_1,2\omega_2)\rangle)
  \tag{\rm Ellip${}^d$-$B_2$-$S$}.
\end{gather*}
Since $4\wp(2t;2\omega_1,2\omega_2)=\wp(t;\omega_1,\omega_2)$
etc., (Ellip${}^d$-$B_2$-$S$) coincides with (Ellip-$B_2$-$S$) by
replacing $(\omega_1,\omega_2)$ by $(2\omega_2,\omega_1)$.
\end{remark}
Then we have the following diagrams and their duals, where the
arrows with double lines represent specializations of parameters.
For example, ``Trig-$B_2$ $\overset{5:3}\Rightarrow$
Trig-$BC_2$-reg'' means that 2 parameters (coupling constants
$C_2$ and $C_3$) out of 5 in the potential function (Trig-$B_2$)
are specialized to get the potential function (Trig-$BC_2$-reg)
with 3 parameters. For the normal cases see
Def\/inition~\ref{def:restrict} and the diagrams in the last part
of the next section (type $B_n$).
\begin{gather*}
\text{\bf Hierarchy of Normal Integrable Potentials of type $\boldsymbol{B_2}$}\notag\\
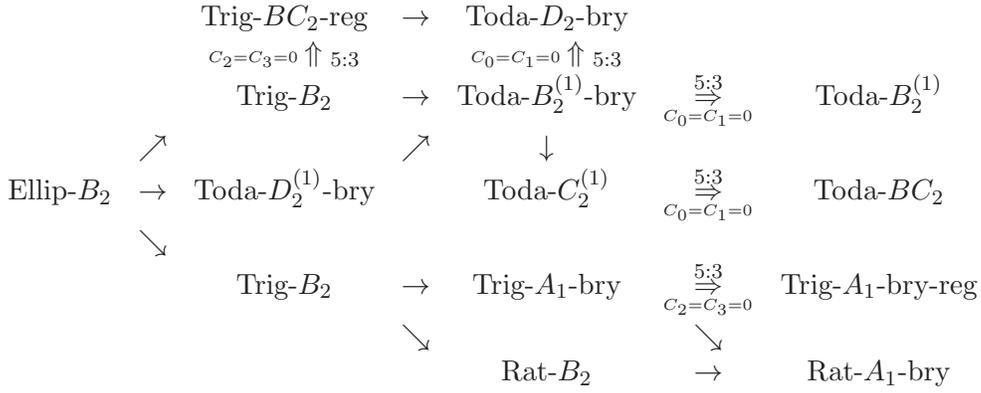
\begin{matrix}
 & &\text{\rm Trig-$BC_2$-reg}&\to & \text{\rm Toda-$D_2$-bry}\\
 & &{}_{{}^{C_2=C_3=0}\!}\Uparrow{}_{5:3}&&
   {}_{{}^{C_0=C_1=0}\!}\Uparrow{}_{5:3}\\
 & &\text{\rm Trig-$B_2$}&\to&\text{\rm Toda-$B_2^{(1)}$-bry}
  & \underset{{}^{C_0=C_1=0}}{\overset{5:3}\Rightarrow}
  & \text{\rm Toda-$B_2^{(1)}$}\\
 &\nearrow & & \nearrow& \downarrow\\
 \text{\rm Ellip-$B_2$}&\to&\text{\rm Toda-$D_2^{(1)}$-bry}&&
 \text{\rm Toda-$C_2^{(1)}$} &
  \underset{{}^{C_0=C_1=0}}{\overset{5:3}\Rightarrow}&\text{\rm Toda-$BC_2$}\\
 &\searrow\\
 & &\text{\rm Trig-$B_2$}&\to&\text{\rm Trig-$A_1$-bry}
  &\underset{{}^{C_2=C_3=0}}{\overset{5:3}\Rightarrow}
  &\text{\rm Trig-$A_1$-bry-reg}\\
 & & &\searrow & &\searrow\\
 & & & &\text{\rm Rat-$B_2$} &\to &\text{\rm Rat-$A_1$-bry}\\
\end{matrix}
\end{gather*}
%%%%  Special B_2 %%%%%
\begin{gather*}
\text{\bf Hierarchy of Special Integrable Potentials of type $\boldsymbol{B_2}$}\notag\\
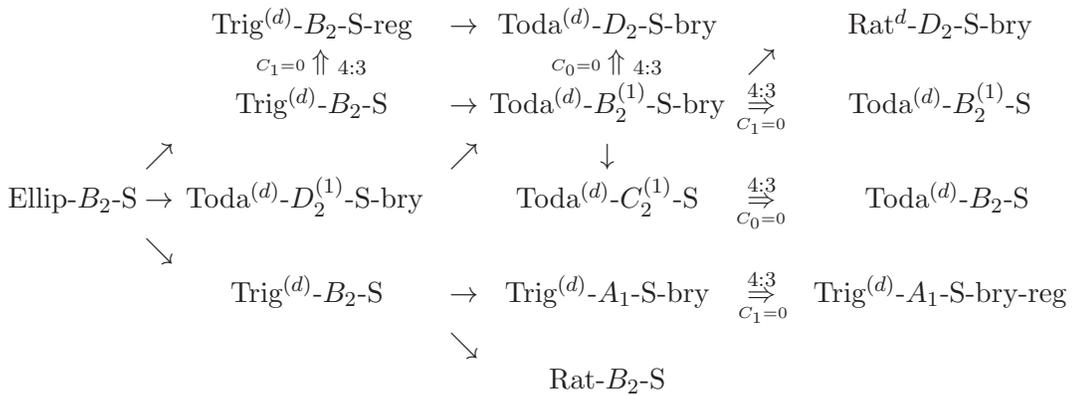
\begin{matrix}
 & &\text{\makebox[0pt]{\rm Trig$^{(d)}$-$B_2$-S-reg}}&
   \to & \text{\makebox[0pt]{\rm Toda$^{(d)}$-$D_2$-S-bry}}
 & &\text{\rm Rat${}^d$-$D_2$-S-bry}\\
 & &{}_{{}^{C_1=0}\!}\Uparrow{}_{4:3}&&{}_{{}^{C_0=0}\!}\Uparrow{}_{4:3}
 & \nearrow\\
 & &\text{\rm Trig$^{(d)}$-$B_2$-S}&\to&\text{\makebox[0pt]
 {\rm Toda$^{(d)}$-$B_2^{(1)}$-S-bry}} & \underset{{}^{C_1=0}}{\overset{4:3}\Rightarrow}
  & \text{\rm Toda$^{(d)}$-$B_2^{(1)}$-S}\\
 &\nearrow & & \nearrow&\downarrow\\
 \text{\rm Ellip-$B_2$-S\!\!\!\!}&\to&\text{
{\rm \!\!\!\!\!Toda$^{(d)}$-$D_2^{(1)}$-S-bry}}&
  &\text{\rm Toda$^{(d)}$-$C_2^{(1)}$-S}
  &\underset{{}^{C_0=0}}{\overset{4:3}\Rightarrow}
  &\text{\rm Toda$^{(d)}$-$B_2$-S\!\!\!}\\
 &\searrow\\
 & &\text{\rm \!\!Trig$^{(d)}$-$B_2$-S}&\to&\text{\rm Trig$^{(d)}$-$A_1$-S-bry}
 &  \underset{{}^{C_1=0}\!}{\overset{4:3}\Rightarrow} &\text{
{\rm \!\!\!\! Trig$^{(d)}$-$A_1$-S-bry-reg}}\\
 & & &\searrow\\
 & & & &\text{\rm Rat-$B_2$-S}& &
\end{matrix}
\end{gather*}

\begin{definition}\label{def:restrictS}
We define some potential functions as specializations.
\begin{description}\itemsep=0pt
\item[\rm(Trig-$B_2$-S-reg)] {\it Trigonometric special potential
of type $B_2$ with regular boundary conditions} is (Trig-$B_2$-S)
with $C_1=0$. %%
\item[\rm(Toda-$D_2$-S-bry)] {\it Toda special potential of type
$D_2$ with boundary conditions} is (Toda-$B_2^{(1)}$-S-bry) with
$C_0=0$. \item[\rm(Toda-$B_2^{(1)}$-S)] {\it Toda special
potential of type $B_2^{(1)}$} is (Toda-$B_2^{(1)}$-S-bry) with
$C_1=0$. \item[\rm(Toda-$B_2$-S)] {\it Toda special potential of
type $B_2$} is (Toda-$C_2^{(1)}$-S-bry) with $C_0=0$.
\item[\rm(Trig-$A_1$-S-bry-reg)] {\it Trigonometric special
potential of type $A_1$ with regular boundary conditions} is
(Trig-$A_1$-S-bry) with $C_1=0$.
\end{description}
\end{definition}
\begin{remark}
We have some equivalences as follows:
\begin{gather*}
\text{\rm (Ellip-$B_2$-S) } = \text{\rm (Ellip${}^{d}$-$B_2$-S)},
\\[.3ex]
\text{\rm (Rat-$B_2$-S) } = \text{\rm (Rat${}^{d}$-$B_2$-S)},
\\[.3ex]
\text{\rm (Trig-$BC_2$-reg) } = \text{\rm
(Trig${}^{d}$-$B_2$-S-reg)},
\\
\text{\rm (Trig-$A_1$-bry-reg) } = \text{\rm
(Toda${}^d$-$D_2$-S-bry)},
\\[.3ex]
\text{\rm (Toda-$D_2$-bry)} = \text{\rm
(Trig${}^d$-$A_1$-S-bry-reg)},
\\[.3ex]
\text{\rm (Toda-$B_2^{(1)}$)} = \text{\rm
(Toda${}^d$-$B_2^{(1)}$-S)},
\\[.3ex]
\text{\rm (Toda-$BC_2$) } = \text{\rm (Toda${}^d$-$B_2$-S)}.
\end{gather*}
\end{remark}
%%%%%%%%%%%%%%%%%%%%%%%%%%%%%%%%%%%%%%%%%%%%%%%%%%%%%%%%%%%%%%%%%%%%%
\section[Type $B_n$ ($n\ge 3$)]{Type $\boldsymbol{B_n}$ ($\boldsymbol{n\ge 3}$)}\label{sec:Bn}

In this section we construct integrals of the completely
integrable systems of type $B_n$ appearing in the following
diagram. The diagram is also given in \cite[Figure III.1]{Diejen},
where (Toda-$B_n^{(1)}$-bry) is missing. The most general system
(Ellip-$B_n$) is called Inozemtzev model
(cf.~\cite{Inozemtsev1989-1}).
\begin{gather*}
\text{\bf Hierarchy of Integrable Potentials with 5 parameters
($\boldsymbol{n\ge2}$)}
\notag\\
\begin{matrix}
&&\text{\!\!\rm Toda-$D_n^{(1)}$-bry\!}&\to
  &\text{\!\rm Toda-$B_n^{(1)}$-bry\!}&\to&\text{\rm Toda-$C_n^{(1)}$} \\
 &\nearrow& &\nearrow\\
\text{\rm Ellip-$B_n$} &\to &\text{\rm Trig-$B_n$}&\to&\text{\rm Rat-$B_n$}\\
 & &&\searrow&&\searrow\\
 & &&&\text{\!\rm Trig-$A_{n-1}$-bry\!}&\to&\text{\!\rm Rat-$A_{n-1}$-bry}
\end{matrix}
\end{gather*}
\subsection{Integrable potentials}
%%%%%%%%%%%%%%%%%%%%%%%%%%%%%%%%%%%%%%%%%%%%%%%%%%%%%%%%%%%%%%
\begin{definition}\label{def:BnP}
The potential functions $R(x)$ of \eqref{eq:Shr} are as follows:\\
Here $A$, $C_0$, $C_1$, $C_2$ and $C_3$ are any complex
numbers.\vspace{1ex}

\noindent (Ellip-$B_n$) {\it Elliptic potential of type} $B_n$:
\begin{gather*}
\sum\limits_{1\le i<j\le n}A(\wp(x_i-x_j;2\omega_1,2\omega_2)
+\wp(x_i+x_j;2\omega_1,2\omega_2))
+\sum\limits_{k=1}^n\sum\limits_{j=0}^3C_j\wp(x_k+\omega_j;2\omega_1,2\omega_2),
%\label{eq:EBn}
\end{gather*}
(Trig-$B_n$) {\it Trigonometric potential of type} $B_n$:
\begin{gather*}
\sum\limits_{1\le i<j\le n}A(\sinh^{-2}\lambda(x_i-x_j)
+\sinh^{-2}\lambda(x_i+x_j))
\\  \qquad
{}+\sum\limits_{k=1}^n\left(C_0\sinh^{-2}\lambda x_k +
C_1\cosh^{-2}\lambda x_k + C_2\sinh^2\lambda x_k +
\frac{C_3}4\sinh^22\lambda x_k\right),
%\label{eq:Bn}
\end{gather*}
(Rat-$B_n$) {\it Rational potential of type} $B_n$:
\begin{gather*}
\sum\limits_{1\le i<j\le n}\biggl(\frac{A}{(x_i-x_j)^2}
+\frac{A}{(x_i+x_j)^2}\biggr) +\sum\limits_{k=1}^n(C_0x_k^{-2} +
C_1x_k^2 + C_2x_k^4 + C_3x_k^6),
%\label{eq:RBn}
\end{gather*}
(Trig-$A_{n-1}$-bry) {\it Trigonometric potential of type
$A_{n-1}$ with boundary conditions}:
\begin{gather*}
\sum\limits_{1\le i<j\le n}A\sinh^{-2}\lambda(x_i-x_j) +
\sum\limits_{k=1}^n(C_0e^{-2\lambda x_k} + C_1e^{-4\lambda
x_k}+C_2e^{2\lambda x_k} + C_3e^{4\lambda x_k}),
%\label{eq:Anb}
\end{gather*}
(Toda-$B_n^{(1)}$-bry) {\it Toda potential of type $B_n^{(1)}$
with boundary conditions}:
\begin{gather*}
\sum\limits_{i=1}^{n-1}\!A e^{-2\lambda(x_i-x_{i+1})}
 \!+\! Ae^{-2\lambda(x_{n-1}\!+\!x_n)} \!+\! C_0e^{2\lambda x_1} \!+\! C_1e^{4\lambda x_1}
 \!+\! C_2\sinh^{-2}\lambda x_n \!+\!  C_3\sinh^{-2}2\lambda x_n,
\end{gather*}
(Toda-$C_n^{(1)}$) {\it Toda potential of type} $C_n^{(1)}$:
\begin{gather*}
 \sum\limits_{i=1}^{n-1}A e^{-2\lambda(x_i-x_{i+1})}
 + C_0e^{2\lambda x_1} + C_1e^{4\lambda x_1}
 + C_2e^{-2\lambda x_n}+ C_3e^{-4\lambda x_n},
 %\label{eq:ECn}
\end{gather*}
(Toda-$D_n^{(1)}$-bry) {\it Toda potential of type $D_n^{(1)}$
with boundary conditions}:
\begin{gather*}
\sum\limits_{i=1}^{n-1} Ae^{-2\lambda(x_i-x_{i+1})}
 + Ae^{-2\lambda(x_{n-1}+x_n)} + Ae^{2\lambda(x_1+x_2)}
 \\ \qquad
 {}+ C_0\sinh^{-2}\lambda x_1 + C_1\sinh^{-2}{2\lambda x_1}
  + C_2\sinh^{-2}\lambda x_n + C_3\sinh^{-2}{2\lambda x_n},
  %\label{eq:EDnb}
 \end{gather*}
(Rat-$A_{n-1}$-bry) {\it Rational potential of type $A_{n-1}$ with
boundary conditions}:
\begin{gather*}
 \sum\limits_{1\le i<j\le n}\frac{A}{(x_i-x_j)^2}
 + \sum\limits_{k=1}^n(C_0x_k+C_1x_k^2+C_2x_k^3+C_3x_k^4).
 %\label{eq:Anbry}
\end{gather*}
\end{definition}
%%%%%%%%%%%%%%%%%%%%%%%%%%%%%%%%%%%%%%%%%%%%%%%%%
\begin{remark}
In these cases the Schr\"odinger operator $P$ is of the form
\begin{gather*}
%\label{eq:BnPR}
 P = -\frac12\sum\limits_{j=1}^n\frac{\partial^2}{\partial x_j^2}+R(x),
 \\
 R(x) = \sum\limits_{1\le i<j\le n}(u_{e_i-e_j}(x)+u_{e_i+e_j}(x))
 + \sum\limits_{j=0}^3C_j v^j(x),
 \qquad
 v^j(x) = \sum\limits_{k=1}^n v_{e_k}^j(x).
\end{gather*}
Here
\begin{gather*}
 \partial_a u_\alpha(x)=\partial_b v_\beta^j(x)=0\qquad\text{if}\qquad a,b\in\mathbb R^n
 \qquad\text{satisfy}\qquad
 \langle a,\alpha\rangle=\langle b,\beta\rangle=0.
\end{gather*}
\end{remark}
%%%

The complete integrability of the invariant systems (Ellip-$B_n$),
(Trig-$B_n$) and (Rat-$B_n$) is established by \cite{Oshima1998}.
We review their integrals and then we prove that the other f\/ive
systems are also completely integrable by constructing enough
integrals, which is announced by~\cite{Oshima2005}. The complete
integrability of (Trig-$A_{n-1}$-bry), (Toda-$C_n^{(1)}$),
(Toda-$D_n^{(1)}$-bry) and \mbox{(Rat-$A_{n-1}$-bry)} is presented
as an unknown problem by \cite{Diejen} and then that of
(Toda-$B_n^{(1)}$-bry), (Toda-$C_n^{(1)}$) and
(Toda-$D_n^{(1)}$-bry) are established by \cite{Kuznetsov1997,
Kuznetsov1995, Kuznetsov1994} using $R$-matrix method. The compete
integrability of (Trig-$A_{n-1}$-bry) and (Rat-$A_{n-1}$-bry)
seems to have not been proved.
%%%%%%%%%%%%%%%%%%%%%%%%%%%%%%%%%%%%%%%%%%%%%%%%%
\begin{definition}[\cite{Ochiai1994, Oshima1998}]\label{def:Bn}
Let $u_\alpha(x)$ and $T_I^o(v^j)$ are functions given for
$\alpha\in\Sigma(D_n)$ and subsets
$I=\{i_1,\dots,i_k\}\subset\{1,\dots,n\}$ such that
\begin{gather*}
 u_\alpha(x) = u_{-\alpha}(x)\qquad\text{and}\qquad
 \partial_y u_\alpha=0\qquad\text{for}\qquad y\in\mathbb R^n\qquad\text{with}\qquad
 \langle\alpha,y\rangle=0.
\end{gather*}
Def\/ine a dif\/ferential operator
\begin{gather*}
 P_n(C) = \sum\limits_{k=0}^n\frac1{k!(n-k)!}\sum\limits_{w\in\mathfrak S_n}
       \big(q_{\{w(1),\ldots,w(k)\}}(C)\cdot\Delta_{\{w(k+1),\ldots,w(n)\}}^2\big)
\end{gather*}
by
\begin{gather}
\Delta_{\{i_1,\ldots,i_k\}} = \sum\limits_{0\le j\le [\frac
k2]}\frac1{2^kj!(k-2j)!}\label{eq:BnDelta}
\\ \phantom{\Delta_{\{i_1,\ldots,i_k\}} =}
{}\times\sum\limits_{w\in
W(B_k)}\!\varepsilon(w)w(u_{e_{i_1}-e_{i_2}}(x)
u_{e_{i_3}-e_{i_4}}(x)\cdots u_{e_{i_{2j-1}}-e_{i_{2j}}}(x)\cdot
 \partial_{i_{2j+1}}\partial_{i_{2j+2}}\cdots\partial_{i_k}),
 \nonumber
 \\
q_{\{i_1,\ldots,i_k\}}(C)= \sum\limits_{\nu=1}^{k}\
\sum\limits_{I_1\amalg\cdots\amalg I_\nu=\{i_1,\ldots,i_k\}}
T_{I_1}\cdots T_{I_\nu}, \label{eq:Bnq}
\\
T_I = (-1)^{\# I -1}\Biggl(CS_{I}^o -
\sum\limits_{j=0}^3C_jT_{I}^o(v^j)\Biggr), \label{eq:BnTI}
\end{gather}
where
\begin{gather*}
  S_{\{i_1,i_2,\dots,i_k\}}^o =\frac12\sum\limits_{w\in W(B_k)}
  w(u_{e_{i_1}-e_{i_2}}(x)u_{e_{i_2}-e_{i_3}}(x)\cdots u_{e_{i_{k-1}}-e_{i_k}}(x)),
  \\
  S_\varnothing^o = 0,\ S_{\{k\}}^o = 1,\qquad
  S_{\{i,j\}}^o = 2u_{e_i-e_j}(x)+2u_{e_i+e_j}(x),
  \\
  T_{\varnothing}^o(v^j) = 0,\qquad
  T_{\{k\}}^o(v^j) = 2v_{e_k}^j(x)\qquad\text{for}\qquad 1\le k\le n,
  \\
  q_{\varnothing} = 1,\qquad
  q_{\{i\}} = T_{\{i\}},\qquad
  q_{\{i_1i_2\}} = T_{\{i_1\}}T_{\{i_2\}} + T_{\{i_1,i_2\}}, \qquad \ldots
\end{gather*}
In the above, we identify $W(B_k)$ with  the ref\/lection group
generated by $w_{e_{i_k}}$ and $w_{e_{i_\nu} - e_{i_{\nu+1}}}$
($\nu=1,\dots,k-1$). The sum in \eqref{eq:Bnq} runs over all the
partitions of the set $I$ and the order of the subsets
$I_1,\dots,I_\nu$ is ignored.

Replacing $\partial_j$ by $\xi_j$ for $j=1,\ldots,n$ in the
def\/inition of $\Delta_{\{i_1,\ldots,i_k\}}$ and $P_n(C)$, we
def\/ine functions $\bar \Delta_{\{i_1,\ldots,i_k\}}$ and $\bar
P_n(C)$ of $(x,\xi)$, respectively.
\end{definition}

We will def\/ine $u_\alpha(x)$ and $T^o_I(v^j)$ so that
\begin{gather}
\label{eq:BnC}
 [P_n(C),P_n(C')] = 0\qquad \text{for}\qquad C,\,C'\in\mathbb C.
\end{gather}
Then putting
\begin{gather}
 q^o_I = q_I\big|_{C=0},\nonumber
 \\
 P_n = P_n(0) = \sum\limits_{k=0}^n\frac1{k!(n-k)!}\sum\limits_{w\in\mathfrak S_n}
(q_{\{w(1),\ldots,w(k)\}}^o\Delta_{\{w(k+1),\ldots,w(n)\}}^2)
 \label{eq:BnPn},
 \\
 P_{n-j} =\! \sum\limits_{i=j}^n\sum\limits_{k=i}^n\frac{(-1)^{i-j}}{i!(k\!-\!i)!(n\!-\!k)!}
\sum\limits_{w\in\mathfrak S_n} \sum\limits_{I_1\amalg\cdots\amalg
I_j=w(\{1,\ldots,i\})} \!\!\!\!\!\!\!\!\!\!\!\!\!\!\!\!
 S_{I_1}^o\cdots S_{I_j}^o
     q_{w(\{i+1,\ldots,k\})}^o\Delta_{w(\{k+1,\ldots,n\})}^2,
 \label{eq:BnPj}
\end{gather}
we have $P_n(C)=\sum\limits_{j=0}^nC^jP_{n-j}$ and
\eqref{eq:typeBnI} and then
\begin{gather}
 [P_i,P_j] = 0\qquad\text{for}\qquad  1\le i<j\le n,\nonumber
 \\
 -\frac{P_1}2 = -\frac12\sum\limits_{j=1}^n\frac{\partial^2}{\partial x_j^2}
  + \sum\limits_{1\le i<j\le n}(u_{e_i-e_j}(x)+ u_{e_i+e_j}(x))
  + \sum\limits_{j=0}^3 C_jv^j(x).
\label{eq:BnPnCom}
\end{gather}
\begin{remark}
\label{rem:Tbilin} {\rm i)} When $n=2$, $T(x,y)$ in the last
section corresponds to $T_{12}$, namely
\begin{gather*}
  T(x_1,x_2) = T_{12}\big|_{C=0}.
\end{gather*}

{\rm ii)} Put
\begin{gather*}
  U(x)=\sum\limits_{1\le i<j\le n}(u_{ij}^-(x_i-x_j)+u_{ij}^+(x_i+x_j))
   \qquad\text{and}\qquad
 V(x)=\sum\limits_{k=1}^n v_k(x_k)
\end{gather*}
in \eqref{eq:typeBn} and let $T_{I}(U;V)$ be the corresponding
$T_I$ given by \eqref{eq:BnTI}. Then \cite[Remark 4.3]{Oshima1998}
says
\begin{gather}
 T_{I}(c_0 U;c_1 V+c_2 W) =
  c^{\# I-1}_0c_1 T_{I}(U;V)+ c^{\# I-1}_0c_2 T_{I}(U;W)\qquad\text{for}\qquad
  c_i\in\mathbb C.
\label{eq:Bnlinear}
\end{gather}

{\rm iii)} The def\/inition  \eqref{eq:BnTI} may be replaced by
\begin{gather}
\label{eq:BnConst} T_I = (-1)^{\# I
-1}\Biggl(C\sum\limits_{\nu=1}^{\# I}
\sum\limits_{I_1\amalg\cdots\amalg I_\nu=I}
(-2\lambda)^{\nu-1}(\nu-1)!\cdot S_{I_1}^o\cdots S_{I_\nu}^o -
\sum\limits_{j=0}^3C_jT_{I}^o(v^j)\Biggr)
\end{gather}
because $\lambda$ can be any complex number in \cite[Lemma 5.2
ii)]{Oshima1998} when $v=C$. Note that we f\/ixed $\lambda=1$ in
\cite{Oshima1998}. Combining \eqref{eq:BnConst} and
\eqref{eq:Bnlinear}, we may put
\begin{gather*}
T_I  = (-1)^{\# I -1}\Biggl(C
S_{I}^o+c'\sum\limits_{\nu\ge2}c^{\nu-1}
\sum\limits_{I_1\amalg\cdots\amalg I_\nu=I}S_{I_1}^o\cdots
S_{I_\nu}^o
 - \sum\limits_{j=0}^3C_jT_{I}^o(v^j)\Biggr)
\end{gather*}
for any $c$, $c'\in\mathbb C$ and hence
\begin{gather*}
T_I  = (-1)^{\# I -1}\Biggl(C S_{I}^o + \sum\limits_{\nu\ge
2}c_\nu \sum\limits_{I_1\amalg\cdots\amalg I_\nu=I}S_{I_1}^o\cdots
S_{I_\nu}^o
 - \sum\limits_{j=0}^3C_jT_{I}^o(v^j)\Biggr)
\end{gather*}
for any $c_2,c_3,\ldots\in\mathbb C$.
\end{remark}
%%%
\begin{theorem}[Ellip-$\boldsymbol{B_n}$, {\cite{Ochiai1994},
\cite[Theorem~7.2]{Oshima1998}}]\label{thm:BnEllip} Put
\begin{alignat*}{3}
& u_{e_i\pm e_j}(x) = A\wp_0(x_i\pm x_j;2\omega_1,2\omega_2)
\qquad &&\text{for}\quad 1\le i<j\le n,&
\\
& v_{e_k}^j(x) = \wp_0(x_k+\omega_j;2\omega_1,2\omega_2) \qquad &&
\text{for}\quad 1\le k\le n\quad \text{and}\quad 0\le j\le 3&
\end{alignat*}
and
\begin{gather}
 T_{I}^o(v^j) = \sum\limits_{\nu=1}^{\# I}\sum\limits_{I_1\amalg\cdots\amalg I_\nu=I}
(-A)^{\nu-1}(\nu-1)!\cdot S_{I_1}(v^j)\cdots S_{I_\nu}(v^j),
\label{eq:BnT}
\\
 S_{\{i_1,\ldots,i_k\}}(v^j) = \sum\limits_{w\in W(B_k)}
v_{w(e_{i_1})}^j(x)u_{w(e_{i_1}-e_{i_2})}(x)u_{w(e_{i_2}-e_{i_3})}(x)
\cdots u_{w(e_{i_{k-1}}-e_{i_k})}(x). \label{eq:BnS}
\end{gather}
Then \eqref{eq:BnC} holds.
\end{theorem}
%%%
\begin{example}
Put $v_k^j= v_{e_k}^j$, $\tilde v_k = \sum\limits_{j=0}^3C_j
v^j_k$, and $w^\pm_{ij}=u^-_{ij}\pm u^+_{ij}$. Then
\begin{gather*}
\Delta_{\{1\}} = \partial_1,
\\
\Delta_{\{1,2\}} = \partial_1\partial_2+u^-_{12}-u^+_{12} =
\partial_1\partial_2+w^-_{12},
\\
\Delta_{\{1,2,3\}} =
\partial_1\partial_2\partial_3+w^-_{12}\partial_3+w^-_{23}\partial_1+w^-_{13}\partial_2,
\\
\Delta_{\{1,2,3,4\}}=\partial_1\partial_2\partial_3\partial_4+w^-_{34}\partial_1\partial_2+
w^-_{24}\partial_1\partial_3+w^-_{23}\partial_1\partial_4+w^-_{14}\partial_2\partial_3
+ w^-_{13}\partial_2\partial_4
\\ \phantom{\Delta_{\{1,2,3,4\}}=}
{}+w^-_{12}\partial_3\partial_4
+w^-_{12}w^-_{34}+w^-_{13}w^-_{24}+w^-_{14}w^-_{23},
\\
S_{1}(v^j) = 2v^j_1,
\\
S_{\{1,2\}}(v^j) = 2v^j_1u^-_{12}+2v^j_1u^+_{12}
+2v^j_2u^-_{12}+2v^j_2u^+_{12} =2(v^j_1+v^j_2)(u^-_{12}+u^+_{12}),
\\
 S_{\{1,2,3\}}(v^j) =  2v^j_1u^-_{12}u^-_{23} +2v^j_1u^-_{12}u^+_{23}
 +2v^j_1u^+_{12}u^-_{23} +2v^j_1u^+_{12}u^+_{23} +\cdots
 \\ \phantom{S_{\{1,2,3\}}(v^j)}
 {} = 2(v^j_1+v^j_3)w^+_{12}w^+_{23} +2(v^j_1+v^j_2)w^+_{23}w^+_{13} +2(v^j_2+v^j_3)w^+_{12}w^+_{13},
\\
T_{\{1\}} = CS_{\{1\}} - \sum\limits_{j=0}^3C_jT^o_{\{1\}}(v^j)=
C-2\tilde v_1,
\\
T_{\{1,2\}} = -CS_{\{1,2\}}^o + \sum\limits_{j=0}^3
C_jT^o_{\{1,2\}}(v^j),
\\
q_{\{1\}} = T_{\{1\}},
\\
q_{\{1,2\}} = T_{\{1\}}T_{\{2\}} + T_{\{1,2\}},
\\
q_{\{1,2,3\}} = T_{\{1\}}T_{\{2\}}T_{\{3\}} + T_{\{1,2\}}T_{\{3\}}
 + T_{\{1,3\}}T_{\{2\}}+T_{\{1\}}T_{\{2,3\}}+T_{\{1,2,3\}}.
\end{gather*}
If $T^o(v^j)$ and $S_{i_1,\dots,i_k}(v^j)$ are given by
\eqref{eq:BnT} and \eqref{eq:BnS}, then
\begin{gather*}
T^o_{\{1\}}(v^j) = 2v^j_1,
\\
T^o_{\{1,2\}}(v^j) =S_{\{1,2\}}(v^j) - AT^o_{\{1\}}T^o_{\{2\}} =
2(v_1^j+v_2^j)w_{12}^+ -4Av^j_1v^j_2,
\\
T^o_{\{1,2,3\}}(v^j) = S_{\{1,2,3\}}(v^j) -
2A(v^j_1S_{\{2,3\}}(v^j)+v^j_2S_{\{1,3\}}(v^j)+v^j_3S_{\{1,2\}}(v^j))
 + 16A^2v^j_1v^j_2v^j_3.
\end{gather*}
In  particular, if $n=2$, then
\begin{gather*}
P_2(C) = \Delta^2_{\{1,2\}}
+q_{\{1\}}\Delta^2_{\{2\}}+q_{\{2\}}\Delta^2_{\{1\}}+q_{\{1,2\}} =
\bigl(\partial_1\partial_2+u^-_{12}-u^+_{12}\bigr)^2
\\ \phantom{P_2(C) =}
{}+T_{\{1\}}\partial_2^2+T_{\{2\}}\partial_1^2 +T_{\{1\}}T_{\{2\}}
-CS_{\{1,2\}}^o + \sum\limits_{j=0}^3C_jT^o_{\{1,2\}}(v^j)
\\ \phantom{P_2(C)}
{}= \bigl(\partial_1\partial_2+u^-_{12}-u^+_{12}\bigr)^2 +
(C-2\tilde v_1)\partial^2_2+(C-2\tilde v_2)\partial^2_1
 + (C-2\tilde v_1)(C-2\tilde v_2)
\\ \phantom{P_2(C) =}
{} -2C(u^-_{12}+u^+_{12})+ 2(\tilde v_1+\tilde
v_2)(u^-_{12}+u^+_{12}) -4A\sum\limits_{j=0}^3C_jv^j_1v^j_2 = C^2
-2P\cdot C + P_2
\end{gather*}
with
\begin{gather*}
P=-\frac12(\partial_1^2+\partial_2^2) + \tilde v_1 + \tilde v_2 +
u^-_{12} + u^+_{12},
\\
P_2=\bigl(\partial_1\partial_2\!+\!u^-_{12}\!-\!u^+_{12}\bigr)^2
\!-\! 2\tilde v_1\partial_2^2 \!-\! 2\tilde
v_2\partial^2_1\!+\!4\tilde v_1\tilde v_2 \!+\!2(\tilde
v_1\!+\!\tilde
v_2)(u^-_{12}\!+\!u^+_{12})\!-\!4A\sum\limits_{j=0}^3C_jv^j_1v^j_2,
\end{gather*}
which should be compared with \eqref{eq:b2-1}, \eqref{eq:defQ} and
\eqref{eq:B2Q}.

In general
\begin{gather}
 P_1 = \sum\limits_{k=1}^n(\Delta^2_{\{k\}}+q^o_{\{k\}})
 -\sum\limits_{1\le i<j\le n}S_{\{i,j\}}^o
= \sum\limits_{k=1}^n(\partial_k^2-2\tilde v^j_k)
-2\sum\limits_{1\le i<j\le n}w^+_{ij},\nonumber
\\
P_2 = \sum\limits_{1\le i<j\le n}\Delta^2_{\{i,j\}}
+\sum\limits_{\substack{1\le i\le n\\ 1\le j\le n,\,i\ne
j}}\sum\limits_{j=1}^n
q^o_{\{i\}}\Delta^2_{\{j\}}+\sum\limits_{1\le i<j\le
n}q^o_{\{i,j\}}\nonumber
\\ \phantom{P_2=}
{} -\sum\limits_{\substack{1\le i< j\le n\\1\le k\le n,\,k\ne
i,\,j}}
 S_{\{i,j\}}^o(\Delta^2_{\{k\}}+q^o_{\{k\}})
+\sum\limits_{\substack{1\le i< j\le n\\ i< k< \ell\le n\\ j\ne
k,\,\ell}} S_{\{i,j\}}^oS_{\{k,\ell\}}^o +\sum\limits_{1\le i< j<
k\le n}S_{\{i,j,k\}}^o\nonumber
\\ \phantom{P_2}
{}=\sum(\partial_i\partial_j+w_{ij}^-)^2 -\sum2\tilde
v_i\partial_j^2+\sum4\tilde v_i\tilde v_j +\sum
C_jT_{\{k,\ell\}}^o(v^j)\nonumber
\\ \phantom{P_2=}
{}-2\sum w_{ij}^+(\partial_k^2-2\tilde v_k)
 + 4\sum w_{ij}^+w_{k\ell}^+  + 2\sum w_{ij}^+w_{jk}^+.
 \label{eq:BnP22}
\end{gather}
Here if \eqref{eq:BnT} and  \eqref{eq:BnS} are valid, then
\begin{gather}
T^o_{\{k,\ell\}}(v^j) = 2(v_k^j+v_\ell^j)w_{k\ell}^+
-4Av_k^jv_\ell^j. \label{eq:BnP23}
\end{gather}
The commuting operator $P_3$ of the 6-th order is
\begin{gather*}
P_3 = \sum\Delta^2_{\{i,j,k\}}+\sum q^o_{\{k\}}\Delta^2_{\{i,j\}}
+\sum q^o_{\{i,j,k\}} -\sum S_{\{k,\ell\}}^o\Delta^2_{\{i,j\}}
\\ \phantom{P_3 =}
{}-\sum S_{\{k,\ell\}}^oq^o_{\{i\}}\Delta^2_{\{j\}} -\sum
S_{\{k,\ell\}}^oq^o_{\{i,j\}}+\sum S_{\{i,j,k,\ell,m\}}^o +\sum
S_{\{i,j,k,\ell\}}^oS_{\{\mu,\nu\}}^o
\\ \phantom{P_3 =}
{}+\sum S_{\{i,j,k\}}^oS_{\{\ell,\mu,\nu\}}^o +\sum
S_{\{i,j,k\}}^oS_{\{\ell,m\}}^o S_{\{\mu,\nu\}}^o
 +\sum S_{\{i_1,i_2\}}^o S_{\{j_1,j_2\}}^o  S_{\{k_1,k_2\}}^o S_{\{\ell_1,\ell_2\}}^o.
\end{gather*}
In Theorem~\ref{thm:BnEllip} the Schr\"odinger operator is
\begin{gather*}
 P = -\frac12\sum\limits_{k=0}^n\frac{\partial^2}{\partial x_k^2}
       +A\sum\limits_{1\le i<j\le n}(\wp(x_i-x_j)+\wp(x_i+x_j))
       + \sum\limits_{j=0}^3C_j\sum\limits_{k=1}^n \wp(x_k+\omega_j)
\end{gather*}
and the operator $P_2$ satisfying $[P,P_2]=0$ is given by
\eqref{eq:BnP22} and \eqref{eq:BnP23} with
\begin{gather*}
 \tilde v_k = \sum\limits_{\nu=0}^3 C_\nu\wp(x_k+\omega_\nu),\qquad
 v^j_k = \wp(x_k+\omega_j),\qquad
 w_{ij}^\pm = A(\wp(x_i-x_j)\pm\wp(x_i+x_j)).
\end{gather*}
\end{example}

\subsection{Analytic continuation of integrals}
\begin{theorem}[Toda-$\boldsymbol{D_n^{(1)}}$-bry]\label{thm:TodaDn1b}
For
\begin{gather}
  u_{e_i-e_j}(x) =
  \begin{cases}
   Ae^{-2\lambda(x_i-x_{i+1})} & (j=i+1),
   \\
   0                      & (|j-i|>1),
  \end{cases}
 \nonumber \\
   u_{e_i+e_j}(x) =
  \begin{cases}
    Ae^{2\lambda(x_1+x_2)}&(i+j=3),\\
    Ae^{-2\lambda(x_{n-1}+x_n)}&(i+j=2n-1),\\
    0&(i+j\notin\{3,2n-1\}),
  \end{cases} \label{eq:TodaDn1b}
  \\
 v^0_k(x) = \delta_{1k}\sinh^{-2}\lambda x_1,\qquad
 v^1_k(x) = \delta_{1k}\sinh^{-2}2\lambda x_1,
 \nonumber\\
 v^2_k(x) = \delta_{nk}\sinh^{-2}\lambda x_n,\qquad
 v^3_k(x) = \delta_{nk}\sinh^{-2}2\lambda x_n,
\nonumber
\end{gather}
we have commuting integrals $P_j$ by \eqref{eq:BnPn},
\eqref{eq:BnDelta}, \eqref{eq:Bnq}, \eqref{eq:BnTI} and
\begin{gather*}
  S_{\{k\}}^o  = 1\qquad\text{for}\qquad 1\le k\le n,\nonumber
  \\
  S_I^o = 0   \qquad\text{if}\qquad I\ne\{k,k+1,\dots,\ell\}
  \qquad\text{for}\qquad 1\le k<\ell\le n,\nonumber
  \\
  S_{\{k,k+1,\dots,\ell\}}^o  =2A^{\ell-k+1}(e^{-2\lambda(x_k-x_\ell)}
    +\delta_{1k}e^{2\lambda(x_1+x_\ell)}+\delta_{\ell n}e^{-2\lambda(x_k+x_n)}),\nonumber
  \\
  T^0_{\{k\}}(v^j)  = 2v^j_k(x)\qquad\text{for}\qquad 0\le j\le 3,\qquad k=1,\dots,n,\nonumber
  \\
  T^0_I(v^0) = 0\qquad \text{if}\qquad I\ne\{1,\dots,k\}\qquad\text{for}\qquad k=1,\dots,n,\nonumber
  \\
  T^0_I(v^2) = 0\qquad\text{if}\qquad I\ne\{k,\dots,n\}\qquad \text{for}\qquad k=1,\dots,n,\nonumber
  \\
  T^0_I(v^1)  = T^0_I(v^3) = 0\qquad\text{if}\qquad \#I>1,\nonumber
 \\
  T^0_{\{1,\dots,k\}}(v^0) = 8A^{k-1}(e^{2\lambda x_k}+\delta_{kn}e^{-2\lambda x_n})
  \qquad\text{for}\qquad k\ge 2,\nonumber
  \\
  T^0_{\{n-k+1,\dots,n\}}(v^2)= 8A^{k-1}(e^{-2\lambda x_{n-k+1}}+\delta_{kn}e^{2\lambda x_1})
   \qquad\text{for}\qquad k\ge 2.
%\label{eq:TodaDn1b2}
\end{gather*}
\end{theorem}
\begin{proof}
Put
\begin{gather*}
 \tilde x  = \biggl(x_1-\frac{1-1}{n-1}\omega_2,\dots,x_k-\frac{k-1}{n-1}\omega_2,
 \ldots,x_n-\frac{n-1}{n-1}\omega_2\biggr),
 \\
 \tilde u_{e_i\mp e_j}(\tilde x) =A\frac{e^{{2\lambda \omega_2}/(n-1)}}{4\lambda^2}
\wp_0\biggl(x_i-\frac{i-1}{n-1}\omega_2\mp\biggl(x_j-\frac{j-1}{n-1}\omega_2\biggr);
2\omega_1,2\omega_2\biggr),
\\
 \tilde v^j_k(\tilde x) = \frac{(-1)^j}{\lambda^2}
 \wp_0\biggl(x_k-\frac{k-1}{n-1}\omega_2+\omega_j; 2\omega_1,2\omega_2\biggr)
 \qquad\text{for}\qquad 0\le j\le3,\qquad 1\le k\le n.
\end{gather*}
When $\omega_2\to\infty$, $\tilde u_{e_i\mp e_j}(\tilde x)$ and
$\tilde v_k^\ell$ ($\ell=0,1,2,3$) converge to $u_{e_i\mp e_j}(x)$
in \eqref{eq:TodaDn1b} and
\begin{gather*}
 v^0_k(x) = \delta_{1k}\sinh^{-2}\lambda x_1,\quad
 v^1_k(x) = \delta_{1k}\cosh^{-2}\lambda x_1,\\
 v^2_k(x) = \delta_{nk}\sinh^{-2}\lambda x_n,\quad
 v^3_k(x) = \delta_{nk}\cosh^{-2}\lambda x_n,
\end{gather*}
respectively. Under the notation in Theorem~\ref{thm:BnEllip}, let
$\tilde S_I(\tilde v^\ell)$ and $\tilde T_I^o(\tilde v^\ell)$ be
the functions def\/ined in the same way as $S_I(v^\ell)$ and
$T_I^o(v^\ell)$, respectively, where $(u_{e_i\mp e_j}(x)$,
$v^\ell_k(x))$ are replaced by $(\tilde u_{e_i\mp e_j}(x)$,
$\tilde v^\ell_k(\tilde x))$. Then by taking the limits for
$\omega_2\to\infty$, $\tilde T_I^o(\tilde v^\ell)$ converge to the
follo\-wing~$\bar T_I^o(v^\ell)$
\begin{gather*}
  \bar T^0_I(v^0)  = \bar T^0_I(v^1) =  0\qquad
  \text{if}\qquad I\ne\{1,\dots,k\}\qquad\text{for}\qquad k=1,\dots,n,
  \\[.5ex]
  \bar T^0_I(v^2)  = \bar T^0_I(v^3) = 0\qquad
  \text{if}\qquad I\ne\{k,\dots,n\}\qquad \text{for}\qquad k=1,\dots,n.
\end{gather*}
If $k\ge 2$, then
\begin{gather*}
  \bar T^0_{\{1,\dots,k\}}(v^0)
   =\lim_{\omega_2\to\infty}\sum\limits_{\nu=1}^{\# I}
   \sum\limits_{I_1\amalg\cdots\amalg I_\nu=I}
  \biggl(-A\frac{e^{{2\lambda \omega_2}/(n-1)}}{4\lambda^2}\biggr)^{\nu-1}
  (\nu-1)!\tilde S_{I_1}(\tilde v^0)\cdots \tilde S_{I_\nu}(\tilde v^0)
\\ \phantom{\bar T^0_{\{1,\dots,k\}}(v^0)}
   {} = \lim_{\omega_2\to\infty}\tilde S_{\{1,\dots,k\}}(\tilde v^0)
    -\lim_{\omega_2\to\infty}  \tilde S_{\{1\}}(\tilde v^0)
     A\frac{e^{{2\lambda \omega_2}/(n-1)}}{4\lambda^2}
     \tilde S_{\{2,\dots,k\}}(\tilde v^0)
\\ \phantom{\bar T^0_{\{1,\dots,k\}}(v^0)}
   = 2A^{k-1}\sinh^{-2}\!\lambda x_1(e^{-2\lambda(x_1-x_k)}\!+e^{2\lambda(x_1+x_k)}
     +\delta_{kn}e^{2\lambda(x_1-x_n)} + \delta_{kn}e^{-2\lambda(x_1+x_n)})
\\ \phantom{\bar T^0_{\{1,\dots,k\}}(v^0)=}
 {}- 2\sinh^{-2}\lambda x_1\cdot 2A^{k-1}( e^{2\lambda x_k}+\delta_{kn}e^{-2\lambda x_n})
   = 8A^{k-1}(e^{2\lambda x_k}+\delta_{kn}e^{-2\lambda x_n}),
\\
 \bar T^0_{\{1,\dots,k\}}(v^1)
   = 2A^{k-1}\cosh^{-2}\!\lambda x_1(e^{-2\lambda(x_1-x_k)}\!+e^{2\lambda(x_1+x_k)}
    \! +\delta_{kn}e^{2\lambda(x_1-x_n)}\! + \delta_{kn}e^{-2\lambda(x_1+x_n)})
\\ \qquad
   {}+ 4A^{k-1}\cosh^{-2}\lambda x_1(e^{2\lambda x_k} +\delta_{kn}e^{-2\lambda x_n})
   = 8A^{k-1}(e^{2\lambda x_k}+\delta_{kn}e^{-2\lambda x_n}),
\\
  \bar T^0_{\{n-k+1,\dots,n\}}(v^2) = 8A^{k-1}(e^{-2\lambda x_{n-k+1}}+\delta_{kn}e^{2\lambda x_1}),
\\
  \bar T^0_{\{n-k+1,\dots,n\}}(v^3) = 8A^{k-1}(e^{-2\lambda x_{n-k+1}}+\delta_{kn}e^{2\lambda x_1}).
\end{gather*}
Replacing $v^1$ and $v^3$ by $(1/4)(v^0-v^1)$ and
$(1/4)(v^2-v^3)$, respectively, we have the theorem by the
analytic continuation given in Lemma~\ref{lem:wpcont}.
\end{proof}
As is proved by \cite{Oshima1998}, suitable limits of the
functions in Theorem~\ref{thm:BnEllip} give the following theorem.
%%%%
\begin{theorem}[Trig-$\boldsymbol{B_n}$, {\cite[Proposition~6.1]{Oshima1998}}]
\label{thm:TrigBn} For complex numbers $\lambda$, $C$,
$C_0,\ldots$, $C_3$ and $A$ with $\lambda\ne0$, we have
\eqref{eq:BnPnCom} by putting
\begin{gather*}
   u_{e_i\pm e_j}(x) = A\sinh^{-2}\lambda(x_i\pm x_k),
   \\
   v^0_{e_k}(x) =  \sinh^{-2}\lambda x_k, \qquad v^1_{e_k}(x)  =  \cosh^{-2}\lambda x_k,
   \\
   v^2_{e_k}(x)  =  \sinh^2\lambda x_k, \qquad v^3_{e_k}(x)  =  \dfrac14\sinh^22\lambda x_k
\end{gather*}
and
\begin{gather*}
T_I  = (-1)^{\#I-1}(CS_I^o - C_0T_I^o(v^0) - C_1T_I^o(v^1) -
C_2S_I(v^2) - C_3S_I(v^3)
\\ \phantom{T_I  =}
{}+ 2C_3\sum\limits_{I_1\amalg I_2=I}( S_{I_1}(v^2) S_{I_2}(v^2) +
S_{I_1}(v^2) S_{I_2}^o + S_{I_1}^o S_{I_2}(v^2))),
\\
T^o_I(v^0)  = \sum\limits_{\nu=1}^{\#
I}\sum\limits_{I_1\amalg\cdots\amalg I_\nu=I} (-A)^{\nu-1}(\nu-1)!
S_{I_1}(v^0)\cdots S_{I_\nu}(v^0),
\\
T^o_I(v^1)  = \sum\limits_{\nu=1}^{\#
I}\sum\limits_{I_1\amalg\cdots\amalg I_\nu=I} A^{\nu-1}(\nu-1)!
S_{I_1}(v^1)\cdots S_{I_\nu}(v^1).
\end{gather*}
\end{theorem}
%%%
\begin{theorem}[Trig-$\boldsymbol{A_{n-1}}$-bry]
For
\begin{gather*}
 u_{e_i-e_j}(x) = A\sinh^{-2}\lambda(x_i-x_j),\qquad
 u_{e_i+e_j}(x) = 0,
 \\
 v^0_{e_k}(x) = e^{-2\lambda x_k},\qquad
 v^1_{e_k}(x) = e^{-4\lambda x_k},\qquad
 v^2_{e_k}(x) = e^{2\lambda x_k},\qquad
 v^3_{e_k}(x) = e^{4\lambda x_k}
\end{gather*}
we have \eqref{eq:BnPnCom} by putting
\begin{gather*}
 T_I \!=\! (-1)^{\#I-1}\Biggl(\!CS_I^o-\!\!\sum\limits_{j=0}^3C_jS_I(v^j)\!+
 2\!\!\!\!\!\sum\limits_{I_1\amalg I_2=I}\!\!\!(C_1S_{I_1}(v^0)S_{I_2}(v^0)
     \! +\!C_3S_{I_1}(v^2)S_{I_2}(v^2))\!\Biggr).
\end{gather*}
\end{theorem}
%%%
\begin{proof}
Putting
\begin{gather*}
 \tilde u_{e_i\pm e_j} =A\sinh^{-2}\lambda\bigl((x_i+N)\pm (x_j+N)\bigr),
\qquad
 \tilde v^0_k = \frac14 e^{2\lambda N}\sinh^{-2}\lambda(x_k+N),
\\[.5ex]
 \tilde v^1_k = \frac14 e^{4\lambda N}\sinh^{-2}2\lambda(x_k+N)
= \frac1{16} e^{4\lambda N}\bigl(\sinh^{-2}2\lambda(x_k+N) -
\cosh^{-2}2\lambda(x_k+N)\bigr),
\\[.5ex]
 \tilde v^2_k = 4 e^{-2\lambda N}\sinh^{2}\lambda(x_k+N),
\qquad
 \tilde v^3_k = 4 e^{-4\lambda N}\sinh^{2}2\lambda(x_k+N),
\\[.5ex]
 \tilde x  = (x_1+N, x_2+N,\dots,x_n+N)
\end{gather*}
under the notation in Theorem~\ref{thm:TrigBn}, we have
\begin{gather*}
 (\bar u_{e_i-e_j}, \bar u_{e_i+e_j}, \bar v^0_k, \bar v^1_k, \bar v^2_k, \bar v^3_k)
  :=  \lim_{N\to\infty}(\tilde u_{e_i-e_j}, \tilde u_{e_i+e_j},
   \tilde v^0_k, \tilde v^1_k, \tilde v^2_k, \tilde v^3_k)
 \\ \qquad
  {}=(A\sinh^{-2}\lambda( x_i-x_k), 0, e^{-2\lambda x_k}, e^{-4\lambda x_k}e^{2\lambda x_k}, e^{4\lambda x_k}),
\\
 \lim_{N\to\infty}\frac14 e^{2\lambda N}T^o_I(v^0)(\tilde x) = \bar S_I(\bar v^0),
\\
  \lim_{N\to\infty}\frac1{16} e^{4\lambda N}\bigl(T^o_I(v^0)(\tilde x) -T^o_I(v^1)(\tilde x))
   = \bar S_I(\bar v^1)-2\sum\limits_{I_1\amalg I_2=I} \bar S_{I_1}(\bar v^0)\bar S_{I_2}(\bar v^0),
\\
 \lim_{N\to\infty}4 e^{-2\lambda N}T^o_I(v^2)(\tilde x) = \bar S_I(\bar v^2),
\\
  \lim_{N\to\infty}4 e^{-2\lambda N}T^o_I(v^3)(\tilde x)
   =\bar S_I(\bar v^3) - 2\sum\limits_{I_1\amalg I_2=I}\bar S_{I_1}(\bar v^2)\bar S_{I_2}(\bar v^2).
\end{gather*}
Here $\bar S_I(\bar v^\ell)$ are def\/ined by \eqref{eq:BnS} with
$u_{e_i\pm e_j}$ and $v^\ell_{e_k}$ replaced  by $\bar u_{e_i\pm
e_j}$ and $\bar v^\ell_{e_k}$, respectively. Then the theorem is
clear.
\end{proof}
\begin{theorem}[Toda-$\boldsymbol{B_n^{(1)}}$-bry]\label{thm:TodaBn1b}
For the potential function defined by
\begin{gather}
    u_{e_i-e_j}(x) =
  \begin{cases}
    Ae^{-2\lambda(x_i-x_{i+1})} &\text{if}\qquad j=i+1,
    \\
    0&\text{if}\qquad 1\le i<i+1<j\le n,
  \end{cases}\nonumber
  \\
    u_{e_i+e_j}(x) =
  \begin{cases}
    Ae^{-2\lambda(x_{n-1}+x_n)}&\text{if}\qquad i=n-1,\qquad j=n,
  \\
    0 &\text{if}\qquad 1\le i<j\le n\qquad \text{and}\qquad i\ne n-1,
  \end{cases}\label{eq:TodaBn1b}
  \\
  v^0_k(x) = \delta_{k1}e^{2\lambda x_1},\qquad
  v^1_k(x) = \delta_{k1}e^{4\lambda x_1},
 \nonumber \\
  v^2_k(x) = \delta_{kn}\sinh^{-2}\lambda x_n,\qquad
  v^3_k(x) = \delta_{kn}\sinh^{-2}2\lambda x_n,
  \nonumber
\end{gather}
we have \eqref{eq:BnPnCom} with
\begin{gather*}
 S_{\{k\}}^o  = 1\qquad\text{for}\qquad 1\le k\le n,
 \\[.2ex]
S_I^o = 0 \qquad\text{if}\qquad I\ne\{k,k+1,\dots,\ell\}\qquad
  \text{for}\qquad 1\le k<\ell\le n,
\\[.2ex]
  S_{\{k,k+1,\dots,\ell\}}^o  =2A^{\ell-k+1}(e^{-2\lambda(x_k-x_\ell)}
    +\delta_{\ell n}e^{-2\lambda(x_k+x_n)}),
\\[.2ex]
  T^0_{\{k\}}(v^j)  = 2v^j_k(x)\qquad\text{for}\qquad 0\le j\le 3,\qquad k=1,\dots,n,
\\[.2ex]
  T^0_I(v^0) = 0\qquad\text{if}\qquad I\ne\{1,\dots,k\}\qquad\text{for}\qquad k=1,\dots,n,
\\[.2ex]
  T^0_I(v^2) = 0\qquad\text{if}\qquad I\ne\{k,\dots,n\}\qquad\text{for}\qquad k=1,\dots,n,
\\[.2ex]
  T^0_I(v^1)  = T^0_I(v^3) = 0\qquad\text{if}\qquad\#I>1,
\\[.2ex]
  T^0_{\{1,\dots,k\}}(v^0) = 2A^{k-1}(e^{2\lambda x_k}+\delta_{kn}e^{-2\lambda x_n})
  \qquad\text{for}\qquad k\ge 2,
\\[.2ex]
  T^0_{\{n-k+1,\dots,n\}}(v^2) = 8A^{k-1}e^{-2\lambda x_{n-k+1}}\qquad\text{for}\qquad k\ge 2.
\end{gather*}
%%%%%%%%%%%%%
\end{theorem}
\begin{proof}
Suppose $\mathrm{Re}\,\lambda>0$. In (Toda-$D_n^{(1)}$-bry) put
\begin{gather*}
 \tilde x = (x_1-(n-1)N,\dots,x_k-(n-k)N,\dots,x_n-(n-n)N),
 \\
  \tilde u_{e_i-e_j} =  \begin{cases}
   Ae^{-2\lambda N} e^{-2\lambda(x_i-(n-i)N - x_{i+1}+(n-i-1)N)} & (j=i+1),
 \\
   0    & (|j-i|>1),
  \end{cases}
 \\
   \tilde u_{e_i+e_j} =  \begin{cases}
    Ae^{-2\lambda N} e^{2\lambda(x_1-(n-1)N+x_2-(n-2)N)}&(i+j=3),
    \\
    Ae^{-2\lambda N} e^{-2\lambda(x_{n-1}-N+x_n)}&(i+j=2n-1),
    \\
    0&(i+j\notin\{3,2n-1\}),
  \end{cases}\\
 \tilde v^0_k = \delta_{1k}\frac{e^{2\lambda(n-1)N}}4 \sinh^{-2}\lambda(x_1-(n-1)N),
 \\
 \tilde v^1_k = \delta_{1k}\frac{e^{4\lambda(n-1)N}}4\sinh^{-2}2\lambda(x_1-(n-1)N),
 \\
 \tilde v^2_k = \delta_{nk}\sinh^{-2}\lambda x_n,
 \qquad
 \tilde v^3_k = \delta_{nk}\sinh^{-2}2\lambda x_n
\end{gather*}
and we have \eqref{eq:TodaBn1b} by the limit $N\to\infty$.
Moreover for $k\ge2$, it follows from Theorem~\ref{thm:TodaDn1b}
and~\eqref{eq:Bnlinear} that
\begin{gather*}
\tilde T_{\{1,\dots,k\}}(\tilde v^1)  = \tilde
T_{\{1,\dots,k\}}(\tilde v^3)=0,
\\
\tilde T_{\{1,\dots,k\}}(\tilde v^0)  =
  \frac14e^{2\lambda(n-1) N}(Ae^{-2\lambda N})^{k-1}
    (8e^{2\lambda(x_k-(n-k)N)} +8\delta_{kn}e^{-2\lambda x_n})
\\ \phantom{\tilde T_{\{1,\dots,k\}}(\tilde v^0)}
  {} = 2A^{k-1}(e^{2\lambda x_k}+\delta_{kn}e^{-2\lambda x_n}),
\\
\tilde T_{\{n-k+1,\dots,n\}}(\tilde v^2) =(Ae^{-2\lambda N})^{k-1}
 (8e^{-2\lambda(x_{n-k+1}- (k-1)N)}+ 8\delta_{1k}e^{2\lambda(x_1-(n-1)N)})
\\ \phantom{\tilde T_{\{n-k+1,\dots,n\}}(\tilde v^2)}
 {} = 8A^{k-1}e^{-2\lambda x_{n-k+1}},
\end{gather*}
which implies the theorem.
\end{proof}
\begin{theorem}[Toda-$\boldsymbol{C_n^{(1)}}$]\label{thm:TodaCn1}
For the potential function defined by
\begin{gather*}
%\label{eq:TodaCn1}
    u_{e_i-e_j}(x) =
\begin{cases}
    Ae^{-2\lambda(x_i-x_{i+1})} &\text{if}\qquad j=i+1,
\\
    0&\text{if}\qquad 1\le i<i+1<j\le n,
\end{cases}
\\
  u_{e_i+e_j}(x) = 0 \qquad\text{for}\qquad 1\le i<j\le n,
\\
  v^0_k(x) = \delta_{k1}e^{2\lambda x_1},\qquad
  v^1_k(x) = \delta_{k1}e^{4\lambda x_1},
\\
  v^2_k(x) = \delta_{kn}e^{-2\lambda x_n},\qquad
  v^3_k(x) = \delta_{kn}e^{-4\lambda x_n},
\end{gather*}
we have \eqref{eq:BnPnCom} with
\begin{gather*}
  S_{\{k\}}^o  = 1\qquad\text{for}\qquad 1\le k\le n,
\\[.2ex]
  S_I^o = 0\qquad \text{if}\qquad I\ne\{k,k+1,\dots,\ell\}\qquad
  \text{for}\qquad 1\le k<\ell\le n,
\\[.2ex]
  S_{\{k,k+1,\dots,\ell\}}^o  =2A^{\ell-k+1}e^{-2\lambda(x_k-x_\ell)},
\\[.2ex]
  T^0_{\{k\}}(v^j)  = 2v^j_k(x)\qquad \text{for}\qquad 0\le j\le 3,\qquad  k=1,\dots,n,
\\[.2ex]
  T^0_I(v^0) = 0\qquad \text{if}\qquad I\ne\{1,\dots,k\}\qquad \text{for}\qquad k=1,\dots,n,
\\[.2ex]
  T^0_I(v^2) =  0\qquad \text{if}\qquad  I\ne\{k,\dots,n\}\qquad \text{for}\qquad k=1,\dots,n,
\\[.2ex]
  T^0_I(v^1) = T^0_I(v^3) = 0\qquad \text{if}\qquad \#I>1,
\\[.2ex]
  T^0_{\{1,\dots,k\}}(v^0) = 2A^{k-1}e^{2\lambda x_k}\qquad \text{for}\qquad k\ge 2,
\\[.2ex]
  T^0_{\{n-k+1,\dots,n\}}(v^2) = 2A^{k-1}e^{-2\lambda x_{n-k+1}}\qquad \text{for}\qquad k\ge 2.
\end{gather*}
\end{theorem}
\begin{proof}
Substituting $x_k$ by $x_k+R$ for $k=1,\dots,n$ and multiplying
$v_k^0$, $v_k^1$, $v_k^2$ and $v_k^3$  by $e^{-2\lambda R}$,
$e^{-4\lambda R}$, $(1/4)e^{2\lambda R}$ and $(1/4) e^{4\lambda
R}$, respectively, we have the claim from
Theorem~\ref{thm:TodaBn1b}.
\end{proof}
%%%%
\begin{theorem}[Rat-$\boldsymbol{A_{n-1}}$-bry]
We have \eqref{eq:BnPnCom} if
\begin{gather*}
 u_{e_i-e_j}(x) = \dfrac{A}{(x_i-x_j)^2},\qquad
 u_{e_i+e_j}(x) = 0,
 \qquad
 v^j_k(x) = x_k^{j+1},
\\
 T_I  = (-1)^{\#I-1}\Biggl(CS_I^o-\sum\limits_{j=0}^3C_jS_I(v^j) +
  \sum\limits_{I_1\amalg I_2=I}C_1(S_{I_1}(v^0) S_{I_2}^o
       + S_{I_1}^o S_{I_1}(v^0))
\\ \phantom{T_I  =}
   {} +  \sum\limits_{I_1\amalg I_2=I}C_3(S_{I_1}(v^1) S_{I_2}^o
       + S_{I_1}(v^0) S_{I_2}(v^0)+ S_{I_1}^o S_{I_2}(v^1))\Biggr).
\end{gather*}
\end{theorem}
%%%
\begin{proof} Put
\begin{gather*}
 \tilde u_{e_i-e_j}  =\lambda^2\sinh^{-2}\lambda(x_i-x_j),\qquad
 \tilde u_{e_i+e_j}  =0,
\\
 \tilde v^0_k  =  \frac1{2\lambda}(e^{2\lambda x_k}-1),\qquad
 \tilde v^1_k  = \frac1{4\lambda^2} (e^{2\lambda x_k}+e^{-2\lambda x_k}-2),
 \\
 \tilde v^2_k  =\frac1{8\lambda^3}(e^{4\lambda x_k}\! -\! 3e^{2\lambda x_k}\!-\!e^{-2\lambda x_k}+3),\quad
 \tilde v^3_k  =\frac1{16\lambda^4}(e^{4\lambda x_k}\!+\!e^{-4\lambda x_k}
 \! -\!4e^{2\lambda x_k}\!-\!4e^{-2\lambda x_k}\!+\!6).
\end{gather*}
Then taking $\lambda\to0$ we have the required potential function.

Owing to  (Trig-$A_{n-1}$-bry) and Remark~\ref{rem:Tbilin}, we
have
\begin{gather*}
 \lim_{\lambda\to0}\tilde S_I\Bigl(\sum \tilde v^j_k\Bigr)=\bar S_I\Bigl(\sum x_k^{j+1}\Bigr),
\\
 \lim_{\lambda\to0}\lambda^2\frac1{8\lambda^3}
 \Bigl(\tilde S_{I_1}\Bigl(\sum e^{2\lambda x_k}\Bigr) \tilde S_{I_2}\Bigl(\sum e^{2\lambda x_k}\Bigr)
 - 4\tilde S_{I_1}^o\tilde S_{I_2}^o\Bigr)
 \\ \qquad
{}=\frac12 \Bigl(\bar S_{I_1}\Bigl(\sum x_k\Bigr)\bar S_{I_2}^o
  + \bar S_{I_1}^o \bar S_{I_2}\Bigl(\sum x_k\Bigr)\Bigr),
\\
 \lim_{\lambda\to0}\lambda^2\frac1{16\lambda^4}
 \Bigl(\tilde S_{I_1}\Bigl(\sum e^{2\lambda x_k}\Bigr) \tilde S_{I_2}\Bigl(\sum e^{2\lambda x_k}\Bigr)
 +\tilde S_{I_1}\Bigl(\sum e^{-2\lambda x_k}\Bigr) \tilde S_{I_2}\Bigl(\sum e^{-2\lambda x_k}\Bigr)
 - 8\tilde S_{I_1}^o\tilde S_{I_2}^o\Bigr)
\\ \qquad
{} =\frac12\Bigl(\bar S_{I_1}\Bigl(\sum x_k^2\Bigr) \bar S_{I_2}^o
  +\bar S_{I_1}\Bigl(\sum x_k\Bigr)\bar S_{I_2}\Bigl(\sum x_k\Bigr)
  +\bar S_{I_1}^o \bar S_{I_2}\Bigl(\sum x_k\Bigr)\Bigr).
\end{gather*}
and thus the theorem.
\end{proof}
%%%%
As is proved by \cite{Oshima1998}, suitable limits of the
functions in Theorem~\ref{thm:TrigBn} give the following theorem.
\begin{theorem}[Rat-$\boldsymbol{B_n}$, {\cite[Proposition~6.3]{Oshima1998}}]
Put
\begin{gather*}
  u_{e_i-e_j}(x)=\dfrac{A}{(x_i-x_j)^2},\qquad
  u_{e_i+e_j}(x)=\dfrac{A}{(x_i+x_j)^2}
  \\
  v^0_k(x) = x_k^{-2},\qquad
  v^1_k(x) = x_k^2,\qquad
  v^2_k(x) = x_k^4,\qquad
  v^3_k(x) = x_k^6.
\end{gather*}
Then \eqref{eq:BnPnCom} holds with
\begin{gather*}
T_I = (-1)^{\#I-1}\Biggl(CS_I^o - C_0T_I^o(v^0)- C_1S_I(v^1)-
C_2S_I(v^2)
\\ \phantom{T_I =}
{}+2C_2\sum\limits_{I_1\amalg I_2=I}(S_{I_1}(v^1)
S_{I_2}^o+S_{I_1}^o S_{I_2}(v^1)) -  2C_3S_I^o(v^3)
\\ \phantom{T_I =}
{}+C_3\sum\limits_{I_1\amalg I_2=I}(S_{I_1}(v^1) S_{I_2}(v^1)
+2S_{I_1}(v^2) S_{I_2}^o+ 2S_{I_1}^o S_{I_2}(v^2))
\\ \phantom{T_I =}
{}-24C_3\sum\limits_{I_1\amalg I_2 \amalg I_3=I}(S_{I_1}(v^1)
S_{I_2}^o S_{I_3}^o +S_{I_1}^o S_{I_2}(v^1) S_{I_3}^o+S_{I_1}^o
S_{I_2}^o S_{I_3}(v^1))\Biggr),
\\
 T^o_I(v^0) =  \sum\limits_{\nu=1}^{\# I}\sum\limits_{I_1\amalg\cdots\amalg I_\nu=I}
        (-A)^{\nu-1}(\nu-1)!\cdot S_{I_1}(v^0)\cdots S_{I_\nu}(v^0).
\end{gather*}
\end{theorem}

\begin{definition}\label{def:restrict}
We def\/ine some potential functions as specializations of
potential functions in Def\/inition~\ref{def:BnP}.
\begin{description}\itemsep=0pt
{\item[\rm(Trig-$A_{n-1}$-bry-reg)] {\it Trigonometric potential
of type $A_{n-1}$ with regular boundary conditions} is
(Trig-$A_{n-1}$-bry) with $C_2=C_3=0$. %%
\item[\rm(Trig-$A_{n-1}$)] {\it Trigonometric potential of type
$A_{n-1}$} is (Trig-$A_{n-1}$-bry) with $C_0=C_1=C_2=C_3=0$. %%
\item[\rm(Trig-$BC_n$-reg)] {\it Trigonometric potential of type
$BC_n$ with regular boundary conditions} is (Trig-$B_n$) with
$C_2=C_3=0$. %%
\item[\rm(Toda-$D_n$-bry)] {\it Toda potential of type $D_n$ with
boundary conditions} is (Toda-$B_n^{(1)}$-bry) with $C_0=C_1=0$.
\item[\rm(Toda-$B_n^{(1)}$)] {\it Toda potential of type
$B_n^{(1)}$} is (Toda-$B_n^{(1)}$-bry) with $C_2=C_3=0$. %%
\item[\rm(Toda-$D_n^{(1)}$)] {\it Toda potential of type
$D_n^{(1)}$} is (Toda-$D_n^{(1)}$-bry) with $C_0=C_1=C_2=C_3=0$.
\item[\rm(Toda-$A_{n-1}$)] {\it Toda potential of type $A_{n-1}$}
is (Toda-$C_n^{(1)}$) with $C_0=C_1=C_2=C_3=0$. %%
\item[\rm(Toda-$BC_n$)] {\it Toda potential of type $B_n$} is
(Toda-$C_n^{(1)}$) with $C_0=C_1=0$. %%
\item[\rm(Ellip-$D_n$)] {\it Elliptic potential of type $D_n$} is
(Ellip-$B_n$) with $C_0=C_1=C_2=C_3=0$. %%
\item[\rm(Trig-$D_n$)] {\it Trigonometric potential of type $D_n$}
is (Trig-$B_n$) with $C_0=C_1=C_2=C_3=0$. %%
\item[\rm(Rat-$D_n$)] {\it Rational potential of type $D_n$} is
(Rat-$B_n$) with $C_0=C_1=C_2=C_3=0$. %%
\item[\rm(Toda-$D_n$)] {\it Toda potential of type $D_n$} is
(Toda-$B_n^{(1)}$-bry) with $C_0=C_1=C_2=C_3=0$. %%
\item[\rm(Rat-$B_n$-2)] {\it Rational potential of type $B_n${\rm
-2}} is (Rat-$B_n$) with $C_2=C_3=0$. %%
\item[\rm(Rat-$A_{n-1}$-bry2)] {\it Rational potential of type
$A_{n-1}$ with $2$-boundary conditions} is (Rat-$A_{n-1}$-bry)
with $C_2=C_3=0$. In this case, we may assume $C_0=0$ or $C_1=0$
by the transformation $x_k\mapsto x_k+c$ ($k=1,\ldots,n$) with a
suitable $c\in\mathbb C$.}
\end{description}
\end{definition}
Then we have the following diagrams for $n\ge3$. Note that we
don't write all the arrows in the diagrams (ex.\ (Toda-$D_n$-bry)
$\to$ (Toda-$BC_n$)) and the meaning of the symbol \
$\overset{5:3}\Rightarrow$ \ is same as in the diagram for type
$B_2$. Namely, 5 parameters (coupling constant) in the potential
are reduced to 3 parameters by a certain restriction.
\begin{gather*}
\text{\bf Hierarchy of Elliptic-Trigonometric-Rational Integrable Potentials}\notag\\
\begin{matrix}
 & &\text{\rm Rat-$B_n$-2} & & \text{\rm Ellip-$D_n$}\\
 & &\Uparrow_{5:3} &  & \downarrow\\
 & &\text{\rm Rat-$B_n$}& \nwarrow & \text{\rm Trig-$D_n$} & \rightarrow & \text{\rm Rat-$D_n$}\\
 & &\uparrow& &\Uparrow_{3:1} \\
\text{\rm Ellip-$B_n$}\!& \to\!\! & \text{\rm Trig-$B_n$}
&\overset{5:3}\Rightarrow  \!
&\text{\rm Trig-$BC_n$-reg} & & \!\text{Ellip-$A_{n-1}$}\\
 & & \downarrow& &\downarrow& & \downarrow\\
 & &\! \text{\rm Trig-$A_{n-1}$-bry} & \overset{5:3}\Rightarrow & \text{\rm Trig-$A_{n-1}$-bry-reg} &
 \overset{3:1}\Rightarrow &
\text{\rm Trig-$A_{n-1}$}\\
 & & \downarrow & &\downarrow& & \downarrow\\
 & & \text{\rm Rat-$A_{n-1}$-bry} &\overset{5:3}\Rightarrow &\text{\rm Rat-$A_{n-1}$-bry2}  &\overset{3:1}\Rightarrow & \text{\rm Rat-$A_{n-1}$}
\end{matrix}
\end{gather*}

\begin{gather*}
\text{\bf Hierarchy of Toda Integrable Potentials}\notag\\
 \begin{matrix}
&&\text{\rm Trig-$BC_n$-reg} & \rightarrow & \text{\rm
Toda-$D_n$-bry}
& \overset{3:1}\Rightarrow & \text{\rm Toda-$D_n$}\\
&&\Uparrow_{5:3} & & \Uparrow_{5:3} & & \Uparrow_{3:1}\\
&&\text{\rm Trig-$B_n$}& \rightarrow &\! \text{\rm
Toda-$B_n^{(1)}$-bry}\! &
  \overset{5:3}\Rightarrow & \text{\rm Toda-$B_n^{(1)}$}\\
&\nearrow& &\nearrow & &\searrow\\
\text{\rm Ellip-$B_n$}&\rightarrow&\!\text{\rm
Toda-$D_n^{(1)}$-bry}\!
&\overset{5:1}\Rightarrow& \text{\rm Toda-$D_n^{(1)}$}&&\text{\rm Toda-$C_n^{(1)}$}\\
&\,\rotatebox{-45}{$\Rightarrow$}\!{}_{\text{5:1}}\!\!\!\!\!
&&\nearrow&&&\Downarrow_{5:3}\\
&&\text{\rm Ellip-$D_n$}&&\text{\rm Trig-$A_{n-1}$} && \text{\rm Toda-$BC_n$}\\
&&& \nearrow& &\searrow & \Downarrow_{3:1}\\
&&\text{\rm Ellip-$A_{n-1}$}&\rightarrow&\text{\rm
Toda-$A_{n-1}^{(1)}$} &\rightarrow &\!\! \text{\rm Toda-$A_{n-1}$}
 \end{matrix}
\end{gather*}
%%%
\section[Type $D_n$ ($n\ge 3$)]{Type $\boldsymbol{D_n}$ ($\boldsymbol{n\ge 3}$)}\label{sec:Dn}

\begin{theorem}[Type $D_n$]
The Schr\"odinger operators {\rm (Ellip-$D_n$)}, {\rm
(Trig-$D_n$)}, {\rm (Rat-$D_n$)}, {\rm (Toda-$D_n^{(1)}$)} and
{\rm (Toda-$D_n$)} are in the commutative algebra of differential
operators generated by $P_1,P_2,\ldots, P_{n-1}$ and
$\Delta_{\{1,\dots,n\}}$ which are the corresponding operators for
{\rm (Ellip-$B_n$)}, {\rm (Trig-$B_n$)}, {\rm (Rat-$B_n$)}, {\rm
(Toda-$D_n^{(1)}$-bry)}, {\rm (Toda-$D_n$-bry)} with
$C_0=C_1=C_2=C_3=0$, respectively.
\end{theorem}
\begin{proof}
This theorem is proved by \cite{Oshima1998} in the cases
(Ellip-$D_n$), (Trig-$D_n$), (Rat-$D_n$). Other two cases have
been def\/ined by suitable analytic continuation and therefore the
claim is clear.
\end{proof}
\begin{remark}
In the above theorem we have $P_n=\Delta_{\{1,\dots,n\}}^2$
because $q^o_I=0$ if $I\ne\varnothing$. Then $[P_j,P_n]=0$ implies
$[P_j,\Delta_{\{1,\dots,n\}}]=0$.
{\samepage \begin{gather*}
\text{{\bf Hierarchy of Integrable Potentials of Type
$\boldsymbol{D_n}$ $\boldsymbol{(n\ge3)}$}}
\\
\begin{matrix}
&&\text{\rm Toda-$D_n^{(1)}$}&\to
  &\text{\rm Toda-$D_n$} \\
 &\nearrow& &\text{$\nearrow$}&\\%llap{$\searrow$}} &\\
\text{\rm Ellip-$D_n$} &\to &\text{\rm Trig-$D_n$}&\to
  &\text{\rm Rat-$D_n$}
\end{matrix}
\end{gather*}}
\end{remark}
%%%
\section{Classical limits}\label{sec:dynam}
For functions $f(\xi,x)$ and $g(\xi,x)$ of
$(\xi,x)=(\xi_1,\dots,\xi_n,x_1,\dots,x_n)$, we def\/ine their
Poisson bracket by
\begin{gather*}
 \bigl\{f,g\bigr\}=\sum\limits_{k=1}^n\biggl(
 \frac{\partial f}{\partial \xi_k}\frac{\partial g}{\partial x_k}
  - \frac{\partial g}{\partial\xi_k}\frac{\partial f}{\partial x_k}\biggr).
\end{gather*}
\begin{theorem}
Put
\begin{gather*}
  \bar P(\xi,x) = -\frac12 \sum\limits_{k=1}^n\xi_k^2 +R(x).
\end{gather*}
Then for the integrable potential function $R(x)$ given in this
note, the functions $\bar P_k(\xi,x)$ and $\bar
\Delta_{\{1,\dots,n\}}(\xi,x)$ of $(\xi,x)$ defined by replacing
$\partial_\nu$ by $\xi_\nu$ $(\nu=1,\ldots,n)$ in the definitions
of $P_k$ and $\Delta_{\{1,\dots,n\}}$ in Sections~$\ref{sec:An}$,
$\ref{sec:B2}$ and $\ref{sec:Bn}$ satisfy
\begin{gather*}
 \bigl\{\bar P_i(\xi,x), \bar P_j(\xi,x)\bigr\}
 =\bigl\{\bar P(\xi,x), \bar P_k(\xi,x)\bigr\}=0
 \qquad\text{for}\quad 1\le i<j\le n\quad \text{and}\quad 1\le k\le n.
\end{gather*}
Hence $\bar P(\xi,x)$ are Hamiltonians of completely integrable
dynamical systems.

Moreover if the potential function $R(x)$ is of type $D_n$, then
\begin{gather*}
 \bigl\{\bar \Delta_{\{1,\dots,n\}}(\xi,x),\bar P_k(\xi,x)\bigr\}
 =\bigl\{\bar \Delta_{\{1,\dots,n\}}(\xi,x),\bar P(\xi,x)\bigr\}=0
 \qquad\text{for}\qquad 1\le k\le n.
\end{gather*}
\end{theorem}
\begin{proof}
If $R(x)$ is a potential function of (Ellip-$A_{n-1}$),
(Ellip-$B_n$) or (Ellip-$D_n$), the claim is proved in
\cite{Oshima1998, Oshima1995}. Since the claim keeps valid under
suitable holomorphic continuations with respect to the parameters
which are given in the former sections, we have the theorem.
\end{proof}
\begin{remark}
Since our operators $P_k$ are expressed by operators $P_k^\nu =
\sum\limits_i p_{k,i}^\nu(x)q_{k,i}^\nu(\partial)$ such that the
polynomials $q_{k,i}^\nu(\partial)$ satisfy
$[p_{k,i}^\nu(x),q_{k,i}^\nu(\partial)]=0$, there is no ambiguity
in the def\/inition of the classical limits by replacing
$\partial_\nu$ by $\xi_\nu$. In another word, if we have given the
above integrals $\bar P_j(x,\xi)$ of the classical limit, we have
a natural unique quantization of them.
\end{remark}

\section{Analogue for one variable}\label{sec:ODE}
Putting $n=1$ for the Schr\"odinger operator $P$ of type $A_n$ in
Section~\ref{sec:An} or of type $B_n$ in Section~\ref{sec:Bn}, we examine the
ordinary dif\/ferential equation $Pu=Cu$ with $C\in\mathbb C$
(cf.\ \cite[\S~10.6]{Whittaker}). We will write the operators
$Q=P-C$.

\noindent (Ellip-$B_1$)\quad The Heun equation
(cf.~\cite[\S~8]{Oshima1995}, \cite[pp. 576]{Whittaker})
\begin{gather*}
-\frac12\frac{d^2}{dt^2}+\sum\limits_{j=0}^3C_j\wp(t+\omega_j)-C.
\end{gather*}
(Ellip-$A_1$)\quad The Lam\'e equation
\begin{gather*}
-\frac12\frac{d^2}{dt^2}+A\wp(t)-C.
\end{gather*}
(Trig-$BC_1$-reg)\quad The Gauss hypergeometric equation
\begin{gather*}
-\frac12\frac{d^2}{dt^2}+\frac{C_0}{\sinh^{2}\lambda
t}+\frac{C_1}{\sinh^{2}2\lambda t} -C.
\end{gather*}
(Trig-$A_1$)\quad The Legendre equation
\begin{gather*}
-\frac12\frac{d^2}{dt^2}+\frac{C_0}{\sinh^{2}\lambda t}-C.
\end{gather*}
(Trig-$B_1$) with $C_0=C_1=C_3=0$.\quad The (Modif\/ied) Mathieu
equation
\begin{gather*}
-\frac12\frac{d^2}{dt^2}+C_2\cosh2\lambda t-C.
\end{gather*}
(Rat-$B_1$-2)\quad  Equation of the paraboloid of revolution
\begin{gather*}
-\frac12\frac{d^2}{dt^2}+\frac{C_0}{t^2}+{C_1}{t^2}-C.
\end{gather*}
This is the Weber equation if $C_0=0$. With $s=t^2$ and using the
unknown function $t^{\frac12}u$, the above equation is reduced to
the Whittaker equation:
\begin{gather*}
-\frac12\frac{d^2}{ds^2}+\frac{C_0'}{s^{2}}+\frac{C_1'}{s}-C'.
\end{gather*}
(Rat-$A_0$-bry2) with $C_2=C_3=0$:
\begin{gather*}
 -\frac12\frac{d^2}{dt^2}+C_0 t + C_1t^2-C.
\end{gather*}
If $C_1\ne0$, this is transformed into the Weber equation under
the coordinate $s= t+{C_2}/({2C_1})$. If $C_1=0$, this is the
Stokes equation which is reduced to the  Bessel equation. In
particular, the Airy equation corresponds to $C=C_1=0$.

\noindent (Toda-$BC_1$)
\begin{gather*}
-\frac12\frac{d^2}{dt^2}+C_0e^{-2t}+C_1e^{-4t}-C,
\end{gather*}
which is transformed into (Rat-$B_1$-2) by putting $s=e^{-t}$. In
particular

\noindent (Toda-$A_1$)
\begin{gather*}
-\frac12\frac{d^2}{dt^2}+C_0e^{-2t}-C
\end{gather*}
is reduced to the Bessel equation.

\noindent (Rat-$A_1$) the Bessel equation
\begin{gather*}
-\frac12\frac{d^2}{dt^2}+\frac{C_0}{t^2}-C.
\end{gather*}
In fact, the equation $-{u''}/2+{C_0u}/{t^2}=Cu$ is equivalent to
\begin{gather*}
 \biggr(\frac{d^2}{dt^2}+\frac1t\frac d{dt}
 -\frac{2C_0+1/4}{t^2}+2C\biggl)t^{-1/2}u=0
\end{gather*}
since
\begin{gather*}
t^{-\frac12}\circ\frac{d^2}{dt^2}\circ t^{\frac12}=
\frac{d^2}{dt^2}+\frac1t\frac d{dt}-\frac1{4t^2}.
\end{gather*}
Hence if $C\ne0$, the function $v=t^{-1/2}u$ satisf\/ies the
following Bessel equation with $s=\sqrt{-2C}t$:
\begin{gather*}
 \frac{d^2v}{ds^2}+\frac1s\frac{dv}{ds}-\biggl(1-\frac{C_0+1/8}{Cs^2}\biggr)v=0.
\\[2ex]
\text{\bf Hierarchy of ordinary dif\/ferential equations}
\\
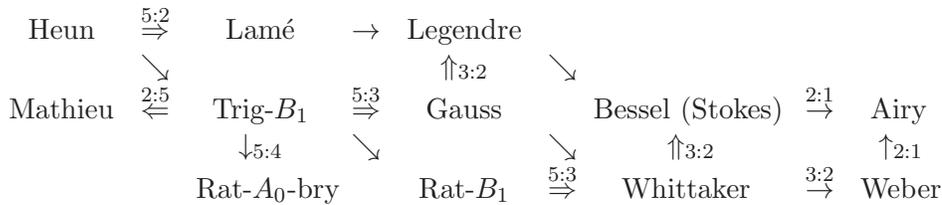
\begin{matrix}
 \text{\rm Heun}&\overset{5:2}\Rightarrow&\text{\rm Lam\'e}&\to&\text{\rm Legendre}
 \\
 &\searrow&&&\Uparrow_{3:2}&\searrow
 \\
 \text{\rm Mathieu}&\overset{2:5}\Leftarrow&\text{\rm Trig-$B_1$}&\overset{5:3}\Rightarrow&
 \text{\rm Gauss}&&\!\!\text{\rm Bessel (Stokes)}\!
 &\overset{2:1}\to&\!\text{\rm Airy}
 \\
 &&\downarrow_{5:4}&\searrow&&\searrow&\Uparrow_{3:2}&&\uparrow_{2:1}
 \\
 &&\text{\rm Rat-$A_0$-bry}\!\!\!
 &&\text{\rm Rat-$B_1$}&\overset{5:3}\Rightarrow
  &\!\!\text{\rm Whittaker}&\overset{3:2}\to&\text{\rm Weber}
\end{matrix}
\end{gather*}
\section{A classification}\label{sec:class}
We present a conjecture which characterizes the systems listed in
this note.
%%%%

Let $P$ be the Schr\"odinger operator with the expression
\eqref{eq:Shr} and consider the condition
\begin{gather}
\label{eq:conj} \text{there exist } P_1,\dots,P_n\text{ such that
}
\begin{cases}
P\in\mathbb C[P_1,\dots,P_n],
\\[1ex]
[P_i,P_j]=0\qquad(1\le i<j\le n),
\\[1ex]
\sigma(P_k)=\!\sum\limits_{1\le j_1<\dots<j_k\le
n}\!\!\xi_{j_1}^2\cdots\xi_{j_k}^2 \qquad(1\le k\le n).
\end{cases}
\end{gather}
Note that all the completely integrable systems given in
Sections~\ref{sec:An}, \ref{sec:B2} or \ref{sec:Bn} satisfy this
condition.

\smallskip

\noindent {\bf Conjecture.} \textit{Suppose $P$ satisfies
\eqref{eq:conj}. Under a suitable affine transformation of the
coordinate $x\in\mathbb C^n$ which keeps the algebra $\mathbb
C\Bigl[\sum\limits_{k=1}^n\partial_k^2,
\sum\limits_{k=1}^n\partial_k^4,\dots,\sum\limits_{k=1}^{n}\partial_k^{2n}\Bigr]$
invariant, $P$ is transformed into an integrable Schr\"odinger
operator studied in Sections~$\ref{sec:An}$, $\ref{sec:B2}$ or
$\ref{sec:Bn}$, (namely $u^\pm_{ij}$, $v_k$ and $w_k$ in~\eqref{eq:typeBn} are suitable analytic continuations of the
corresponding functions of the invariant elliptic systems) or in
general a {\it direct sum} of such operators and/or trivial
operators
\begin{gather*}
(A_1) \qquad\qquad \dfrac{d^2}{dx^2}+v(x)
\end{gather*}
with arbitrary functions $v(x)$ of one variable.}

\smallskip

Here the direct sum of the two operators
$P_j(x,\partial_x)=\sum\limits_{\alpha\in\mathbb\{0,1,\ldots\}^{n_j}}
a_\alpha(x)
\partial_x^\alpha$
of $x\in\mathbb C^{n_j}$ for $j=1,2$ means the operator
$P_1(x,\partial_x)+P_2(y,\partial_y)$ of $(x,y)\in\mathbb
C^{n_1+n_2}$.

We review known conditions assuring this conjecture and give
another condition (cf.~Theo\-rem~\ref{thm:Period} and
Remark~\ref{rem:periodic}). We also review related results on the
classif\/ication of completely integrable quantum systems
associated with classical root systems.
\begin{remark}
The condition
\begin{gather}
\label{eq:P2} \text{there exists $P_2$ such that $[P,P_2]=0$ and
$\sigma(P_2)=\sum\limits_{1\le i<j\le n}\xi_i^2\xi_j^2$}
\end{gather}
may be suf\/f\/icient to assure the claim of the conjecture.
\end{remark}
\begin{remark}[Type $\boldsymbol{A_2}$]\label{rem:A2}
{\samepage If $n=2$ and if there exists $P_3$ satisfying
\begin{gather*}
  \sigma(P_3)=\xi_1\xi_2+\xi_2\xi_3+\xi_3\xi_1 \quad\text{and}\quad
  [P,P_3]=[\partial_1+\partial_2+\partial_3,P]=[\partial_1+\partial_2+\partial_3,P_3]=0,
\end{gather*}
then Conjecture is true.

}

In fact this case is reduced to solving the equation
\begin{gather}
\label{eq:A2}
 \left|\begin{matrix}
  u(x) & u'(x) & 1\\[.5ex]
  v(y) & v'(y) & 1\\[.5ex]
  w(z) & w'(z) & 1
 \end{matrix}\right|=0\qquad\text{for}\qquad x+y+z=0
\end{gather}
for three unknown functions $u(t)$, $v(t)$ and $w(t)$, which is
solved by \cite{Braden, Buchstaber}. Here $u(t)=u_{e_1-e_2}(t)$,
$v(t)=u_{e_2-e_3}(t)$ and $w(t)=u_{e_1-e_3}(-t)$.
\end{remark}
\subsection{Pairwise interactions and meromorphy}
\begin{theorem}[\cite{Wakida}]
The potential function  $R(x)$ of $P$ satisfying \eqref{eq:conj}
is of the form
\begin{gather}
\label{eq:BnWa0}
 R(x) = \sum\limits_{\alpha\in\Sigma(B_n)^+}u_\alpha(\langle\alpha,x\rangle)
\end{gather}
with meromorphic functions $u_\alpha(t)$ of one variable.
\end{theorem}
\begin{remark} {\rm i)}
The condition \eqref{eq:P2} assures
\begin{gather*}
  R(x) = \sum\limits_{\alpha\in\Sigma(B_n)^+}u_\alpha(\langle\alpha,x\rangle)
         +\sum\limits_{1\le i<j<k\le n}C_{ijk}x_ix_jx_k
\end{gather*}
with $C_{ijk}\in\mathbb C$ and thus the above theorem is proved in
the invariant case (cf.~Section~\ref{sec:inv})  by~\cite{Oshima1995} or
in the case of Type $B_2$ by~\cite{Ochiai1996} or in the case of
Type $A_{n-1}$. This theorem is proved in~\cite{Wakida} by using
$[P,P_2]=[P,P_3]=0$.

{\rm ii)} Suppose $n=2$ and the operators
\begin{gather*}
 P  =-\dfrac12\biggl(\dfrac{\partial^2}{\partial x_1^2}+\dfrac{\partial^2}{\partial x_2^2}\biggr)+R(x_1,x_2),
 \\
T=\displaystyle\sum\limits_{i=0}^mc_j\dfrac{\partial^m}{\partial
x_1^i\partial x_2^{m-i}}
   +\sum\limits_{i+j\le m-2}T_{i,j}(x_1,x_2)\frac{\partial^{i+j}}{\partial x_1^i\partial x_2^j}
\end{gather*}
satisfy $[P,T]=0$ and $\sigma_m(T)\notin\mathbb C[\sigma(P)]$.
Here $c_j\in\mathbb C$. Then \cite[Theorem~8.1]{Oshima2007} shows
that there exist functions $u_{\nu,i}(t)$ of one variable such
that
\begin{gather*}
 R(x_1,x_2) = \sum\limits_{\nu=1}^L\sum\limits_{i=0}^{m_\nu-1}
 (b_\nu x_1 +a_\nu x_2)^i  u_{\nu,i}(a_\nu x_1-b_\nu x_2)
\end{gather*}
by putting
\begin{gather*}
 \biggl(\xi\frac{\partial}{\partial\tau}-\tau\frac{\partial}{\partial\xi}\biggr)
 \sum\limits_{i=0}^m c_i\xi^{m-i}\tau^i =
 \prod_{\nu=1}^{L}(a_\nu\xi-b_\nu\tau)^{m_\nu}.
\end{gather*}
Here $(a_\nu,b_\nu)\in\mathbb C^2\setminus\{(0,0)\}$ and $a_\nu
b_\mu\ne a_\mu b_\nu$ if $\mu\ne\nu$.
\end{remark}
\begin{definition}\label{def:support}
By the expression \eqref{eq:BnWa0}, put
\begin{gather*}
 S=\{\alpha\in\Sigma(B_n)^+\,;\, u_\alpha'\ne0\}
\end{gather*}
and let $W(S)$ be the Weyl group generated by
$\{w_\alpha\,;\,\alpha\in S\}$ and moreover put $\bar S=W(S)S$.
\end{definition}
\begin{theorem}[\cite{Ochiai1996} for Type $\boldsymbol{B_2}$, \cite{Wakida} in general]\label{thm:mero}
If the root system $\bar S$ has no irreducible component of rank
one, then \eqref{eq:P2} assures that any function $u_\alpha(t)$
extends to a meromorphic function on $\mathbb C$.
\end{theorem}
\begin{remark}[{\cite[(6.4)--(6.5)]{Oshima1995}, \cite[\S~3]{Wakida}}]\label{rem:Bncomp}
The condition \eqref{eq:P2} is equivalent to
\begin{gather}
\label{eq:bncomp} S^{ij}=S^{ji}\qquad(1\le i<j\le n)
\end{gather}
with
\begin{gather*}
S^{ij}  = \Biggl(\!\partial_i^2 v_i(x_i) + \sum\limits_{\nu\in
I(i,j)}\partial_i^2\bigl(u^+_{i\nu}(x_i+x_\nu)
 +u^-_{i\nu}(x_i-x_\nu)\bigr)\!\Biggr)(u^+_{ij}(x_i+x_j)-u^-_{ij}(x_i-x_j))
 \\ \phantom{S^{ij}  =}
{} + 3\Biggl(\!\!\partial_i v_i(x_i) + \!\!\!\sum\limits_{\nu\in
I(i,j)}\!\!\!
 \partial_i\bigl(u^+_{i\nu}(x_i+x_\nu)+u^-_{i\nu}(x_i-x_\nu)\bigr)\!\!\Biggr)
 (\partial_iu^+_{ij}(x_i+x_j)-\partial_iu^-_{ij}(x_i-x_j))
 \\ \phantom{S^{ij}  =}
 {}+ 2\Biggl(\!v_i(x_i) + \sum\limits_{\nu\in I(i,j)}(u^+_{i\nu}(x_i+x_\nu)+u^-_{i\nu}(x_i-x_\nu))\!\Biggr)
(\partial_i^2u^+_{ij}(x_i+x_j)-\partial_i^2u^-_{ij}(x_i-x_j))
\\ \phantom{S^{ij}  =}
{} + \sum\limits_{\nu\in
I(i,j)}(\partial_i^2u^+_{i\nu}(x_i+x_\nu)-\partial_i^2u^-_{i\nu}(x_i-x_\nu))
  (u^+_{j\nu}(x_j+x_\nu) - u^-_{j\nu}(x_j-x_\nu)).
\end{gather*}
Here $I(i,j)=\{1,2,\dots,n\}\setminus\{i,j\}$.
\end{remark}
\begin{lemma}\label{lem:decomp}
Suppose $P$ satisfies \eqref{eq:P2} and \eqref{eq:BnWa0}. Let
$S_0$ be a subset of $\bar S$ such that
\begin{gather*}
 S_0\subset\sum\limits_{i=1}^m\mathbb R e_i
 \qquad\text{and}\qquad
 \bar S\setminus S_0\subset\sum\limits_{i=m+1}^n\mathbb R e_i
\end{gather*}
with a suitable $m$. Then the Schr\"odinger operator
\begin{gather*}
   P'=-\frac12\sum\limits_{i=1}^m\partial_i^2
      +\sum\limits_{\alpha\in S_0\cap S}u(\langle\alpha,x\rangle)
\end{gather*}
on $\mathbb R^m$ admits a differential operator $P'_2$ on $\mathbb
R^m$ satisfying $[P',P'_2]=0$ and
$\sigma(P'_2)=\!\!\!\sum\limits_{1\le i<j\le
n}^m\!\!\!\xi_i^2\xi_j^2$, that is, the condition \eqref{eq:P2}
with replacing $P$ by  $P'$.
\end{lemma}
\begin{proof}
This lemma clearly follows from the equivalent condition
\eqref{eq:bncomp} given in Remark~\ref{rem:Bncomp}.
\end{proof}
\subsection{Invariant case}\label{sec:inv}
\begin{theorem}[\cite{Ochiai2003, Ochiai1994, Oshima1998, Oshima1995}]\label{thm:inv}
Assume that $P$ in \eqref{eq:Shr} is invariant under the Weyl
group $W=W(A_{n-1})$, $W(B_n)$ or $W(D_n)$ with $n\ge 3$, or
$W=W(B_2)$. If we have \eqref{eq:CI} with
\begin{gather*}
 P_1 = \partial_1+\partial_2+\cdots+ \partial_n\qquad \text{if}\qquad W=W(A_{n-1}),
 \\
 \sigma(P_k) = \begin{cases}
   \displaystyle
   \sum\limits_{1\le j_1<j_2<\cdots<j_k\le n}\xi_{j_1}\xi_{j_2}\cdots\xi_{j_k}
   \qquad\text{if}\qquad W=W(A_{n-1})\quad\text{and}\quad 1\le k\le n,
   \\[3ex]
   \displaystyle
   \sum\limits_{1\le j_1<j_2<\cdots<j_k\le n}\xi_{j_1}^2\xi_{j_2}^2\cdots\xi_{j_k}^2
   \qquad\text{if}\qquad W=W(B_n) \quad\text{and}\quad 1\le k\le n,
   \\[3ex]
   \displaystyle
   \sum\limits_{1\le j_1<j_2<\cdots<j_k\le n}\xi_{j_1}^2\xi_{j_2}^2\cdots\xi_{j_k}^2
   \qquad\text{if}\qquad W=W(D_n) \quad\text{and}\quad 1\le k< n,
\end{cases}\\[1ex]
\sigma(P_n) = \xi_1\xi_2\cdots\xi_n \qquad\text{if}\qquad
W=W(D_n),
\end{gather*}
Conjecture is true.
\end{theorem}
\begin{remark}
The condition
\begin{gather*}
  [P,P_1]=[P,P_3]= 0\qquad \text{if}\qquad W=W(A_{n-1}),
  \\
  [P,P_2]=0\qquad \text{if}\qquad W=W(B_n)\qquad \text{or}\qquad W=W(D_n)
\end{gather*}
together with \eqref{eq:BnWa0} is suf\/f\/icient for the proof of
this theorem.
\end{remark}
\subsection{Enough singularities}
Put $\Xi=\{\alpha\in\Sigma(B_n)^+\,;\,u_\alpha(t) \text{ is not
entire}\}.$
\begin{theorem}\label{thm:pole}
{\rm i)} {\rm (\cite{Ochiai1996})} Suppose $n=2$ and let $\bar S$
be of type $B_2$. If $\#\Xi\ge 2$, then Conjecture is true.

{\rm ii)} {\rm (\cite{Wakida})} If $\bar S$ is of type $A_{n-1}$
or of type $B_n$ and moreover the reflections $w_\alpha$ for
$\alpha\in\Xi$ generate $W(A_{n-1})$ or $W(B_n)$, respectively,
then Conjecture is true.
\end{theorem}
This theorem follows from the following key Lemma.
\begin{lemma}[\cite{Ochiai1996, Taniguchi, Wakida}]\label{lem:pole}
Suppose \eqref{eq:conj} and moreover that there exist $\alpha$ and
$\beta$ in $S$ such that $\alpha\ne\beta$,
$\langle\alpha,\beta\rangle\ne0$ and $u_{\alpha}(t)$ has a
singularity at $t=t_0$. Then $u_\alpha(t-t_0)$ is an even function
with a pole of order two at the origin and
\begin{gather}
  u_{w_\alpha(\beta)}\biggl(t-2t_0\dfrac{\langle\alpha,\gamma\rangle}{\langle\alpha,\alpha\rangle}\biggr)
   = u_\beta(t)\qquad \text{if}\qquad w_\alpha(\beta)\in\Sigma(B_n)^+,
\nonumber \\
  u_{-w_\alpha(\beta)}\biggl(-t+2t_0\dfrac{\langle\alpha,\gamma\rangle}{\langle\alpha,\alpha\rangle}\biggr)
  = u_\beta(t)\qquad \text{if}\qquad -w_\alpha(\beta)\in\Sigma(B_n)^+.\label{eq:pole}
\end{gather}
\end{lemma}
\begin{corollary}\label{cor:poles}
Suppose the assumption in Lemma~$\ref{lem:pole}$.

{\rm i)} If $u_\alpha(t)$ has another singularity at $t_1\ne t_0$,
then
\begin{gather}
\label{eq:poles}
  u_\gamma\biggl(t+ 2(t_1-t_0) \frac{\langle\alpha,\gamma\rangle}{\langle\alpha,\alpha\rangle}\biggr)
  = u_\gamma(t)\qquad\text{for}\qquad\gamma\in S.
\end{gather}

{\rm ii)} Assume that $u_\alpha$ has poles at $0$, $t_0$ and $t_1$
such that $t_0$ and $t_1$ are linearly independent over~$\mathbb
R$. Then $u_\beta(t)$ is a doubly periodic function and therefore
$u_\beta(t)$ has poles and hence $u_\alpha(t)$ is also a doubly
periodic function. We may moreover assume that $u_\beta$ has a
pole at $0$ by a parallel transformation of the variable $x$.

{\rm Case I:} Suppose $\alpha=e_i-e_j$, $\beta=e_j-e_k$ with $1\le
i<j<k\le n$.
\begin{gather*}
 u_{e_i-e_j}(t)=u_{e_j-e_k}(t)=u_{e_i-e_k}(t)=C\wp(t;2\omega_1,2\omega_2)+C'
\end{gather*}
with suitable $C,\,C'\in\mathbb C$, which corresponds to {\rm
(Ellip-$A_2$)}.

{\rm Case II:} Suppose $\alpha=e_i-e_j$ and $\beta=e_j$ with $1\le
i<j\le n$.

Then $\bigl(u_{e_i - e_j}(t), u_{e_i+e_j}(x), u_{e_i}(t),
u_{e_j}(t)\bigr)$ is
 {\rm (Ellip-$B_2$)}, {\rm (Ellip-$B_2$-S)} or {\rm (Ellip${}^d$-$B_2$)}.
\end{corollary}

{\rm iii)} If $\bar S$ is of type $A_{n-1}$ or $B_n$ or $D_n$ and
one of $u_\alpha(t)$ is a doubly periodic function with poles,
then $P$ transforms into (Ellip-$A_{n-1}$) or (Ellip-$B_n$) or
(Ellip-$D_n$) under a suitable parallel transformation on $\mathbb
C^n$.
\begin{proof}[Proof of Corollary~\ref{cor:poles}] i) is a direct consequence of Lemma~\ref{lem:pole}.
iii) follows from ii). We have only to show ii).

Case I: It follows from \eqref{eq:pole} that
$u_\alpha(t)=u_\beta(t)=u_{e_i-e_k}(t)$ and they are even
functions. Let
\begin{gather*}
  \Gamma_{2\omega_1,2\omega_2}=
   \{2m_1\omega_1+2m_2\omega_2\,;\,m_1,m_2\in\mathbb Z\}
\end{gather*} be the set of poles of $u_\alpha$.  Then \eqref{eq:poles} implies
$u_\beta(t+2\omega_1)=u_\beta(t+2\omega_2)=u_\beta(t)$. Since
$2\omega_1$ and $2\omega_2$ are periods of $\wp(t)$ and there
exists only one double pole in the fundamental domain def\/ined by
these periods, we have the claim.

Case II: It follows from \eqref{eq:pole} that
$u_{e_i-e_j}(t)=u_{e_i+e_j}(t)$ and $u_{e_i}(t)=u_{e_k}(t)$ and
they are even functions. Let $\Gamma_{2\omega_1,2\omega_2}$ be the
poles of $u_{e_i-e_j}(t)$. Then \eqref{eq:poles} means
$u_{e_i}(t+2\omega_1)=u_{e_i}(t+2\omega_2)=u_{e_i}(t)$.
Considering the poles of $u_{e_i-e_k}(t)$ with \eqref{eq:poles},
we have four possibilities of poles of $u_{e_i}$:
\begin{gather*}
 \text{(Case II-0): }\Gamma_{2\omega_1,2\omega_2},
 \\
 \text{(Case II-1): }\Gamma_{2\omega_1,2\omega_2}\textstyle\bigcup(\omega_1+\Gamma_{2\omega_1,2\omega_2}),
 \\
 \text{(Case II-2): }\Gamma_{2\omega_1,2\omega_2}\textstyle\bigcup(\omega_2+\Gamma_{2\omega_1,2\omega_2}),
 \\
 \text{(Case II-3): }\Gamma_{2\omega_1,2\omega_2} \textstyle\bigcup(\omega_1+\Gamma_{2\omega_1,2\omega_2})
  \textstyle\bigcup(\omega_2+\Gamma_{2\omega_1,2\omega_2}).
\end{gather*}
Here we note that (Case II-1) changes into (Case II-2) if we
exchange $\omega_1$ and $\omega_2$. Then we have
\begin{gather*}
 \text{(Case II-0): } u_{e_i-e_j}(t+4\omega_1)=u_{e_i-e_j}(t+4\omega_2) = u(t),
 \\
 \text{(Case II-2): } u_{e_i-e_j}(t+4\omega_1)=u_{e_i-e_j}(t+2\omega_2)= u(t),
 \\
 \text{(Case II-3): } u_{e_i-e_j}(t+2\omega_1)=u_{e_i-e_j}(t+2\omega_2) = u(t).
\end{gather*}
Thus (Case II-0), (Case II-2) and (Case II-3) are reduced to
(Ellip${}^d$-$B_2$), (Ellip-$B_2$-S) and \mbox{(Ellip-$B_2$)},
respectively.
\end{proof}

Let $\mathcal H$ be a f\/inite set of mutually non-parallel
vectors in $\mathbb R^n$ and suppose
\begin{gather*}
  P=-\frac12\sum\limits_{j=1}^n\partial_j^2+R(x),\qquad
  R(x) = \sum\limits_{\alpha\in\mathcal H}C_\alpha
    \frac{\langle\alpha,\alpha\rangle}{\langle\alpha,x\rangle^2}+\tilde R(x).
\end{gather*}
Here $C_\alpha$ are nonzero complex numbers and $\tilde R(x)$ is
real analytic at the origin. We assume that $\mathcal H$ is {\it
irreducible}, namely,
\begin{gather*}
 \mathbb R^n=\sum\limits_{\alpha\in\mathcal H}\mathbb R\alpha,
 \\
 \varnothing\ne\forall\, \mathcal H'\varsubsetneqq\mathcal H\Rightarrow
 \exists\,\alpha\in\mathcal H'\qquad \text{and}\qquad
 \exists\, \beta\in\mathcal H\setminus\mathcal H'\qquad \text{with}\qquad
 \langle\alpha,\beta\rangle\ne0.
\end{gather*}
\begin{definition}\label{def:irred}
The potential function $R(x)$ of a Schr\"odinger operator is {\it
reducible} if $R(x)$ and $\mathbb R^n$ is decomposed as
$R(x)=R_1(x)+R_2(x)$ and $\mathbb R^n=V_1\oplus V_2$ such that
\begin{gather*}
 0\subsetneq V_1\subsetneq\mathbb R^n,\quad V_2=V_1^\perp,\quad
 \partial_{v_2} R_1(x)=\partial_{v_1} R_2(x)=0\quad\text{for}\quad
 \forall \, v_2\in V_2\quad\text{and}\quad\forall\,  v_1\in V_1.
\end{gather*}
If $R(x)$ is not reducible, $R(x)$ is called to be irreducible.
\end{definition}
\begin{theorem}[\cite{Taniguchi}]\label{thm:Ta}
Suppose $n\ge2$ and there exists a differential operator $Q$ with
$[P,Q]=0$ whose principal symbol does not depend on $x$ and is not
a polynomial of $\sum\limits_{i=0}^n\xi_i^2$. Put
$W=\{w_\alpha;\alpha\in\mathcal H\}$. If
\begin{gather}
\label{eq:generic}
  2C_\alpha\ne k(k+1)\qquad \text{for}\qquad
 k\in\mathbb Z\qquad\text{and}\qquad\alpha\in\mathcal H,
\end{gather}
then $W$ is a finite reflection group and $\sigma(Q)$ is
$W$-invariant.
\end{theorem}
%%%
\subsection{Periodic potentials}
The following theorem is a generalization of the result in
\cite{Oshima2005}.
\begin{theorem}\label{thm:Period}
Assume $R(x)$ is of the form \eqref{eq:BnWa0} with meromorphic
functions $u_\alpha(t)$ on $\mathbb C$ and
\begin{gather}
\label{eq:Period}
 R\biggl(x+\dfrac{2\pi\sqrt{-1}\alpha}{\langle\alpha,\alpha\rangle}\biggr)=R(x)
 \qquad\text{for}\qquad\alpha\in\Sigma(B_n)
\end{gather}
and moreover assume that
\begin{gather*}
\text{the root system $\bar S$ does not contain an irreducible component of}\\
\text{type $B_2$ or even if $\bar S$ contains an irreducible
component}\\
\text{$\bar S_2=\{\pm e_i\pm e_j,\pm e_i,\pm e_j\}$ of type
$B_2$, the origin $s=0$ is not an isolated}\\
\text{essential singularity of $u_\alpha(\log s)$ for
$\alpha\in\bar S_2\textstyle\bigcap \Sigma(D_n)^+$.}
\end{gather*}
Then Conjecture is true.
\end{theorem}
%%%%%%%%%%%%%%%%
\begin{remark}\label{rem:periodic}
{\rm i)} The integrable systems classif\/ied in this note which
satisfy the assumption of Theorem~\ref{thm:Period} under a
suitable coordinate system are (Ellip-$*$) and (Trig-$*$) and
(Toda-$*$), which are the systems given in this note whose
potential functions are not rational.

{\rm ii)} The assumption \eqref{eq:Period} implies that
$u_\alpha(\log s)$ is a meromorphic function on $\mathbb
C\setminus\{0\}$ for any $\alpha\in\Sigma(B_n)^+$. It means that
the corresponding Schr\"odinger operator is naturally def\/ined on
the Cartan subgroup of $Sp(n,\mathbb C)$ with a meromorphic
potential function.
\end{remark}
%%%
\begin{lemma}\label{lem:noess}
Assume $n=2$, \eqref{eq:P2}, \eqref{eq:Period}, $\bar S$ is of
type $B_2$ and moreover $u_\alpha(\log s)$ are holomorphic for
$\alpha\in\Sigma(B_2)^+$ and $0<|s|\ll1$. If the origin is at most
a pole of $u_{\beta}(\log s)$ for $\beta\in\Sigma(D_2)^+$, the
origin is also at most a pole of $u_{\alpha}(\log s)$ for
$\alpha\in\Sigma(B_2)^+$.
\end{lemma}
\begin{proof}
Use the notation as in \eqref{eq:b2-1}. Put
\begin{gather*}
 u^-(\log s) =U_0^-+\sum\limits_{\nu=r}^\infty \nu U^-_\nu s^\nu,\qquad
 u^+(\log s) = U^+_0+\sum\limits_{\nu=m}^\infty \nu U^+_\nu s^\nu,
 \\
 v(\log s) = V_0+\sum\limits_{\nu=-\infty}^\infty \nu V_\nu s^\nu,\qquad
 w(\log s) = W_0+\sum\limits_{\nu=-\infty}^\infty \nu W_\nu s^\nu.
\end{gather*}
with $U^-_\nu,\,U^+_\nu,\,V_\nu,\, W_\nu\in\mathbb C$, $rm\ne 0$
and $(U^-_r,U^+_m)\ne 0$. Then as is shown in \cite{Oshima2005}
the condition for the existence of $T(x,y)$ in \eqref{eq:b2-2} is
equivalent to
\begin{gather}
 pq(2p-q)(p-q)(V_{2p-q}U^+_{q-p}+V_qU^-_{p-q}+W_{q-2p}U^+_p-W_qU^-_p)=0\qquad
\text{for}\quad p, q\in\mathbb Z.\!\!
\end{gather}
Hence if $p<r$ and $p<m$,
\begin{gather*}
  p(p-k)(p+k)k(V_{p+k}U^+_{-k}+V_{p-k}U^-_k)=0\qquad\text{for}\qquad k\in\mathbb Z.
\end{gather*}

Case $U^-_r\ne 0$:  Put $k=r$. Suppose $q$ is negative with a
suf\/f\/iciently large absolute value. Then
$V_{q}=(-{U^+_{-r}}/{U^-_r})V_{q+2r}$, which implies $V_q=0$ since
$\sum\limits_{\nu=-\infty}^\infty \nu V_\nu s^\nu$ converges for
$0<|s|\ll1$.

Suppose $q$ is negative with a suf\/f\/iciently large absolute
value compared to $p$. Then by the relation
$W_{q-2p}U^+_p-W_qU^-_p=0$ we similarly conclude $W_q=0$.

Case $U^+_m\ne 0$: Putting $k=-m$, we have the same conclusion as
above in the same way.
\end{proof}

\begin{proof}[Proof of Theorem~\ref{thm:Period}]
Lemma~\ref{lem:decomp} assures that we may assume $\bar S$ is an
irreducible root system. We may moreover assume that the rank of
$\bar S$ is greater than one.

Suppose that there exists $\gamma\in S$ such that the origin is
neither a removable singularity nor an isolated singularity of
$u_\gamma(\log s)$. Then $u_\gamma(t)$ is a doubly periodic
function with poles. Owing to Corollary~\ref{cor:poles},
$\sum\limits_{\alpha\in S_0}u_\alpha(\langle\alpha,x\rangle)$ is
reduced to the potential function of (Ellip-$A_{n-1}$) or
(Ellip-$B_n$) or (Ellip-$D_n$).

Thus we may assume that the origin is a removable singularity or
an isolated singularity of $u_\alpha(\log s)$ for any $\alpha\in
S$.

Let $\alpha,\beta\in S\textstyle\bigcap \Sigma(D_n)$ with
$\alpha\ne\beta$ and $\langle\alpha,\beta\rangle\ne0$. Put
$\gamma=w_{\alpha}\beta$ or $\gamma=-w_{\alpha}\beta$ so that
$\gamma\in\Sigma(D_n)^+$. Then \cite{Oshima2005} shows that
$u(t)=u_\alpha(t)$, $v(t)=u_\beta(t)$ and $w(t)=u_\gamma(-t)$
satisfy \eqref{eq:A2}. Then Remark~\ref{rem:A2} says that the
origin is at most a pole of $u(\log s)$, $v(\log s)$ and $w(\log
s)$.

Let $\alpha\in S\textstyle\bigcap \Sigma(D_n)$ and $\beta\in
S\setminus \Sigma(D_n)$ with $\langle\alpha,\beta\rangle\ne0$. Let
$W$ be the ref\/lection group generated by $w_\alpha$ and
$w_\beta$ and put
$S^o=W\{\alpha,\beta\}\textstyle\bigcap\Sigma(B_n)$. Then
\cite{Oshima2005} also shows that
\begin{gather*}
  R(x) = \sum\limits_{\gamma\in S^o}
         u_\gamma(\langle\gamma,x\rangle)
\end{gather*}
def\/ines an integrable potential function of type $B_2$. Hence
Lemma~\ref{lem:noess} assures that the origin is at most a pole of
$u_\alpha(\log s)$ for $\alpha\in S^o$.

Since $S$ is irreducible, the origin is at most a pole of
$u_\alpha(\log s)$ for $\alpha\in S$. Then
Theorem~\ref{thm:Period} follows from \cite{Oshima2005}.
\end{proof}

\subsection{Uniqueness}
We give some remarks on the operator which commutes with the
Schr\"odinger operator $P$.
\begin{remark}[{\cite[Lemma~3.1~ii)]{Oshima1995}}]
If dif\/ferential operators $Q$ and $Q'$ satisfy $[Q,Q']=0$,
$\sigma(Q')=\sum\limits_{j=1}^n\xi_j^N$ and $\mathrm{ord}(Q)\le
N-2$, then $Q$ has a constant principal symbol, that is,
$\sigma(Q)$ does not depend on $x$.

Hence if there exist dif\/ferential operators $Q_1,\ldots,Q_n$
with constant principal symbols such that
$\sigma(Q_1),\ldots,\sigma(Q_n)$ are algebraically independent and
moreover they satisfy $[Q_i,Q_j]=0$ for $1\le i<j\le n$, then any
operator $Q$ satisfying $[Q,Q_j]=0$ for $j=1,\dots,n$ has a
constant principal symbol. In particular, if a dif\/ferential
operator $Q$ satisf\/ies $[Q,P_k]=0$ for $P_k$ in \eqref{eq:CI}
and \eqref{eq:typeBnI} with $k=1,\dots,n$, then $\sigma(Q)$ does
not depend on $x$.
\end{remark}
\begin{remark}\label{rem:perioduniq}
Assume that a dif\/ferential operator $Q$ commutes with a
Schr\"odinger operator $P$ and moreover assume that there exist
linearly independent vectors $c_j\in\mathbb C^n$ for $j=1,\dots,n$
such that the operator is invariant under the parallel
transformations $x\mapsto x+c_j$ for $j=1,\dots,n$. Then
$\sigma(Q)$ does not depend on $x$
(cf.~\cite[Lemma~3.1~i)]{Oshima1995}).

Furthermore assume that $P$ is of type (Ellip-$F$) or (Trig-$F$)
or (Rat-$F$) with $F=A_{n-1}$ or~$B_n$ or $D_n$. If the condition
\eqref{eq:generic} holds or $Q$ is $W(F)$-invariant, it follows
from Theorem~\ref{thm:Ta} or \cite[Proposition~3.6]{Oshima1998}
that $Q$ is in the ring $\mathbb C[P_1,\ldots,P_n]$ generated by
the $W(F)$-invariant commuting dif\/ferential operators. If the
condition (8.11) is not valid, $\sigma(Q)$ is not necessarily
$W(F)$-invariant (cf.\ \cite{Chalykh1990, Sergeev, Veselov}).
\end{remark}
\begin{remark}[{\cite[Theorem~3.2]{Oshima1995}}]
Let $P$ be the Schr\"odinger operator in Theorem~\ref{thm:inv}.
Under the notation in Theorem~\ref{thm:inv} suppose $P_k$ are
$W$-invariant for $1\le k\le n$. Then the ring $\mathbb
C[P_1,\dots,P_n]$ is uniquely determined by $P$ and $Q$, where
$Q=P_3$ if $W=W(A_{n-1})$ and $Q=P_2$ if $W=W(B_n)$ or $W(D_n)$.
\end{remark}
\begin{remark} If $P_c=-(1/2)\sum\limits_{j=1}^n \partial_j^2+cR(x)$ is a
Schr\"odinger operator with a coupling constant $c\in\mathbb C$
such that $P_c$ admits a non-trivial commuting dif\/ferential
operator $Q_c$ of order four for any $c\in\mathbb C$, then the
operator $P_c$ may be a system stated in Conjecture under a
suitable coordinate system.

The following example satisf\/ies neither this condition nor the
condition \eqref{eq:generic}. It does not admit commuting
dif\/ferential operators \eqref{eq:CI} satisfying
\eqref{eq:typeBnI} if $m\ne0,\,-1$.
\end{remark}
\begin{example} It is shown in \cite{Chalykh1998, Sergeev} that the Schr\"odinger operator
\begin{gather*}
 P=-\frac12\sum\limits_{j=1}^n\frac{\partial^2}{\partial x_j^2}+\sum\limits_{1\le i<j<n}\frac{m(m+1)}{(x_i-x_j)^2}
   +\sum\limits_{i=1}^{n-1}\frac{m+1}{(x_i-\sqrt m x_n)^2}
\end{gather*}
is completely integrable for any $m$ and {\it algebraically
integrable} if $m$ is an integer.
\end{example}
The following example shows that the Schr\"odinger operator $P$
does not necessarily determine the commuting system $\mathbb
C[P_1,\ldots,P_n]$.
\begin{example}
Let $\alpha$, $\beta$, $\gamma$ and $\lambda$ be complex numbers.
Put $(A_0,A_1,C_0,C_1)=(\alpha,\gamma/2-\lambda/2,\beta,\lambda)$
for (Rat-$B_2$-S) in Theorem~\ref{thm:B2S}
(cf.~\cite[Remark~3.7]{Oshima1995}). Then the Schr\"odinger
operator
\begin{gather*}
  P_{\alpha,\beta,\gamma}
  = -\frac12\biggl(\frac{\partial^2}{\partial x^2}+\frac{\partial^2}{\partial y^2}\biggr)
    +(x^2+y^2)\biggl(\frac{2\alpha}{(x^2-y^2)^2} + \frac\beta{x^2y^2}+\gamma\biggr)
\end{gather*}
commutes with
\begin{gather*}
Q_{\alpha,\beta,\gamma,\lambda} =\biggl(\frac{\partial^2}{\partial
x\partial y}+\frac{4\alpha xy}{(x^2-y^2)^2}
 -2(\gamma-\lambda)xy\biggr)^2\!\!
-2\biggl(\frac{\beta}{y^2}+\lambda
y^2\biggr)\frac{\partial^2}{\partial x^2}
-2\biggl(\frac{\beta}{x^2}+\lambda
x^2\biggr)\frac{\partial^2}{\partial y^2}
\\ \phantom{Q_{\alpha,\beta,\gamma,\lambda}=}
{}+4\biggl(\frac{\beta}{x^2}+\lambda x^2\biggr)
\biggl(\frac{\beta}{y^2}+\lambda y^2\biggr) +\frac{16\alpha\lambda
x^2y^2+16\alpha\beta}{(x^2-y^2)^2} +8\lambda(\gamma-\lambda)x^2y^2
\end{gather*}
for any $\lambda\in\mathbb C$. Note that
$[Q_{\alpha,\beta,\gamma,\lambda},Q_{\alpha,\beta,\gamma,\lambda'}]\ne0$
if $\lambda\ne\lambda'$ and these operators are
$W(B_2)$-invariant. The half of the coef\/f\/icient of the term
$\lambda$ of $Q_{\alpha,\beta,\gamma,\lambda}$ considered as a
polynomial function of $\lambda$ is
\begin{equation*}
 S_{\alpha,\beta,\gamma}=
 -\biggl(y\frac\partial{\partial x}-x\frac{\partial}{\partial y}\biggr)^2
 +2\alpha\biggl(\frac{xy}{(x-y)^2}-\frac{xy}{(x+y)^2}\biggr)
 +2\beta\biggl(\frac{y^2}{x^2}+\frac{x^2}{y^2}\biggr)
 +4\gamma x^2y^2.
\end{equation*}
In particular,
$P=-(1/2)(\partial_x^2+\partial_y^2)+\gamma(x^2+y^2)$ commutes
with $\partial_x\partial_y-2\gamma xy$ and
$x\partial_y-y\partial_x$.

Note that if $R(x)$ is a polynomial function  on $\mathbb C^n$,
the condition
$\Bigl[-(1/2)\sum\limits_{j=1}^n\partial_j^2+R(x),Q\Bigr]=0$ for a
dif\/ferential operator $Q$ implies that the coef\/f\/icients of
$Q$ are polynomial functions (cf.\ \cite[Lemma~3.4]{Oshima1995}).
\end{example}
\subsection{Regular singularities}
\begin{definition}[\cite{Kashiwara}]
Put $\vartheta_k=t_k\partial/{\partial t_k}$ and
$Y_k=\{t=(t_1,\dots,t_n)\in\mathbb C^n\,;\, t_k=0\}$. Then
a~dif\/ferential operator $Q$ of the variable $t$ is said to have
{\it regular singularities}\/ along the set of walls
$\{Y_1,\dots,Y_n\}$ if
\begin{gather*}
  Q = q(\vartheta_1,\dots,\vartheta_n) + \sum\limits_{k=1}^n t_k Q_k(t,\vartheta).
\end{gather*}
Here $q$ is a polynomial of $n$ variables and $Q_k$ are
dif\/ferential operators with the form
\begin{gather*}
 Q_k(t,\vartheta) =
 \sum a_\alpha(t)\vartheta_1^{\alpha_1}\cdots\vartheta_n^{\alpha_n}
\end{gather*}
and $a_\alpha(t)$ are analytic at $t=0$. In this case we def\/ine
\begin{gather*}
 \sigma_*(Q)=q(\xi_1,\ldots,\xi_n)
\end{gather*}
and $\sigma_*(Q)$ is called the {\it indicial polynomial}\/ of
$Q$.
\end{definition}
\begin{theorem}
Let $R(t)$ be a  holomorphic function defined on a neighborhood of
the origin of~$\mathbb C^n$. Let $Q_1$ and $Q_2$ be differential
operators of $t$ which have regular singularities along the set of
walls $\{Y_1,\dots,Y_n\}$. Suppose $\sigma_*(Q_1)=\sigma_*(Q_2)$
and $[Q_1,P]=[Q_2,P]=0$ with the Schr\"odinger operator
\begin{gather*}
  P = -\frac12\Biggl(\vartheta_n^2
      +\sum\limits_{j=1}^{n-1}\bigl(\vartheta_{j+1}-\vartheta_j\bigr)^2\Biggr) + R(t).
\end{gather*}
Then $Q_1=Q_2$.
\end{theorem}
\begin{proof}
Put $t_j=e^{-(x_j-x_{j+1})}$ for $j=1,\ldots,n-1$ and
$t_n=e^{-x_n}$. Then $\partial_j=\vartheta_{j+1}-\vartheta_j$ for
$j=1,\ldots,n-1$ and $\partial_n=-\vartheta_n$. Under the
coordinate system $x=(x_1,\ldots,x_n)$ Remark~\ref{rem:perioduniq}
says that $Q_1-Q_2$ has a constant principal symbol, which implies
$Q_1=Q_2$ because $\sigma_*(Q_1-Q_2)=0$.
\end{proof}
A more general result than this theorem is given in
\cite{Oshima2007}. The following corollary is a direct consequence
of this theorem.
\begin{corollary}
Put $t_j=e^{-\lambda(x_j-x_{j+1})}$ for $j=1,\ldots,n-1$ and
$t_n=e^{-\lambda x_n}$. Suppose $P$ is the Schr\"odinger operator
of type {\rm(Trig-$A_{n-1}$), (Trig-$A_{n-1}$-bry-reg),
(Trig-$BC_n$-reg), (Trig-$D_n$), (Toda-$A_{n-1}$), (Toda-$BC_n$)}
or {\rm (Toda-$D_n$)}.

{\rm i)} $P$ and $P_k$ for $k=1,\dots,n$ have regular
singularities along the set of walls $\{Y_1,\ldots,Y_n\}$.

{\rm ii)} Let $Q$ be a differential operator which has regular
singularities along the set of walls $\{Y_1,\ldots,Y_n\}$ and
satisfies $[Q,P]=0$. If $\sigma_*(Q)=\sigma_*(\tilde Q)$ for an
operator $\tilde Q\in\mathbb C[P_1,\ldots,P_n]$, then $Q=\tilde
Q$.
\end{corollary}
\begin{remark}
{\rm i)} This corollary assures that certain radial parts of
invariant dif\/ferential operators on a symmetric space correspond
to our completely integrable systems with regular singularities
and the map $\sigma_*$ corresponds to the Harish-Chandra
isomorphism (cf.~\cite{Oshima2007}).

{\rm ii)} The system (Trig-$BC_n$-reg) is Heckman--Opdam's
hypergeometric system \cite{Heckman} of type $BC_n$. Since
(Trig-$BC_n$-reg) is a generalization of Gauss hypergeometric
system related to the root system $\Sigma(B_n)$, the systems in
the following diagram are considered to be generalizations of
Gauss hypergeometric system and its limits (cf.~Section~\ref{sec:ODE}).
They form a class whose eigenfunctions should be easier to be
analyzed than those of other systems in this note.
\end{remark}\vspace{-1ex}
\begin{gather*}
\text{\bf Hierarchy starting from (Trig-$\boldsymbol{BC_n}$-reg)}
\\
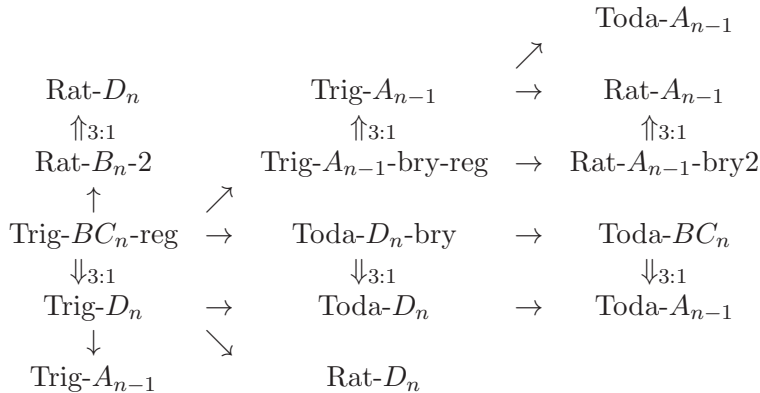
\begin{matrix}
 &&&&\text{\rm Toda-$A_{n-1}$}\\
 & &&\nearrow\\
 \text{\rm Rat-$D_n$}&&\text{\rm Trig-$A_{n-1}$}&\to&\text{\rm Rat-$A_{n-1}$}\\
 \Uparrow_{3:1}&&\Uparrow_{3:1}&&\Uparrow_{3:1}\\
 \text{\rm Rat-$B_n$-2}&&\text{\rm Trig-$A_{n-1}$-bry-reg}&\to&\text{\rm Rat-$A_{n-1}$-bry2}\\
 \uparrow&\nearrow\\
 \text{\rm Trig-$BC_n$-reg}&\to&\text{\rm Toda-$D_n$-bry}&\to &\text{\rm Toda-$BC_n$}\\
 \Downarrow_{3:1}&&\Downarrow_{3:1}&&\Downarrow_{3:1}\\
 \text{\rm Trig-$D_n$}&\to&\text{\rm Toda-$D_n$}&\to&\text{\rm Toda-$A_{n-1}$}\\
 \downarrow&\searrow\\
 \text{\rm Trig-$A_{n-1}$}&&\text{\rm Rat-$D_n$}\\
\end{matrix}
\end{gather*}

\subsection{Other forms}
If a Schr\"odinger operator $P$ is in the commutative algebra
$\mathbb D=\mathbb C[P_1,\ldots,P_n]$, then the dif\/ferential
operator $\tilde P:=\psi(x)^{-1}P\circ\psi(x)$ with a function
$\psi(x)$ is in the commutative algebra $\tilde{\mathbb D}=\mathbb
C[\psi(x)^{-1}P_1\circ\psi(x),
\ldots,\psi(x)^{-1}P_n\circ\psi(x)]$ of dif\/ferential operators.
Then
\begin{gather}
\tilde P = -\frac12\sum\limits_{j=1}^n\frac{\partial^2}{\partial
x_j^2} + \sum\limits_{j=1}^n a_j(x)\frac{\partial}{\partial
x_j}+\tilde R(x), \label{eq:Snsf}
\\
\frac{\partial\psi(x)}{\partial x_j}=a_j(x)\qquad\text{for}\qquad
j=1,\ldots,n. \label{eq:Sauto}
\end{gather}
Conversely, if a function $\psi(x)$ satisf\/ies \eqref{eq:Sauto}
for a dif\/ferential operator $\tilde P$ of the form
\eqref{eq:Snsf}, then $P=\psi(x)\tilde P\circ\psi(x)^{-1}$ is of
the form \eqref{eq:Shr}, which we have studied in this note.

If $\psi(x)$ is a function satisfying
\begin{gather*}
 \frac1{2\psi(x)}\sum\limits_{j=1}^n\frac{\partial^2 \psi}{\partial x_j^2}(x)=R(x),
\end{gather*}
then
\begin{gather*}
\tilde P = \psi(x)^{-1}\Biggl(-\frac12\sum\limits_{j=1}^n
\partial_j^2+R(x)\Biggr)\circ \psi(x)
 = -\frac12\sum\limits_{j=1}^n \partial_j^2-
  \psi(x)^{-1}\sum\limits_{j=1}^n \frac{\partial\psi}{\partial x_j}(x)\partial_j.
\end{gather*}

Note that
\begin{gather*}
 e^{-\phi(x)}\frac{\partial e^{\phi(x)}}{\partial x_j}
= \frac{\partial\phi(x)}{\partial x_j}, \qquad
 e^{-\phi(x)}\sum\limits_{j=1}^n\frac{\partial^2 e^{\phi(x)}}{\partial x_j^2}
=\sum\limits_{j=1}^n\frac{\partial^2\phi(x)}{\partial x_j^2}+
  \sum\limits_{j=1}^n\biggl(\frac{\partial\phi(x)}{\partial x_j}\biggr)^2.
\end{gather*}

Putting
\begin{gather*}
 \phi(x)=m\sum\limits_{1\le i<j\le n}
         \log \sinh \lambda(x_i-x_j),
\end{gather*}
we have
\begin{gather*}
  \frac{\partial\phi(x)}{\partial x_k}=
  \lambda m\sum\limits_{1\le i\le n,\, i\ne k}
  \coth\lambda(x_k-x_i),
\\
 \sum\limits_{j=1}^n\frac{\partial^2\phi(x)}{\partial x_j^2}+
 \sum\limits_{k=1}^n\Bigl(\frac{\partial\phi(x)}{\partial x_k}\Bigr)^2
  = -2\lambda^2 m\sum\limits_{1\le i<j\le n}\sinh^{-2}\lambda(x_i-x_j)
\\ \qquad
{}+2\lambda^2 m^2\sum\limits_{1\le i<j\le
n}\coth^2\lambda(x_i-x_j) +\lambda^2 m^2\frac{n(n-1)(n-2)}3
\\ \qquad
{} =2\lambda^2 m(m-1)\sum\limits_{1\le i<j\le
n}\sinh^{-2}\lambda(x_i-x_j)+\lambda^2m^2\frac{n(n^2-1)}3
\end{gather*}
since
\begin{gather*}
 \coth\alpha\cdot\coth\beta + \coth\beta\cdot\coth\gamma
 + \coth\gamma\cdot\coth\alpha = -1
 \qquad\text{if}\qquad\alpha+\beta+\gamma=0.
\end{gather*}
Hence
\begin{gather}
 \tilde P = -\frac12\sum\limits_{j=1}^n\partial_j^2 -
m\sum\limits_{1\le i<j \le
n}\lambda\coth\lambda(x_i-x_j)(\partial_i-\partial_j),\nonumber
\\
\psi(x) = \prod_{1\le i<j\le
n}\lambda^m\sinh^{m}\lambda(x_i-x_j),\nonumber
\\
 \psi(x)\circ\tilde P\circ\psi^{-1}(x) =
  -\frac12\sum\limits_{j=1}^n\partial_j^2 + \sum\limits_{1\le i<j\le n}
    \frac{m(m-1)\lambda^2}{\sinh^2\lambda(x_i-x_j)}+\frac{m^2n(n^2-1)\lambda^2}{6}
\label{eq:LapA}
\end{gather}
and $\tilde P$ is transformed into the Schr\"odinger operator of
type (Trig-$A_{n-1}$).

Now we put
\begin{gather*}
\phi(x)=m_0\sum\limits_{1\le i<j\le n} (\log \sinh
\lambda(x_i-x_j)+ \log \sinh \lambda(x_i+x_j))
+m_1\sum\limits_{1\le k\le n}\log \sinh \lambda x_k
\\ \phantom{\phi(x)=}
{}+m_2\sum\limits_{1\le k\le n}\log \sinh 2\lambda x_k
\end{gather*}
and we have
\begin{gather*}
\frac{\partial\phi(x)}{\partial x_k}= \lambda
m_0\!\!\!\!\sum\limits_{1\le i\le n,\, i\ne k}\!\!\!\!\!
(\coth\lambda(x_k+x_i)+\coth\lambda(x_k-x_i)) +\lambda m_1\coth
\lambda x_k +2\lambda m_2\coth2\lambda x_k,
\\
\coth \lambda x_k \coth2\lambda x_k =1+\frac12\sinh^{-2}\lambda
x_k,
\\
 \sum\limits_{\{i,j,k\}=I}(  2\coth\lambda(x_k+x_i)\coth\lambda(x_k-x_i)
 +2\coth\lambda(x_k+x_i)\coth\lambda(x_k-x_j)
 \\ \qquad
{}+\coth\lambda(x_k+x_i)\coth\lambda(x_k+x_j)+\coth\lambda(x_k-x_i)\coth\lambda(x_k-x_j))
 \\ \qquad
{}
=\sum\limits_{\{i,j,k\}=I}(\coth\lambda(x_k+x_i)\coth\lambda(x_k+x_j)
+\coth\lambda(x_i-x_j)\coth\lambda(x_i+x_k)
\\ \qquad
{} +\coth\lambda(x_j-x_i)\coth\lambda(x_j+x_k))
 +\sum\limits_{\{i,j,k\}=I}(\coth\lambda(x_k-x_i)\coth\lambda(x_k-x_j))=8
\\
\text{for}\quad I\subset\{1,\ldots,n\}\quad\text{with}\quad\#I=3,
\\
\coth\lambda(x_k+x_i)+\coth\lambda(x_k-x_i) =\frac{\sinh 2\lambda
x_k}{\sinh\lambda(x_k+x_i)\sinh\lambda(x_k-x_i)},
\\
\frac{\cosh2\lambda x_k-\cosh 2\lambda
x_i}{\sinh\lambda(x_k+x_i)\sinh\lambda(x_k-x_i)}
 =\frac{2\cosh^2\lambda x_k-2\cosh^2\lambda x_i}{\sinh\lambda(x_k+x_i)\sinh\lambda(x_k-x_i)}=2,
\\
\sum\limits_{k=1}^n\biggl(\frac{\partial\phi(x)}{\partial
x_k}\biggr)^2
 = 2\lambda^2 m_0^2\sum\limits_{1\le i<j\le n}
(\coth^2\lambda(x_i-x_j)+\coth^2\lambda(x_i+x_j))
\\ \qquad
{}+ \lambda^2 m_1^2\sum\limits_{k=1}^n \coth^2\lambda x_k
   + 4\lambda^2 m_2^2\sum\limits_{k=1}^n \coth^22\lambda x_k
   + 2\lambda^2 m_1m_2\sum\limits_{k=1}^n \sinh^{-2}\lambda x_k
\\ \qquad
{}+\frac{4\lambda^2m_0^2n(n-1)(n-2)}{3}
   + 2\lambda^2m_0(m_1+2m_2)n(n-1)+4\lambda^2m_1m_2n,
\\
\sum\limits_{j=1}^n\frac{\partial^2\phi(x)}{\partial x_j^2}
 + \!\sum\limits_{k=1}^n\biggl(\frac{\partial\phi(x)}{\partial x_k}\biggr)^2
\!\!= 2\lambda^2 m_0(m_0-1)\!\!\!\!\sum\limits_{1\le i<j\le
n}\!\!\!\!
 (\sinh^{-2}\lambda(x_i\!-\!x_j) + \sinh^{-2}\lambda(x_i\!+\!x_j))
\\ \qquad
{} +\lambda^2 m_1(m_1+2m_2-1)\sum\limits_{k=1}^n \sinh^{-2}\lambda
x_k
 +4\lambda^2 m_2(m_2-1)\sum\limits_{k=1}^n \sinh^{-2}2\lambda x_k
\\ \qquad
{}
+\lambda^2\biggl(\biggl(\frac23m_0(2n-1)+2m_1+4m_2\biggr)m_0(n-1)
+(m_1+2m_2)^2\biggr)n.
\end{gather*}
Hence
\begin{gather}
\tilde P = -\frac12\sum\limits_{j=1}^n\partial_j^2
-\sum\limits_{k=1}^n\lambda \Biggl(\sum\limits_{1\le i<j\le n}m_0
(\coth\lambda(x_i-x_k)+\coth\lambda(x_i+x_k))\nonumber
\\ \phantom{\tilde P =}
{}+m_1\coth\lambda x_k + 2m_2\coth2\lambda x_k\Biggr)\partial_k,
\nonumber
\\
\psi(x) = \prod_{1\le i<j\le
n}(\sinh^{m_0}\lambda(x_i-x_j)\sinh^{m_0}\lambda(x_i+x_j))
 \prod_{k=1}^n\sinh^{m_1}\lambda x_k\prod_{k=1}^n\sinh^{m_2}2\lambda x_k
\label{eq:LapB}
\end{gather}
and $\tilde P$ is transformed into the Schr\"odinger operator of
type (Trig-$BC_n$-reg):
\begin{gather*}
\psi(x)\circ\tilde P\circ\psi^{-1}(x) =
-\frac12\sum\limits_{j=1}^n\partial_j^2 +
m_0(m_0-1)\!\!\!\sum\limits_{1\le i<j\le n}\!\!
\biggl(\frac{\lambda^2}{\sinh^2\lambda(x_i\!-\!x_j)}\!+\!\frac{\lambda^2}{\sinh^2\lambda(x_i\!+\!x_j)}\biggr)
\\ \phantom{\psi(x)\circ\tilde P\circ\psi^{-1}(x)}
{} +\sum\limits_{k=1}^n
\frac{m_1(m_1+2m_2-1)\lambda^2}{2\sinh^2\lambda x_k}
 +\sum\limits_{k=1}^n \frac{2m_2(m_2-1)\lambda^2}{\sinh^22\lambda x_k}
\\ \phantom{\psi(x)\circ\tilde P\circ\psi^{-1}(x)}
{}+\lambda^2\biggl(\frac{m_0^2}3(2n-1)(n-1)+m_0(m_1+2m_2)(n-1)+
\frac{(m_1+2m_2)^2}2\biggr)n.
\end{gather*}
\begin{remark}
As is shown in \cite[Theorem~5.24 in Ch.\ II]{Helgason}, the
operator \eqref{eq:LapA} or \eqref{eq:LapB} gives the radial part
of the dif\/ferential equation satisf\/ied by the zonal spherical
function of a Riemannian symmetric space $G/K$ of the non-compact
type which corresponds to the Laplace--Beltrami operator on $G/K$.
Here $G$ is a real connected semisimple Lie group with a f\/inite
center, $K$ is a maximal compact subgroup of $G$ and the numbers
$2m$, $2m_0$, $2m_1$ and $2m_2$ correspond to the multiplicities
of the roots of the restricted root system for $G$.

Similarly the following operator $\tilde P$ is used to
characterize the $K$-f\/ixed Whittaker vector $v$ on
$G=GL(n,\mathbb R)$
\begin{gather*}
 \tilde P = -\frac12\sum\limits_{j=1}^n \partial_j^2 +
  \sum\limits_{j=1}^n \biggl(\frac{n+1}4-\frac j2\biggr)\partial_j
  +C \sum\limits_{j=1}^{n-1}e^{2(x_j-x_{j+1})},
  \qquad
 \psi(x) = e^{\sum\limits_{j=1}^n (j/2- {n+1}/4)x_j},
 \\
 \psi(x)\circ\tilde P\circ\psi^{-1}(x)=-\frac12\sum\limits_{j=1}^n \partial_j^2
  +C \sum\limits_{j=1}^{n-1}e^{2(x_j-x_{j+1})}  +\frac{n(n^2-1)}{48}.
\end{gather*}
Namely $v$ is a simultaneous eigenfunction of the invariant
dif\/ferential operators on $G/K$ and satisf\/ies
$v(nx)=\chi(n)v(x)$ with $n\in N$ and $x\in G/K$. Here $G=KAN$ is
an Iwasawa decomposition of $G$ and $\chi$ is a nonsingular
character of the nilpotent Lie group $N$. Then $v|_A$ is
a~simultaneous eigenfunction of the commuting algebra of
dif\/ferential operators  determined by~$\tilde P$.
\end{remark}
\pdfbookmark[1]{References}{ref}
\LastPageEnding

\end{document}